%% file: bias_allwise.tex
\documentclass[useAMS,usenatbib]{mn2e}
\usepackage{graphicx}
\usepackage{natbib}
\usepackage{lscape}
\usepackage{longtable}
\usepackage{subfig}
\usepackage{amssymb,amsmath}
\usepackage[T1]{fontenc}
\usepackage{hyperref}
\usepackage{amsmath}
\usepackage{breakurl}
\usepackage{times}
\usepackage{morefloats}
\usepackage{bm}
\usepackage{flafter}

\newcommand{\wise}{\textit{WISE}}

\newcommand{\planck}{\textit{Planck}}

\input{commands_mnras.tex}

\title[Updated IR-selected quasar bias]{Updated measurements of the dark matter halo masses of obscured quasars with improved \wise\ and \planck\ data}
\author[DiPompeo et al.]{M.A. DiPompeo$^{1,2}$, R.C. Hickox$^1$, A.D. Myers$^2$ \\
$^1$ Department of Physics and Astronomy, Dartmouth College, 6127 Wilder Laboratory, Hanover, NH 03755, USA  \\
$^2$ Department of Physics and Astronomy 3905, University of Wyoming, 1000 E. University, Laramie, WY 82071, USA}

\begin{document}
\date{Accepted ?; Received 2015 October 22; in original form 2015 October 5}

\pagerange{\pageref{firstpage}--\pageref{lastpage}} \pubyear{2015}

\maketitle

\label{firstpage}

\begin{abstract}
Using the most recent releases of \wise\ and \planck\ data, we perform updated measurements of the bias and typical dark matter halo mass of infrared-selected obscured and unobscured quasars, using the angular autocorrelation function and cosmic microwave background (CMB) lensing cross-correlations.  Since our recent work of this kind, the \wise\ Allwise catalogue was released with improved photometry, and the \planck\ mission was completed and released improved products.  These new data provide a more reliable measurement of the quasar bias and provide an opportunity to explore the role of changing survey pipelines in results downstream.  We present a comparison of IR color-selected quasars, split into obscured and unobscured populations based on optical-IR colors, selected from two versions of the \wise\ data.  Which combination of data is used impacts the final results, particularly for obscured quasars, both because of mitigation of some systematics and because the newer catalogue provides a slightly different sample. We show that Allwise data is superior in several ways, though there may be some systematic trends with Moon contamination that were not present in the previous catalogue.  We opt currently for the most conservative sample that meet our selection criteria in both the previous and new \wise\ catalogues.  We measure a higher bias and halo mass for obscured quasars ($b_{\textrm{obsc}} \sim 2.1$, $b_{\textrm{unob}} \sim 1.8$) --- at odds with simple orientation models --- but at a reduced significance ($\sim$1.5$\sigma$) as compared to our work with previous survey data. 
\end{abstract}


\section{INTRODUCTION}

Large astronomical surveys over a wide range of wavelengths have led to a dramatic increase in public data that is mined by the community for studies of all kinds, especially systematic studies of large samples. Such surveys are generally multi-year efforts, involving multiple data releases as more observations are carried out, sky coverage expanded, and reduction pipelines improved.  These changes can propagate through to the scientific results, and impact findings that need to be revisited.  Further, the way that survey data are handled --- samples selected, regions discarded/weighted, etc. --- can shift results in important ways \citep[e.g.][]{2002ApJ...579...48S, 2006ApJ...638..622M, 2006MNRAS.366..101H, 2011MNRAS.417.1350R, 2012ApJ...761...14H, 2015JCAP...05..040H, 2014JCAP...02..038A, 2014MNRAS.444....2L} 

Quasars, the luminous accreting supermassive black holes (SMBHs) in the nuclei of massive galaxies, are relatively rare objects seen primarily in the early Universe.  Because of this rarity, their study has seen rapid improvements with the dramatic increase in sample sizes from astronomical surveys such as (to name a few) the Large Bright Quasar Survey \citep[LBQS;][]{1995AJ....109.1498H}, Faint Images of the Radio Sky at Twenty Centimeters \citep[FIRST;][]{Becker:1995p345, 2015ApJ...801...26H}, the Sloan Digital Sky Survey \citep[SDSS;][]{2000AJ....120.1579Y}, and the 2dF QSO Redshift Survey \citep[2QZ;][]{2004MNRAS.349.1397C}. Quasars are a key component to studying the growth of black holes over cosmic time \citep[e.g.][]{1994ApJS...95....1E, Richards:2006p3932, 2012NewAR..56...93A}, as well as probing the potential links between those periods of growth, quasar host galaxies, and parent dark matter haloes \citep[e.g.][]{2008ApJS..175..356H, 2010MNRAS.405L...1B, 2011ApJ...737...50V, 2015arXiv150200775S}.

Large samples of both spectroscopic and photometric quasars have permitted analysis of their large-scale distribution and clustering, which reflects their distribution in the underlying dark matter density field (the quasar ``bias'', $b_q$) and provides insight into the masses of their dark matter haloes.  Studies of optically bright (unobscured, or type 1)\footnote{Note that throughout this paper type 1 and type 2 do not refer to spectral classifications based on broad and narrow emission lines, but optical-to-infrared colors as described in more detail in section 2.2} quasars have revealed that they tend to reside in haloes of similar mass ($\sim$3$\times 10^{12} h^{-1}$ M$_{\odot}$) across all redshifts \citep{2004MNRAS.355.1010P, 2005MNRAS.356..415C, 2007ApJ...654..115C, 2007ApJ...658...85M, 2008MNRAS.383..565D, 2009MNRAS.397.1862P, 2009ApJ...697.1634R, 2010ApJ...713..558K, 2012MNRAS.424..933W, 2013ApJ...778...98S, 2015arXiv150708380E}.  This suggests a link between black hole fueling and the growth of large-scale structure.

The dark matter haloes of quasars not only impact their spacial distribution, but deflect the photons of the cosmic microwave background (CMB, which backlights the whole sky) traveling past them via gravitational lensing.  Full-sky maps of the CMB have been steadily improving in depth and resolution, with the current state-of-the-art data being provided by the \planck\ satellite \citep{2011A&A...536A...1P}. The lensing signature of large scale structure has now been detected in numerous studies \citep{2011PhRvL.107b1301D, 2012ApJ...756..142V, 2014A&A...571A..17P, PlanckCollaboration:2015tp}.  Combined with estimates of the intrinsic CMB power spectrum \citep{1999PhRvL..82.2636S, 2001ApJ...557L..79H}, lensing maps of the CMB can trace the projected mass along a given line of sight back to the surface of last scattering at $z \sim 1100$.  

CMB lensing measurements are a particularly powerful tool for studying the haloes of quasars, which peak in number density at $z \sim 2$ \citep{2004MNRAS.349.1397C, 2005MNRAS.360..839R, 2006AJ....131.1203F}, coinciding with the peak of the CMB lensing kernel.  Additionally, CMB lensing measurements are subject to different systematics than clustering, providing independent follow up to such studies.  The first significant detection of a cross-correlation of unobscured quasars and the CMB lensing convergence found a typical halo mass in agreement with clustering results \citep{2012PhRvD..86h3006S}.

However, a subset of the quasar population has remained largely hidden from study because their optical (and in more extreme cases even X-ray) light is dramatically diminished by intervening gas and dust.  The existence of these obscured (type 2) quasars has been known for some time \citep{1989A&A...224L..21S, 1995A&A...296....1C}, but only recently have large IR datasets from \textit{Spitzer} \citep{2004ApJS..154....1W} and the \textit{Wide-Field Infrared Survey Explorer} \citep[\wise;][]{2010AJ....140.1868W} allowed detailed study of their demographics \citep{2004ApJS..154..166L, Stern:2005p2563, 2007ApJ...671.1365H, 2013MNRAS.434..941M, 2012ApJ...753...30S, 2013ApJ...772...26A, 2013ApJS..208...24L, 2015ApJ...804...27A, 2015arXiv150104118L}.  However, the nature of the obscuration in these sources is still unclear \citep[e.g.][]{2015ARA&A..53..365N}, with two prominent models being geometric obscuration by a dusty torus \citep[``unification by orientation'', well supported at low-$L$ and low-$z$; e.g.][]{1993ARA&A..31..473A} or larger, galaxy-scale obscuration \citep[e.g.][]{2012ApJ...755....5G} that may be a product of an evolutionary sequence \citep{1988ApJ...325...74S, 2008ApJS..175..356H, 2009MNRAS.394.1109C, 2010MNRAS.405L...1B}.

A simple test of orientation models for quasars is to compare their dark matter haloes.  If obscured quasars are simply unobscured quasars seen from a dustier line of sight, such as through a torus, then they should have the same halo mass, on average.  Some evolutionary scenarios, however, predict a difference in halo mass between the subclasses as the halo and black hole grow as a product of major galaxy mergers.  \citet{2011ApJ...731..117H}, \citet{2014ApJ...789...44D}, and \citet[][hereafter D14]{2014MNRAS.442.3443D} measured the halo masses of IR-selected quasar samples split into obscured and unobscured populations via their optical-IR colors \citep{2007ApJ...671.1365H}.  All found that obscured quasars seem to cluster more strongly, and thus reside in higher mass haloes, though the levels of significance varied considerably.  \citet[][hereafter D15]{2015MNRAS.446.3492D} followed up on D14 by cross-correlating the \wise-selected quasar samples with a \planck\ CMB lensing map, and found excellent agreement with the clustering results.

However, there have been other recent studies that suggest no difference in the bias and halo masses of obscured and unobscured quasars.  \citet{2013ApJ...776L..41G} cross-correlated \wise-selected quasars with a CMB map from the South Pole Telescope (and \planck\ as well), and found that the bias of obscured and unobscured quasars was roughly consistent.  \citet{2015arXiv150406284M} used a spectroscopic sample over a reduced area ($\sim$10$^{\circ}$) that benefits from individual source redshifts (instead of relying on an estimate of the ensemble average; see section 2.2) compiled from several fields and find no significant difference between obscured and unobscured halo masses.  While the additional redshift information assures that the evolution of the bias with $z$ does not skew the results, D15 illustrated that their mean redshift estimates would need to be offset by an unreasonably large amount to fully account for the difference they measured.

Clearly the relative halo masses of obscured and unobscured quasars is not yet conclusively determined, and so we follow up here on the work of D14 and D15 using updated data from both \wise\ and \planck.  We find that the measurements using the new and old data in various combinations can produce some significant variation in the results. Our goal here is to provide both a quantitative analysis of the difference between the samples selected from the original and new catalogues, as well as to identify the most reliable set of measurements.  We then provide an updated measurement of the obscured and unobscured quasar bias based on the best possible sample.

All models and measured properties use a cosmology of $H_0 = 70.2$ km s$^{-1}$ Mpc$^{-1}$, $\Omega_{\textrm{M}} = \Omega_{\textrm{CDM}} + \Omega_{\textrm{b}} = 0.229 + 0.046 = 0.275$, $\Omega_{\Lambda} = 0.725$, and $\sigma_8 = 0.82$ \citep{2011ApJS..192...18K}.

\section{DATA}
\subsection{\wise}
\subsubsection{Allsky and Allwise catalogues}
The \wise\ mission mapped the entire sky at 3.4, 4.6, 12, and 22 $\mu$m ($W1$, $W2$, $W3$, and $W4$) with angular resolutions of 6.1, 6.4, 6.5, and 12 arcsec, respectively \citep{2010AJ....140.1868W}.  The survey reached at least 0.08, 0.11, 1, and 6 mJy 5$\sigma$ point-source sensitivities in each band (in unconfused regions), with this depth increasing toward the ecliptic poles due to the observing strategy.  The \wise\ full cryogenic mission phase in 2010 resulted in the Allsky (AS) data release.  The AS source catalogue includes objects with SNR $>5$ in any band, at least five good measurements, and not flagged as spurious in at least one band\footnote{See the \wise\ Allsky Release Explanatory Supplement, \url{http://wise2.ipac.caltech.edu/docs/release/allsky/expsup/}}.

After both cryogen tanks were exhausted, The NEOWISE Post-Cryogenic Mission surveyed the entire sky again using the two shorter-wavelength bands, $W1$ and $W2$ \citep{2011ApJ...743..156M}.  Combining data from the original \wise\ survey with the NEOWISE data, along with improved reduction and calibration pipelines, led to the updated Allwise (AW) catalogue release, with improved photometric sensitivity and accuracy, better astrometric precision, and new information on source motions and variability\footnote{See the Allwise Release Explanatory Supplement, \url{http://wise2.ipac.caltech.edu/docs/release/allwise/expsup/}}.

In this work, our goal is to update the analyses of D14 and D15, which used the \wise\ AS catalogue to select quasars and measure their bias and host halo masses, by using the improved AW catalogue (sections 4.1 and 4.2).  We will also provide a detailed comparison of the objects selected as quasars from the two catalogues (section 4.4).

\subsubsection{Quasar selection}
\wise\ is ideal for identifying quasars via their characteristic hot dust emission, which causes a rising power-law spectrum in the mid-IR while stellar populations tend to peak around 1.5 $\mu$m \citep{2004ApJS..154..166L, Stern:2005p2563, 2012ApJ...753...30S, 2007ApJ...660..167D, 2013MNRAS.434..941M, 2013ApJ...772...26A}.  Critically, mid-IR selection is efficient at identifying both optically luminous unobscured \textit{and} optically faint obscured quasars, the latter of which may make up around half of the full quasar population \citep{2007ApJ...671.1365H, 2015ApJ...804...27A}.  

Various methods and criteria using mid-IR data have been developed to select quasars \citep[and lower luminosity active galactic nuclei, e.g.][]{2004ApJS..154..166L, Stern:2005p2563, 2012ApJ...748..142D, 2012MNRAS.426.3271M, 2015MNRAS.452.3124D, 2015arXiv150804472M}, and used in an array of studies of the IR-selected quasar population \citep[e.g.][]{2011ApJ...731..117H, 2014ApJ...783...40G, Smith:2014vd,  2014MNRAS.441.1297S, 2015MNRAS.451L..35E, 2015arXiv150406284M}.  However, even a simple color-cut of $W1-W2 > 0.8$ (along with $W2 < 15.05$, the 10$\sigma$ flux limit in this band, which helps reduce contamination from high-redshift star-forming galaxies as well as faint stars) identifies quasars at 80 per cent completeness and a contamination rate of only 5 per cent \citep{Stern:2005p2563, 2012ApJ...753...30S}.  These cuts have also been used to successfully study unobscured and obscured quasars \citep[][D14; D15]{2014ApJ...789...44D}, and we adopt them here.   \wise\ photometry is not corrected for Galactic extinction, as the rapidly dropping near-IR extinction curves of \citet{2009ApJ...699.1209F} show that this will affect \wise\ minimally, and we avoid the Galactic plane where extinction is more prevalent (see below).

To provide a direct comparison between quasar selection with AW and AS and the resulting effect on bias measurements, we restrict our sample to the same region as \citet{2014ApJ...789...44D}, D14, and D15: $135^{\circ} < \textrm{RA} < 226^{\circ}$ and $1^{\circ} < \textrm{Dec} < 54^{\circ}$.  This area is sufficiently far from the Galactic plane to limit high stellar densities and the majority of Galactic reddening, but is also not affected by depth changes and source confusion in \wise.  An extension beyond this footprint to utilize the full potential of millions of quasars in \wise\ \citep[e.g.][]{Secrest:2015tt} while properly handling the selection function across the whole sky is reserved for a future paper.

In the AW catalogue, 225,303 objects satisfy the selection criteria within this footprint.  This is a reduction of about 10\% from the AS catalogue, which had 250,163 objects satisfying these cuts.

\subsubsection{Cleaning the data}
\begin{figure}
\centering
\vspace{0.3cm}
\hspace{0cm}
   \includegraphics[width=6.5cm]{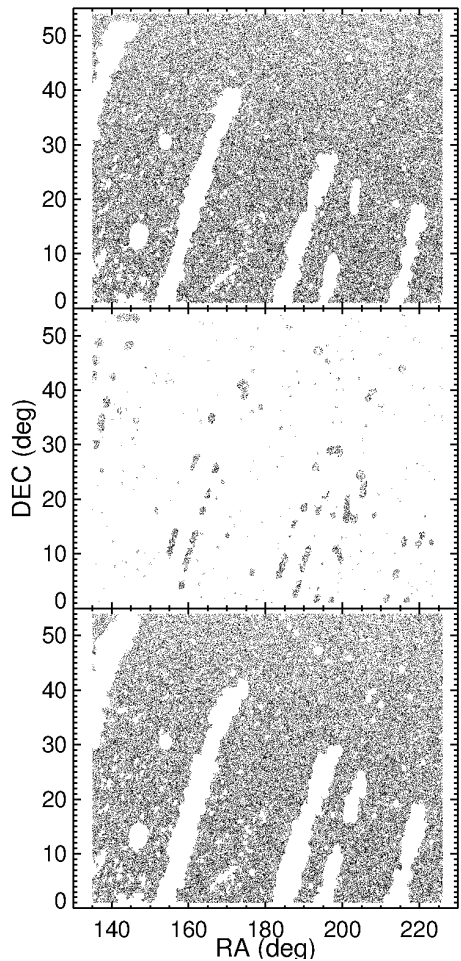}
    \vspace{0cm}
  \caption{\emph{Top:} The distribution on the sky of Allwise-selected quasars using the Allwise mask. \emph{Center:} Allwise-selected quasars that fall within the mask generated for the Allsky data. \emph{Bottom:} The distribution on the sky of Allwise-selected quasars after applying both masks.\label{fig:fields}}
\vspace{0.2cm}
\end{figure}

An accurate data mask is necessary to properly handle the normalization of the angular autocorrelation function by comparison with a randomly distributed sample, as well as to remove pixels from the \planck\ maps where quasar data are discarded.  D14 highlighted the importance of proper masking of the \wise\ data, and by more conservatively removing regions around flagged \wise\ data found a notable drop in the IR-selected obscured quasar bias compared to \citet{2014ApJ...789...44D}.  We develop a similar mask for the AW data here\footnote{This data mask, along with several other files including our samples, is available as a \textsc{mangle} polygon file at \url{http://faraday.uwyo.edu/~admyers/wisemask2015/wisemask.html}}, using the spherical cap utility \textsc{mangle} \citep{2004MNRAS.349..115H, 2008MNRAS.387.1391S}.  Full details of these components can be found in D14, unless a change is mentioned here:

\begin{enumerate}
\item Regions with high Galactic extinction in the $g$-band ($A_g > 0.18$).

\item \wise\ Atlas tiles (the main imaging product of \wise) with contamination from the Moon.  We mask tiles with {\tt moon\_lev} $>1$ in $W4$.  Using $W4$ makes the mask more conservative, as the longer wavelength bands can be affected by scattered light as far away from the Moon as $\sim$30$^{\circ}$, while the shorter bands used for selection can be affected to $\sim$10$^{\circ}$.  The new values of {\tt moon\_lev} in the updated Atlas tiles are used (and are a major source of difference between the two masks; see Figure~\ref{fig:fields}). 

\item Regions around highly grouped sources in the AW flagged data ({\tt cc\_flags} $\ne 0$ in $W1$ or $W2$, {\tt ext\_flag} $\ne 0$, or {\tt n\_b} $>2$; see Figure 1 of D14).  

\item Regions with poor photometric quality based on the \wise\ {\tt ph\_qual} flags, which is not included in the previous AS mask.  We pixelize the sky using \textsc{healpix}\footnote{\url{http://healpix.jpl.nasa.gov}} \citep{2005ApJ...622..759G} with $n_{\textrm{side}} = 64$ (pixel areas of $\sim$0.8 deg$^2$), and mask any pixel that has more than one object with photometric quality not set to `A' (SNR $\ge 10$) in $W1$ and $W2$.  This procedure is tuned based on the \wise\ data over the full sky, primarily to remove prominent strips of low-quality \wise\ data, but does remove some regions in the footprint considered here.  We do not discard other objects with lower quality, but these only make up a small fraction of our sample ($<1$\%), due to the $W2 < 15.05$ cut.  

\item The SDSS bright star mask, which masks circular regions around bright stars from the Yale and Tycho-2 bright star catalogues \citep[][]{1987BAAS...19..733W, 2000A&A...355L..27H}.  The vast majority of bright stars are already masked well by the \wise\ flagged data mask in (iii).

\end{enumerate}

The final region, after all masking is completed, has an area of 3422 deg$^2$, and contains 175,911 quasars from the AW catalogue (Table~\ref{tbl:samples}).  The sample distribution on the sky using this mask is shown in the top panel of Figure~\ref{fig:fields}.  The usable area and sample size is similar to what was found based on the AS sample (3,338 deg$^2$, 180,606 quasars), but is distributed somewhat differently on the sky --- the middle panel of Figure~\ref{fig:fields} shows the objects that are not masked by the AW mask, but fall within the AS mask of D14.  The majority of the difference is due to the Moon level between the two catalogues, as it is clear that the strips contaminated by the Moon in AS are wider than in AW.  

We explore the impact of our masks on our bias measurements by applying various combinations to the data, including the conservative cases of applying \textit{both} the AW and AS masks.  These samples with both masks applied will be labeled with an asterisk ($*$) throughout, and the Allwise$*$ sample is shown in the bottom panel of Figure~\ref{fig:fields}.  The sample labeled ``both'' contains sources that satisfy our selection criteria in both the AW and AS catalogues, and these make up 92.5\% of the full AW$*$ sample.  Table~\ref{tbl:samples} summarizes these samples.

\subsection{Obscured and Unobscured Quasars}
\citet{2007ApJ...671.1365H} used multiwavelength data in the B\"{o}otes field to demonstrate that an optical-IR color cut at $R-$[4.5]$=6.1$ (Vega magnitudes) can robustly separate obscured and unobscured quasar populations.  \citet{2014ApJ...789...44D}, D14, and D15 used AS $W2$ and SDSS $r$-band fluxes in a similar way to study these populations, with $r-W2 = 6$ separating the samples.  We do the same using AW $W2$ magnitudes here.

The SDSS completely covers our footprint, and reaches 50\% completeness at $r=22.6$ \citep[][]{2000AJ....120.1579Y, 2009ApJS..182..543A}.  As in D14/D15, we utilize the $SDSS$ pipeline {\tt psfmag} values, to isolate as much as possible the contribution from the quasar, as compared to that from the host galaxy, in resolved sources.  D14 determined that the use of SDSS {\tt modelMags} did not affect results significantly, and given the general similarity in the morphologies of AW-selected quasars (see below) there is no reason to believe this will change here.  The SDSS $r$-band magnitudes are corrected for Galactic extinction using extinction values supplied with the SDSS data \citep[based on the dust maps of][]{1998ApJ...500..525S}, and converted from AB magnitudes to Vega using $m_{\textrm{r,AB}} = m_{\textrm{r,Vega}} + 0.16$ \citep{2007AJ....133..734B}.

For the obscured and unobscured samples (but not the total IR-selected sample) the SDSS bad fields mask\footnote{See the SDSS-III Data Model at \url{http://data.sdss3.org/datamodel/files/BOSS_LSS_REDUX/reject_mask/MASK.html}} is applied \citep[e.g.][]{2011ApJ...728..126W, 2012MNRAS.424..933W}.  This reduces the AW area to 3,387 deg$^2$ and the total number of sources to 173,834.  We match the AW-selected quasars to SDSS sources with a  radius of 2 arcseconds, only keeping objects with $15 < r < 25$, and find matches for 83\%, the same as was found for AS-selected quasars.  Objects without SDSS matches are placed in the obscured sample, resulting in 72,587 (41.9\%) obscured and 101,247 (58.1\%) unobscured quasars.  

Of the objects with an SDSS counterpart, 68\% are unresolved \citep[based on the {\tt objc\_type} keyword;][]{2002AJ....123..485S}, marginally higher than the 65\% in the AS-selected sample of D14 and significantly higher than the 55\% of \citet{2014ApJ...789...44D}. Note however that \citet{2014ApJ...789...44D} used deeper optical imaging in the COSMOS field to classify source morphologies, which partially explains the lower unresolved fraction. Broken down into subclasses, 36\% and 81\% of the obscured and unobscured sources are unresolved, respectively, broadly consistent with D14.  We note that the additional application of the AS mask to the AW data does not change these ratios.

\begin{figure}
\centering
\vspace{0.3cm}
\hspace{0cm}
%
%
%
   \includegraphics[width=7.cm]{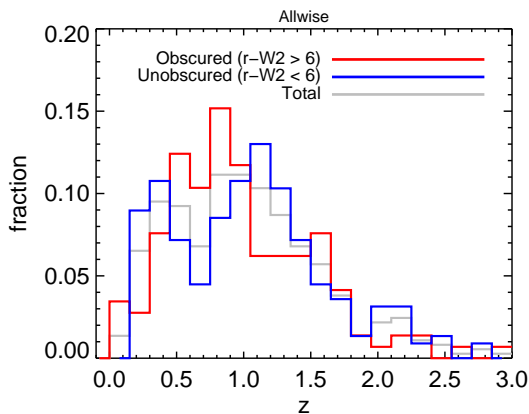}
    \vspace{0cm}
  \caption{Redshift distributions of Allwise selected quasars, generated by applying our mask and selection criteria to objects with multiwavelength and spectroscopic follow-up in the B\"{o}otes field.  These distributions are nearly identical as when using the Allsky catalogue data (D14, D15).\label{fig:z_dist}}
\vspace{0.2cm}
\end{figure}

As in D14/D15, to estimate the redshift distribution ($dN/dz$) of the AW-selected quasars, which is necessary to understand and interpret the bias measurements, we apply our mask and selection criteria to objects in the B\"{o}otes field.  This field has been observed in multiple photometric bands, and has extensive follow-up spectroscopy \citep{2006ApJ...651..791B, 2011ApJ...731..117H, 2012ApJS..200....8K}.  We find 368 with AW data that satisfy our selection in this field, with 145 (39.4\%) and 223 (60.6\%) obscured and unobscured, respectively.  These are consistent with the fractions in our overall sample.  All of the unobscured and all but two of the obscured samples have spectroscopic redshifts, and the distributions are shown in Figure~\ref{fig:z_dist}.  The mean/median/standard deviations of these distributions are 1.02/0.97/0.56 (total), 0.98/0.90/0.54 (obscured), and 1.05/1.04/0.58 (unobscured), consistent with what is found for AS-selected samples.  

We point out that the AW and AS mask do not significantly affect the B\"{o}otes region, and so potential differences in the redshift distribution of sources masked in one catalogue and not the other are not accounted for.  This will be analysed further in section 4.4.1.  Finally, it is worth noting that the AW and AS selected samples have very similar redshift distributions, despite differences in their average photometric properties (see sections 4.4.1 and 4.4.2).

\begin{table*}
  \caption{Summary of samples with various masks applied.}
  \label{tbl:samples}
  \begin{tabular}{lcccccc}
  \hline
                                      &  & Allsky                    & Allwise                & Allsky*                  & Allwise*                & Both                     \\
  \hline
  Area (\wise)                 &  & 3373                      &  3422                  &  3250                    &  3250                   &  3250          \\
  $N$ All IR                    &  & 180,860                 & 175,911               & 174,597               & 167,364               & 154,862                \\
  \\
  Area (\wise$+$SDSS) &  &  3339                     &  3387                   &  3216                   &  3216                   &  3216          \\
  $N$ Obscured             &  & 74,972 (42.1\%)    & 72,587 (41.9\%)   & 72,717 (42.1\%)  & 69,031 (41.7\%)  & 62,715 (41.4\%)    \\
  $N$ Unobscured         &  &  102,858 (57.9\%) & 101,247 (58.1\%) & 99,894 (57.9\%)  & 96,334 (58.3\%)  & 88,834 (58.6\%)   \\
  \\
                                      &  &                            \multicolumn{5}{c}{$+$CMB Lensing Mask}                \\          
                                                                                      \cline{3-7}                      
  Area (\wise)                 &  &  3300                   &  3189                   & 3032                  & 3032                     & 3032           \\
  $N$ All IR                    &  & 176,243               &  165,044              & 163,356             & 157,036               & 145,238                 \\
  \\
  Area (\wise$+$SDSS) &  &   3078                  &  3150                   &  2994                  &  2994                   &  2994          \\
  $N$ Obscured             &  & 70,093 (42.1\%)  & 67,935 (41.7\%)  & 67,839 (42.1\%)  & 64,637 (41.7\%)  & 58,688 (41.4\%)    \\
  $N$ Unobscured         &  &  96,401(57.9\%)  & 94,885 (58.3\%)  & 93,375 (57.9\%)  & 90,256 (58.3\%)  & 83,187 (58.6\%)    \\
\hline
   \end{tabular}
   \\  
{
\raggedright    
 Summary of the areas (in deg$^2$) and number $N$ of each subsample using the two \wise\ catalogues and various masks.  Note that when samples are split to obscured/unobscured, the additional SDSS bad fields mask is applied, reducing the sample size and area further.  An asterisk indicates a sample with the masks from both Allsky and Allwise applied, and ``both'' indicates only objects that meet the \citet{2012ApJ...753...30S} quasar selection criteria in both Allsky and Allwise.  The second half gives the same information but after also applying \planck\ mask and discarding any partially masked HEALPix $n_{\textrm{side}}=2048$ pixels, to limit errors in estimating the used area of partial pixels in the quasar density calculation.  Obscured/unobscured percentages are given in the relevant rows --- the fractions are very consistent across the samples.\\
 }
\end{table*}

\subsection{\planck\ CMB Lensing Maps}
The \planck\ mission \citep{2011A&A...536A...1P} mapped the CMB at nine frequencies, from 30 to 857 GHz, over the entire sky.  The first data release (DR1) in March 2013 (with an update in December 2013) was based on the nominal mission data with 15 months of observations \citep{2014A&A...571A...1P} and included a lensing potential ($\phi$) map with a lensing signature detected at an overall significance $>$25$\sigma$ \citep{2014A&A...571A..17P}.  This map was cross-correlated with the quasar density by \citet{2013ApJ...776L..41G} and D15.

In 2015 a second data set (DR2), using four years of data and improved reduction and calibration pipelines, was released \citep{PlanckCollaboration:2015ue}.  This release included an updated map of the lensing potential, with a lensing signal detected at $>$40$\sigma$ \citep{PlanckCollaboration:2015tp}.  These DR2 data are the highest quality all-sky CMB maps to date, with sensitivities down to $\mu$K and, at the frequencies where most of the lensing information is carried (143 GHz and 217 GHz), resolutions of 7 and 5 arcmin. 

\begin{figure}
\centering
\vspace{0.3cm}
\hspace{0cm}
   \includegraphics[width=7.5cm]{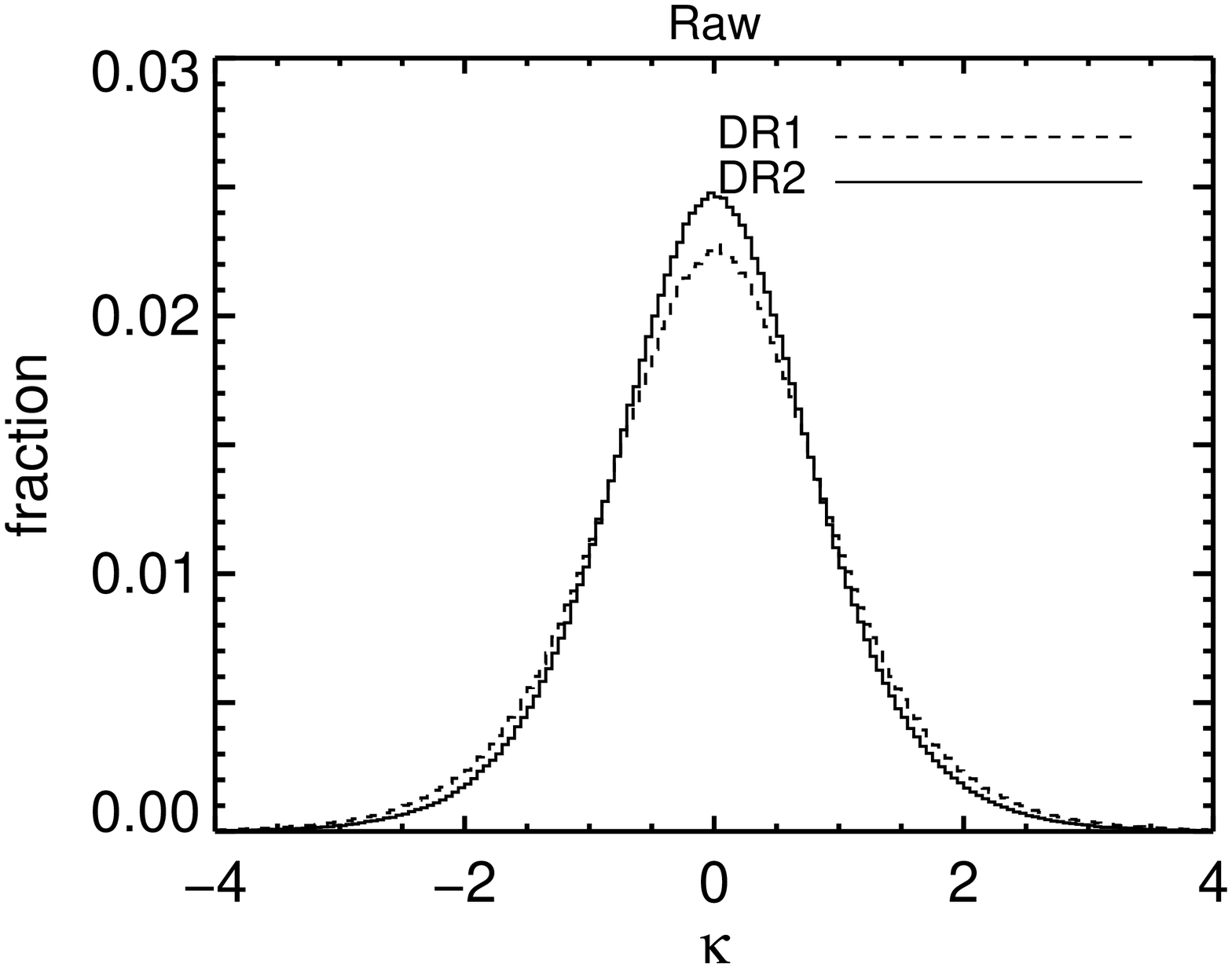}

   \vspace{0.15cm}

    \includegraphics[width=7.5cm]{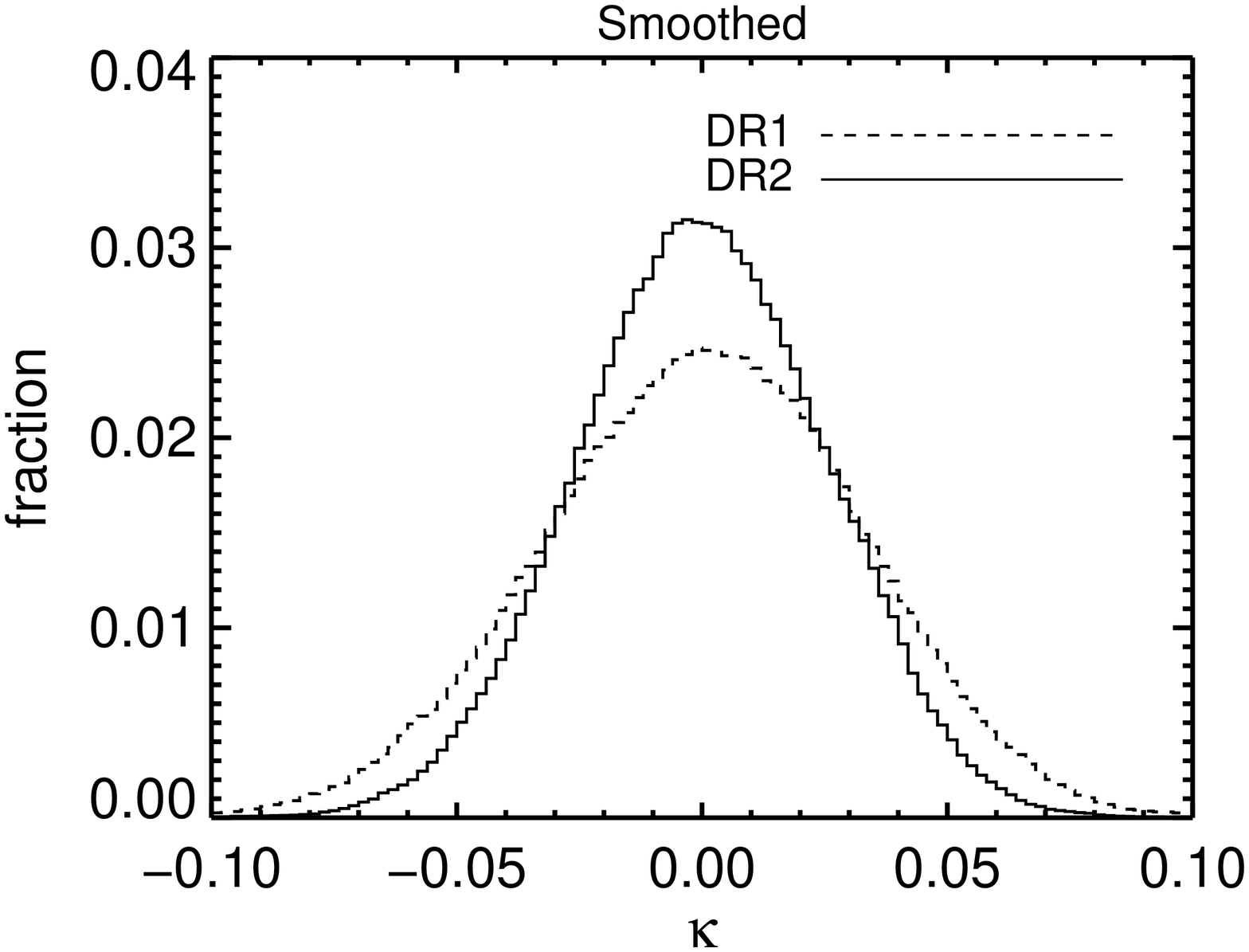}
    \vspace{0cm}
  \caption{Raw (top) and 1$^{\circ}$ Gaussian smoothed (bottom) distributions of $\kappa$ from \planck\ DR1 (dashed) and DR2 (solid). The DR2 distribution is narrower, and has a peak slightly more consistent with zero.\label{fig:kappas}}
\vspace{0.2cm}
\end{figure}

Using the \textsc{healpix} routine \textsc{isynfast}, we convert the \planck\ DR2 ``alm'' ($A_{lm}$) file\footnote{\url{http://irsa.ipac.caltech.edu/data/Planck/release_2/all-sky-maps/}}, which contains the coefficients of the spherical harmonic transform of the data on the sky, to a \textsc{healpix} map of the lensing convergence ($\kappa = \frac{1}{2} \Delta^2 \phi$) with $n_{\textrm{side}}=2048$, which has $\sim$3 arcmin$^2$ pixels.  In Figure~\ref{fig:kappas} we show the distribution of both the raw and 1$^{\circ}$ Gaussian-smoothed $\kappa$ in our region of interest.  The updated DR2 distribution is narrower, and has a peak marginally more consistent with zero, which is expected in the absence of systematic effects.

\section{METHODS}
Our goal is to measure the quasar bias and infer dark mater halo masses using the quasar data alone, as well as cross-correlate the quasar data with CMB lensing maps.  We also explore the \wise\ AS and AW catalogues for possible sources of contamination, to identify which combination of data and masks is the most reliable.  In this section we outline the formalisms for these measurements.

\subsection{Angular autocorrelations}
All of the codes used to calculate the angular autocorrelation functions, including models and bias fitting, are available at \url{https://github.com/mdipompe/angular_clustering}.

\subsubsection{Dark matter autocorrelation function theory}
We utilize the two-point angular (as opposed to real-space, as our sources lack individual redshift measurements) correlation function $\omega(\theta)$ to analyze how \wise\ quasars cluster around themselves.  The angular correlation function is related to the probability that a given pair of objects (dark matter haloes hosting quasars, in our case) with mean number density $n$, separated by a projected angular distance $\theta$, are within a solid angle $d\Omega$ \citep{1969PASJ...21..221T, Peebles:1980vn}:
\begin{equation}
   dP = n(1+\omega(\theta))d\Omega.
\end{equation}

Objects formed in the peaks of a Gaussian random field, such as massive, quasar-hosting galaxies, should cluster more strongly than the underlying typical dark matter distribution \citep{1984ApJ...284L...9K, 1986ApJ...304...15B}.  This excess, or bias, is independent of scale in most models, at least on large scales.  The bias of quasars $b_q$ is related to the underlying dark matter autocorrelation by $\omega_q(\theta) = \omega_{\textrm{dm}}(\theta) b_q^2$.

To calculate $\omega_{\textrm{dm}}(\theta)$, we first generate the non-linear matter power spectrum $P(k,z)$ using CAMB\footnote{\textit{Code for Anisotropies in the Microwave Background} (\url{http://lambda.gsfc.nasa.gov/toolbox/tb_camb_ov.cfm}).  We use our IDL wrapper for CAMB (\textsc{camb4idl}) which can be found at \url{https://github.com/mdipompe/camb4idl}.  Our initialization file for CAMB is available with the other autocorrelation codes linked above, as well as code to parse the CAMB output and generate the model autocorrelation.} \citep{2000ApJ...538..473L}.  We note that this procedure is updated from D14, for more consistency with the CMB lensing cross-correlation measurements.  

For angular scales $\theta << 1$ radian, as we probe here, we can project the matter power spectrum to an angular autocorrelation in a flat Universe via Limber's approximation \citep{1953ApJ...117..134L, Peebles:1980vn, 1991MNRAS.253P...1P}:
\begin{multline}
 \omega_{dm} (\theta) = \pi \int_{z=0}^{\infty} \int_{k=0}^{\infty} \frac{\Delta^2 (k,z)}{k} J_0 [k \theta \chi(z)] ~\times \\
    \left( \frac{dN}{dz} \right)^2 \left(\frac{dz}{d \chi} \right) \frac{dk}{k} dz.
\label{eq:omega_mod}
\end{multline}
In this equation, $\Delta^2 (k,z) = (k^3/2\pi^2)P(k,z)$ is the dimensionless power spectrum, $J_0$ is the zeroth-order Bessel function of the first kind, $\chi$ is the comoving distance along the line of sight, $dN/dz$ is the normalized redshift distribution (taken from a spline fit to the distribution of the appropriate subsample shown in Figure~\ref{fig:z_dist}; see section 5.2), and $dz/d\chi = H_z/c = (H_0/c)[\Omega_{\textrm{m}}(1+z)^3 + \Omega_{\Lambda}]^{1/2}$.  This model of $\omega_{\textrm{dm}}(\theta)$ can then be rescaled to fit the estimated quasar autocorrelation and measure the effective bias, or the bias integrated over our redshift range.  We will explore the effect of other bias models that evolve with $z$ in the discussion section.

\subsubsection{Estimating $\omega_{\textrm{qq}}(\theta)$}
We estimate the quasar autocorrelation $\omega_{\textrm{qq}}(\theta)$ by comparing the number counts of quasar pairs in annuli of increasing radii with what is expected for a random distribution \citep{1993ApJ...412...64L}:
\begin{equation}
   \omega_{\textrm{qq}}(\theta) = \frac{DD - 2DR + RR}{RR}.
   \label{eq:LS}
\end{equation}
In this estimator, $DD$, $DR$, and $RR$ are the normalized numbers of data-data, data-random, and random-random pairs in each bin of $\theta$ (i.e. $DD=DD(\theta)=N_{\textrm{data pairs}}/(N_DN_D)$).  The random sample must follow the same angular selection function as the data, which is simple in our case because the \wise\ selection is uniform over this field with holes described by our mask.  We generate a random catalogue that obeys our mask using the \textsc{mangle} function \textsc{ransack}.  The random catalogue for each measurement is always at least 10 times the size of the data set so that the random counts do not limit the statistical precision. We calculate $\omega_{\textrm{qq}}$ using four bins per dex, beginning at $\sim$12 arcsec and extending to 1.1$^{\circ}$.

Errors on the quasar autocorrelations are estimated using inverse-variance weighted jackknife resampling \citep[e.g.][]{2002ApJ...579...48S, 2005MNRAS.359..741M, 2007ApJ...658...85M}.  We divide our full footprint into $N=16$ equal-area regions, build $N$ subsamples by iteratively removing a single region, and repeating the autocorrelation measurement using the remaining regions.  Denoting each subsample by $L$, the inverse-variance-weighted covariance matrix $\textrm{\textsf{\textbf{C}}}_{ij} = \textrm{\textsf{\textbf{C}}}(\theta_i,\theta_j)$ ($i$ and $j$ denote angular size bins) is
\begin{multline}
\textrm{\textsf{\textbf{C}}}_{ij} = \sum_{L=1}^{N} \sqrt{\frac{RR_L(\theta_i)}{RR(\theta_i)}} [\omega_L(\theta_i) - \omega(\theta_i)] ~\times  \\
      \sqrt{\frac{RR_L(\theta_j)}{RR(\theta_j)}} [\omega_L(\theta_j) - \omega(\theta_j)],
\label{eq:jack}
\end{multline}
where $\omega$ is the angular autocorrelation for all of the quasars and $\omega_L$ is the the angular autocorrelation for subset $L$.  Note here that the $RR$ terms are not normalized by the sizes of the random samples, and account for the different number of counts in each region (though these are very close to the same, given the equal area of each region).  The jackknife errors $\sigma_i$ are taken from the square-root of the diagonal elements of the covariance matrix, though the full matrix is used when performing fits to measure the bias.

\subsection{Cross-correlations with systematics}
In an ideal data set, the quasar density will not correlate with any observational systematics, but in practice such effects can impact target selection.  Below we will cross-correlate \wise\ AS and AW selected quasars with various parameters, to quantify whether one catalogue is superior in this regard, using a pixelization method as in e.g.\ \citet{2002ApJ...579...48S}.  Due to the relatively low density of sources ($\sim$50 deg$^{-2}$, less for the obscured and unobscured sub-samples), we must use somewhat large pixels of 0.66 degrees on a side (0.44 deg$^2$ in area), and are thus limited to exploring cross-correlations on scales of this size or larger.  

Splitting up our region results in 9360 pixels, with an average of $\sim$20 sources per pixel.  The available area of each pixel is calculated by randomly populating it with 1000 sources, applying our mask, and multiplying the full area by the fraction of random points outside the mask.  Only pixels with at least 0.05 deg$^2$ of available area and at least five sources (one in the case of obscured and unobscured sub-samples) are used.

For each pixel $i$, the relative density of quasars is calculated: 
\begin{equation}
\delta_i^q = \frac{\rho_i^q - \langle \rho^q \rangle}{\langle \rho^q \rangle},
\end{equation}
where $\langle \rho^q \rangle$ is the mean quasar density for the whole field.  The relative value of systematics $\delta_i^s$, including $W1$ and $W2$ magnitudes, $W1-W2$ color, Galactic reddening $A_g$, and the \wise\ Moon level (recall that any region with Moon level $>1$ is already discarded) is calculated in a similar way for each pixel, using the values of these parameters at the location of each object.  Cross-correlations are then calculated as:
\begin{equation}
\omega_{qs}(\theta) = \frac{\sum_{i,j} \delta_i^q \delta_j^s \Theta_{ij}}{\sum_{i,j} \Theta_{ij}},
\label{eq:pix_cross}
\end{equation}
where $\Theta_{ij}=1$ if the separation between the centers of pixels $i$ and $j$ is within the bin $\theta$, and zero otherwise (the denominator is then a normalization equal to the number of pixel pairs satisfying this criteria).  The angular binning here is also four bins per dex, beginning at $0.6^{\circ}$ (the smallest scale possible due to the pixel size), and extending to $10^{\circ}$.  

Errors on the angular correlation functions using this method are also estimated via jackknife resampling.  In this case, the region is split into $N=25$ regions, and for a given iteration in the covariance matrix calculation (using Equation~\ref{eq:jack}) any pixel falling within this subregion is excluded.  The square root of the diagonal elements of the covariance matrix are adopted as the 1$\sigma$ errors.

We also use Equation 6 to rapidly calculate the quasar autocorrelation on large scales ($>$1$^{\circ}$), where the pair counting method becomes computationally expensive.

\subsection{CMB Lensing Correlations}
All of our code to work with the CMB lensing maps and perform the CMB cross-correlations with \textsc{healpix} maps can be found at \url{https://github.com/mdipompe/lensing_xcorr}.

\subsubsection{CMB lensing-matter cross-correlation theory}
The quasar bias $b_q$ can also be measured by comparing the cross-correlation of the quasar density with the CMB lensing convergence ($C_l^{\kappa q}$) with theoretical predictions for a given matter distribution.  This formalism is detailed fully elswhere \citep{2012ApJ...753L...9B, 2012PhRvD..86h3006S}, and we provide a brief summary here.

The lensing convergence ($\kappa$) in comoving coordinates ($\chi$) along a line of sight $\mathbf{\hat{n}}$ is the integral over the relative over-density of matter ($\delta(\mathbf{r},z)$) multiplied by the lensing kernel $W^{\kappa}$:

\begin{equation} \kappa(\hat{\mathbf n}) = \int d\chi
W^\kappa(\chi)\delta(\chi{\hat{\mathbf n}},z(\chi)). 
\end{equation}

\noindent The lensing kernel \citep{2000ApJ...534..533C, 2003ApJ...590..664S} is:

\begin{equation} W^\kappa(\chi) = \frac{3}{2}\Omega_{\rm
m}\left(\frac{H_0}{c}\right)^2\frac{\chi}{a(\chi)}\frac{\chi_{\rm CMB} - \chi}{\chi_{\rm CMB}},
\end{equation}
\noindent where $a(\chi) = (1+z(\chi))^{-1}$ is the scale factor, and $\chi_{\rm CMB}$ is the co-moving distance to the CMB.  Fluctuations in the quasar density are given by:

\begin{equation} q(\hat{\mathbf n}) = \int d\chi
W^q(\chi)\delta(\chi{\hat{\mathbf n}},z(\chi)), 
\end{equation}

\noindent where $W^q(\chi)$ is the quasar host distribution kernel:

\begin{equation} W^q(\chi) = \frac{dz}{d\chi}\frac{dN(z)}{dz} b_q(\chi).
\end{equation}

\noindent Here, $dN/dz$ is the normalized redshift distribution of the quasar population (again estimated using a spline interpolation; see section 5.2), which has bias $b_q$, assumed here to be independent of redshift. The cross-power at a Fourier mode $l$ is

\begin{equation} C^{\kappa q}_l = \int dz
\frac{d\chi}{dz}\frac{1}{\chi^2}W^\kappa(\chi)W^q(\chi)P\left(\frac{l}{\chi},z
\right) 
\label{eq:lens_model}
\end{equation}
\noindent where $P(k=l/\chi,z)$ is the matter power spectrum (e.g.\ Eisenstein \& Hu 1999), again generated using CAMB.  Equation~\ref{eq:lens_model} gives us the model cross-power spectrum for the underlying distribution of matter when the effective bias is unity (again we will explore other models of the bias in the discussion section).  We note that the current defaults of CAMB have been slightly updated since D15, and these result in a matter power spectrum with slightly higher amplitude that propagates to the final model, resulting in a lower bias measurement.  A CAMB parameter file is included with our supplied code to aid with consistent future measurements.

\subsubsection{Measuring CMB lensing auto and cross-correlations}
Leveraging the \textsc{healpix} format of the \planck\ maps, and the speed with which cross-correlations can be performed using spherical harmonic transforms, we employ routines in the \textsc{healpix} package to measure auto and cross-correlations with \planck\ data.  The cross-power $C_l^{\kappa X}$ of a $\kappa$ map ($M_{\kappa}$) with map $X$ ($M_X$, which could be another $\kappa$ map or a quasar density map as described below) is measured by taking the Fourier transform of each map and multiplying them:
\begin{equation} 
C_l^{\kappa X} = \left< \mathrm{Re}( \mathcal{F}({\sf
M_{\kappa}})\mathcal{F^*}({\sf M_X}))|_{\mathbf{l}\in l} \right>
\end{equation}
\noindent where $\mathbf{l}\in l$ describes the binning.  We present all results calculated this way with 5 bins in $l$ per dex, beginning at $l=10$.  

To cross-correlate the quasar density with the CMB lensing convergence ($C_l^{\kappa q}$), we first generate a \textsc{healpix} map of the relative quasar density $\delta$ at the same resolution as the \planck\ lensing convergence map ($n_{\textrm{side}} = 2048$).  This requires an estimate of the used area of each pixel, which we find by populating our footprint with 150 million points (for an average of $\sim$30 per pixel), applying our mask, and using the ratio of points inside and outside the mask per pixel.  This area estimation is of course subject to its own errors, and as discussed in D15 can impact the final measurement in several ways.  We therefore discard any pixel that overlaps a mask component, which generally removes less than 100 deg$^2$ (see Table~\ref{tbl:samples}).  The relative quasar density is then calculated with respect to the total number of quasars and area in the remaining complete pixels.

Uncertainties in $C_l^{\kappa X}$ are estimated by repeating the measurement with several rotations of the $\kappa$ map.  D15 illustrated the consistency of this method with others, such as substituting simulated noise maps.  We use 34 rotations, 17 in increments of 20$^{\circ}$ in Galactic longitude, and another 17 with an additional reflection in latitude about the Galactic equator.  We derive covariance matrices ($\textsf{\textbf{C}}(l_i,l_j) = \textsf{\textbf{C}}_{ij}$):
\begin{equation}
\textsf{\textbf{C}}_{ij} = \frac{1}{N-1} \left[ \sum_{k=1}^{N} (C_{l_i,k}^{\kappa X} - C_{l_i}^{\kappa X}) (C_{l_j,k}^{\kappa X} - C_{l_j}^{\kappa X}) \right],
\end{equation}
where $N$ is the number of rotated cross-correlations and $C_{l,k}^{\kappa X} - C_{l}^{\kappa X}$ is the cross-correlation from each rotation.  We adopt the square root of the diagonal elements of $\textsf{\textbf{C}}_{ij}$ as the 1$\sigma$ errors on these correlations.

\subsection{Fitting procedures}
To measure the quasar bias, we fit the models from Equations~\ref{eq:omega_mod} and~\ref{eq:lens_model} to the measured autocorrelations and CMB lensing cross-correlations.  We use the full covariance matrix to scale our given model $f_m(x)$ to the data $f(x)$ with a $\chi^2$ minimization:
\begin{equation}
\chi^2 = \sum_{i,j} [f(x_i) - f_m(x_i)] \textsf{\textbf{C}}_{i,j}^{-1} [f(x_j) - f_m(x_j)].
\end{equation}
Here, $f(x)$ may be $\omega(\theta)$ or $C_l^{\kappa q}$ and the sums are over bins of $\theta$ or $l$.  Both models are only a function of one parameter, the bias $b_q$, and so the errors on the fits are determined by where $\Delta \chi^2 = 1$.

On scales larger than $\sim$1 Mpc/$h$ the clustering of quasars and their parent haloes is sensitive only to the underlying density field and can be fit by the simple models described in the previous sections, while on smaller scales the halo occupation distribution (HOD) and the physics of galaxy formation and evolution become important \citep[e.g.][]{2002ApJ...575..587B, 2003ApJ...593....1B, 2012ApJ...755...30R, Krumpe:2013tv, 2015arXiv150708380E}. At $z =1$ (the approximate mean for all of our samples) this linear scale corresponds to $\sim$0.04$^{\circ}$.  Therefore, we restrict our fits on the autocorrelation to $0.04^{\circ} < \theta < 0.4^{\circ}$.  For CMB lensing cross-correlations we restrict our fits to $40 < l < 400$, which is above the peak of the model cross-correlation and below a possible correlated feature in the \planck\ DR2 lensing map \citep[which lies in the range $638 < l < 732$;][]{PlanckCollaboration:2015tp}.  This is also the range with the smallest errors on the cross-correlation (section 4.2).

\subsection{Dark matter halo masses}
All of the code used to convert biases to halo masses is provided at \url{https://github.com/mdipompe/halomasses}.

Once the quasar bias is determined, it can be converted into a typical dark matter halo mass ($M_h$) for a given sample.  This is done using the model fits to cosmological simulations of \citet{2010ApJ...724..878T}\footnote{Note that here we use the more recent \citet{2010ApJ...724..878T} model, while in D14/D14 we used the \citet{2001MNRAS.323....1S} model for more direct comparison with previous results. In the context of searching for differences in obscured and unobscured halo masses (the main goal of D14/D15), the choice of model is less critical; however, moving forward we prefer the most precise halo masses for future modeling purposes}:
\begin{equation}
b(\nu) = 1 - A \frac{\nu^a}{\nu^a + \delta_c^a} + B\nu^b + C\nu^c,
\label{eq:bias_mh}
\end{equation}
where $\nu = \delta_c/\sigma(M)$.  The numerator, $\delta_c$, is the critical density for collapse of a dark matter halo, and is defined for a Universe containing matter and a cosmological constant $\Lambda$ by:
\begin{equation}
\delta_c = 0.15(12\pi)^{2/3} \Omega_{m,z}^{0.0055},
\end{equation}
where $\Omega_{m,z}$ is the density parameter for matter at the redshift under consideration \citep{1997ApJ...490..493N}.  The denominator, $\sigma(M)$, is the linear matter variance at the size scale of the halo ($R_h = (3M_h/4\pi\bar{\rho}_m)^{1/3}$, with $\bar{\rho}_m$ the mean density of matter) calculated by:
\begin{equation}
\sigma^2(M) = \frac{1}{2\pi^2} \int P(k,z) \hat{W}^2(k,R_h)k^2 dk.
\end{equation}
We again use CAMB to obtain the linear $P(k,z)$, and $\hat{W}(k,R_h)$ is the Fourier transform of a top-hat window function of radius $R_h$:
\begin{equation}
\hat{W}(k,R_h)=\frac{3}{kR_h)^3}[\sin(kR_h)-kR_h\cos(kR_h)].
\end{equation}
The constants $A$, $a$, $B$, $b$, $C$, and $c$ for Equation~\ref{eq:bias_mh} are taken from Table 2 of \citet{2010ApJ...724..878T} and defined for the overdensity parameter $\Delta = 200$ ($y \equiv \log_{10} \Delta$):
\begin{gather}
A = 1.0+0.24y e^{-(4/y)^4} \nonumber \\
a = 0.44y - 0.88 \nonumber \\
B =0.183 \nonumber \\
b = 1.5 \nonumber \\
C = 0.019-0.107y+0.19e^{-(4/y)^4} \nonumber \\
c=2.4
\end{gather}

\section{RESULTS \& ANALYSIS}
\subsection{The quasar autocorrelation function}
\begin{figure*}
\centering
\vspace{0.3cm}
\hspace{0cm}
   \includegraphics[width=16cm]{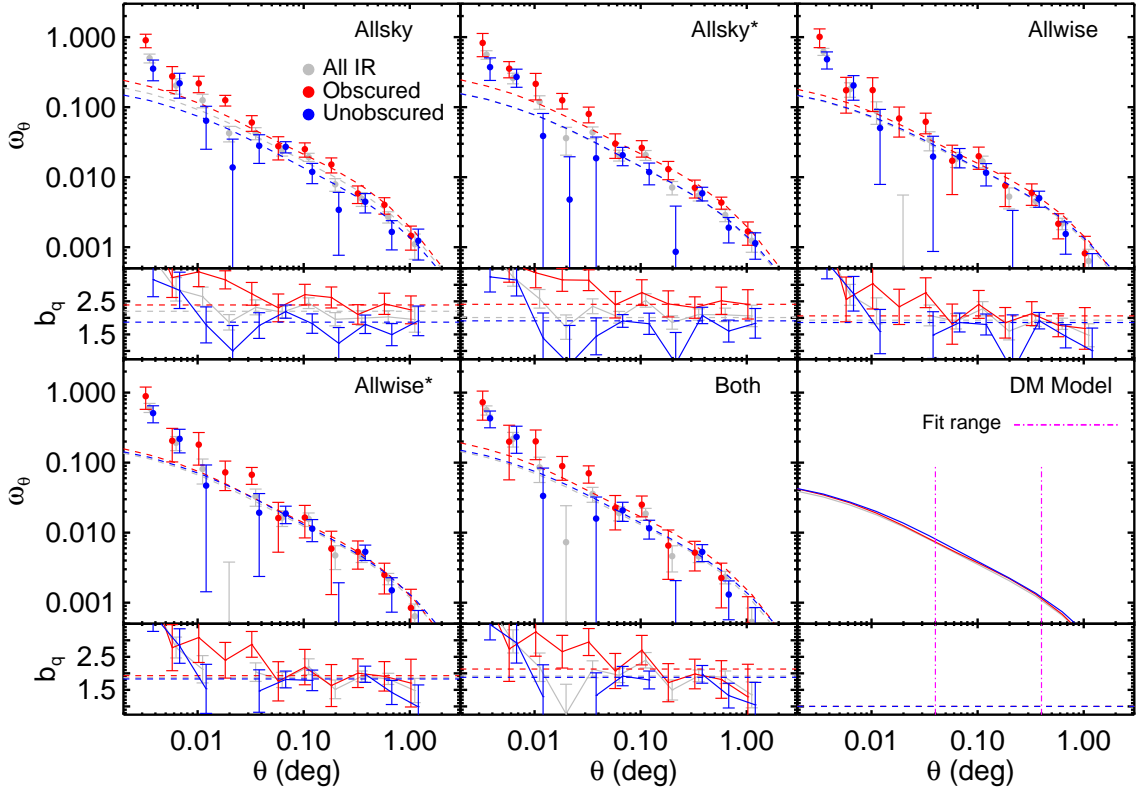}
    \vspace{0cm}
  \caption{The angular autocorrelation for various data and mask combinations.  Note that a sample marked with an asterisk has both the AW and AS masks applied.  The final panel shows the fiducial models for each sample, from Equation~\ref{eq:omega_mod}, as well as the bias fitting range.  The dashed lines in each panel show the model rescaled by measured biases.  Small panels under each show the measurements divided by the models, to highlight any scale dependencies.\label{fig:autocorr}}
\vspace{0.2cm}
\end{figure*}

In Figure~\ref{fig:autocorr} we show the quasar angular autocorrelation $\omega(\theta)$ for various combinations of samples and masks.  As a reminder, samples marked with an asterisk have both the AW and AS generated masks applied.  The final panel of this plot shows our fiducial model, as generated with Equation~\ref{eq:omega_mod} for each of the sample redshift distributions (note that the subtle differences in $dN/dz$ do not strongly affect the models).  The final panel also shows the fitting ranges for the bias measurement. 

In each panel, the models rescaled by the measured bias are shown as dashed lines.  The panels below each plot show the autocorrelations divided by the models, to highlight any scale dependencies of the bias.  Bias measurements for these samples are given in the top half of Table~\ref{fig:all_bias}, and shown in the right side of Figure~\ref{fig:all_bias}.

In all cases, over the chosen fitting range the bias is independent of scale.  Below $\sim$0.04$^{\circ}$, for the obscured sample particularly, clustering amplitudes anti-correlate with angular scale.  This is true for the unobscured sample as well, but generally only below $\sim$0.01$^{\circ}$.  The bias of the unobscured sample is very consistent across all samples and mask combinations (Figure~\ref{fig:all_bias}).  On the other hand, there is a marked decrease in the obscured bias when switching to AW-selected sources, even when the AS mask is applied.  

The difference in bias between obscured and unobscured quasars identified by D14 remains present when using the AS data with the additional AW mask components\footnote{Note that there is a decrease in the measurements presented here compared to D14/D15 due to our updated model procedure as well as the more restricted fitting range.}.  It is also still marginally present in the AW-selected samples, as well as the sample selected from both catalogues --- however, in these cases the error bars overlap significantly, reducing the difference to at best 1$\sigma$.

\begin{table*}
  \caption{Summary of all bias measurements.}
  \label{tbl:biases}
  \begin{tabular}{lcccccc}
  \hline
                                 &  & Allsky                                            & Allwise                                         & Allsky*                 & Allwise*           & Both                     \\
  \hline
                              &  &                            \multicolumn{5}{c}{Quasar Autocorrelation}                \\          
                                                                                      \cline{3-7}                      \\
  $b_q$ All IR             &  & 2.20$\pm$0.06 & 1.93$\pm$0.12  & 2.01$\pm$0.10  & 1.87$\pm$0.12  & 1.92$\pm$0.12  \\
  $b_q$ Obscured      &  & 2.39$\pm$0.14 & 2.06$\pm$0.19  & 2.41$\pm$0.17  & 1.93$\pm$0.21  & 2.13$\pm$0.21   \\
  $b_q$ Unobscured  &  & 1.87$\pm$0.15 & 1.86$\pm$0.12  & 1.91$\pm$0.15  & 1.83$\pm$0.13  & 1.88$\pm$0.15    \\
\\
                                 &  &                            \multicolumn{5}{c}{CMB Lensing cross-correlation}                \\          
                                                                                      \cline{3-7}                      \\
  $b_q$ All IR            &  & (2.18$\pm$0.17) 1.94$\pm$0.14 & (1.95$\pm$0.15) 1.82$\pm$0.14 & (2.10$\pm$0.13) & 1.86$\pm$0.14 & 1.87$\pm$0.14 \\
  $b_q$ Obscured     &  & (2.27$\pm$0.25) 2.06$\pm$0.21 & (2.22$\pm$0.21) 1.82$\pm$0.20 & (2.25$\pm$0.22) & 2.12$\pm$0.22 & 2.06$\pm$0.22 \\
  $b_q$ Unobscured &  & (1.85$\pm$0.24) 1.78$\pm$0.17 & (1.74$\pm$0.20) 1.79$\pm$0.19 & (1.93$\pm$0.16) & 1.70$\pm$0.19 & 1.72$\pm$0.18 \\ 
\hline
   \end{tabular}
   \\  
{
\raggedright    
 Bias measurements for the various sample and mask combinations, using the quasar autocorrelation (top half) and CMB lensing cross-correlations (bottom half).  Measurements are made by fitting the model to the data over the range $40 < l < 400$ or $0.04^{\circ} < \theta < 0.4^{\circ}$. Samples with an asterisk have had the masks developed with both Allsky and Allwise samples applied. Values in parentheses in the bottom half are from measurements with the \planck\ DR1 data, and all others use the DR2 data.  These results are shown in Figure~\ref{fig:all_bias}.\\
 }
\end{table*}

\subsection{Quasar CMB lensing cross-correlation}
\begin{figure*}
\centering
\vspace{0.3cm}
\hspace{0cm}
   \includegraphics[width=16cm]{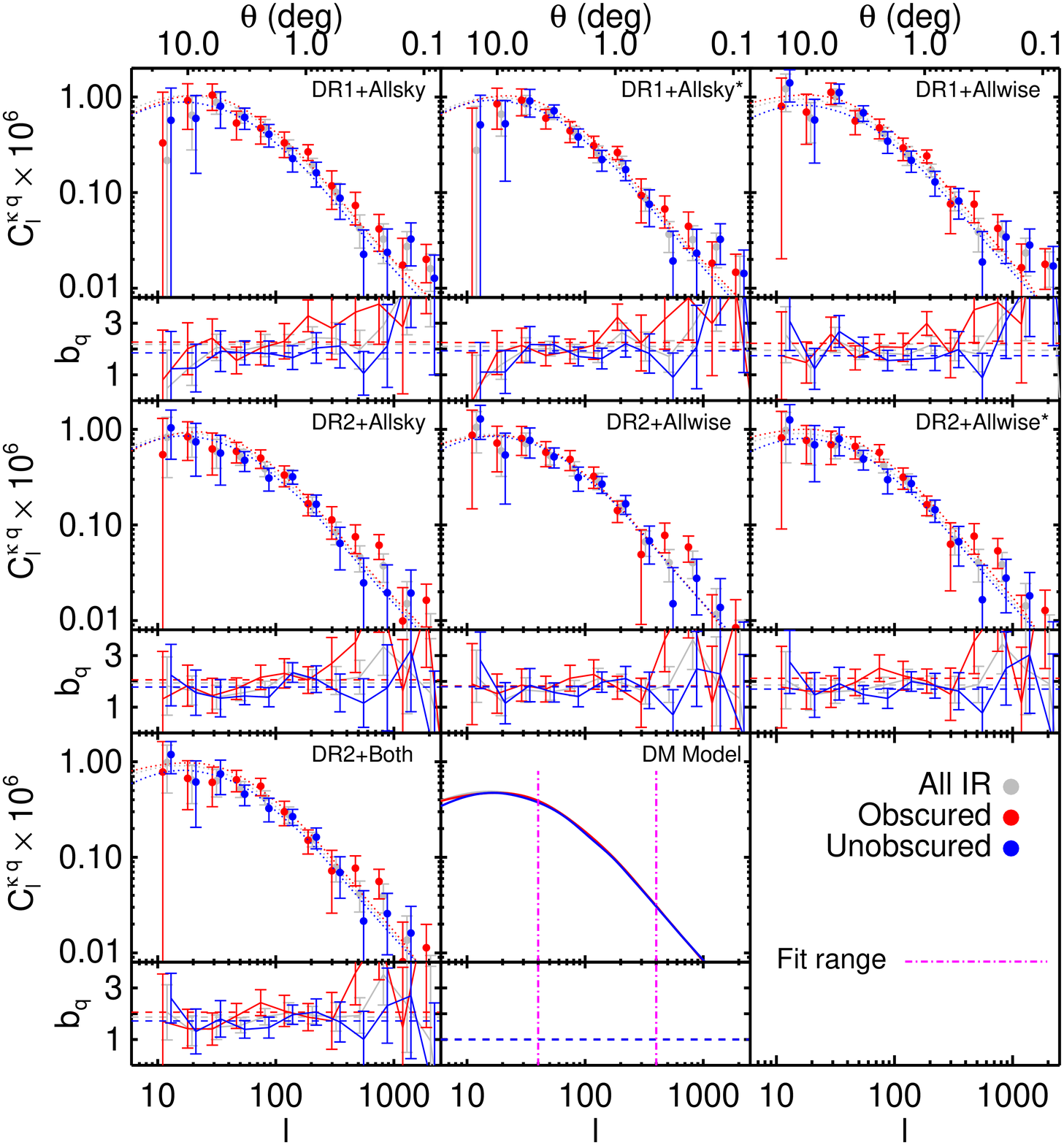}
    \vspace{0cm}
  \caption{The CMB-lensing cross-correlations for various \wise, \planck, and mask combinations.  Note that \wise\ samples marked with an asterisk have both the AW and AS masks applied.  The bottom-center panel shows the fiducial model for each subsample, as well as the fitting range for the bias measurement.  The dotted lines in each panel show the model rescaled by the measured bias.  Small panels below each measurement show the measurement divided by the model, to highlight any scale dependencies.\label{fig:lensing}}
\vspace{0.2cm}
\end{figure*}

\begin{figure*}
\centering
\vspace{0.3cm}
\hspace{0cm}
   \includegraphics[width=16cm]{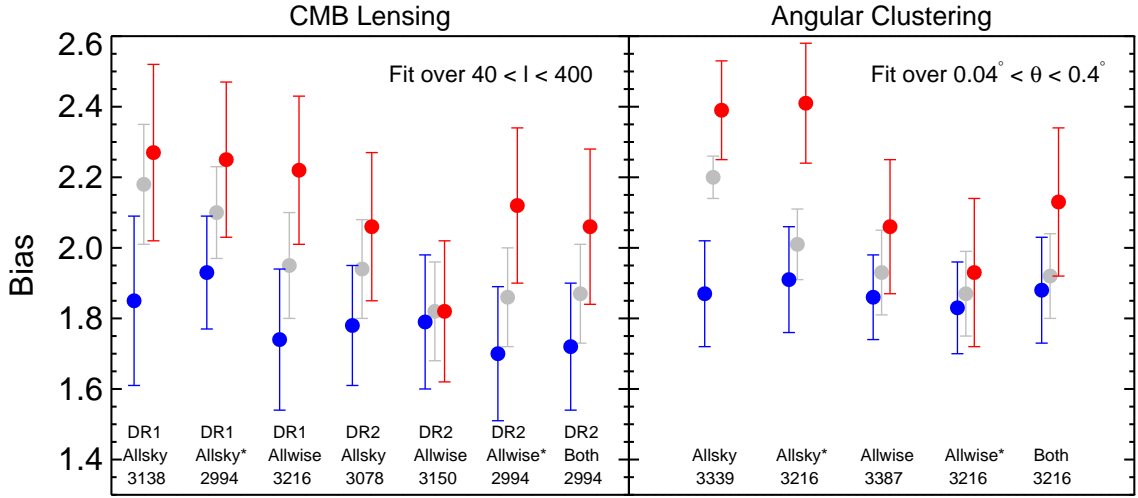}
    \vspace{0cm}
  \caption{A comparison of the measured bias via CMB lensing cross-correlations (left) and the quasar autocorrelation (right), using various data sets.  In the CMB lensing panel, ``DR1'' and ``DR2'' refer to the \planck\ data release.  An asterisk on the \wise\ catalogue indicates that the masks derived from both catalogues have been applied, and ``Both'' is the measurement for objects satisfying our criteria in both catalogues. The numbers under each measurement indicate the total area used, in deg$^2$.\label{fig:all_bias}}
\vspace{0.2cm}
\end{figure*}

Figure~\ref{fig:lensing} shows the quasar-CMB lensing cross-correlations for various combinations of the data, including the AW and AS samples as well as \planck\ DR1 and DR2 lensing maps.  The bottom-center panel shows the model generated from Equation~\ref{eq:lens_model} for each $dN/dz$, and highlights the bias fitting region.  Again, the model does not depend strongly on the subtle differences in redshift distribution for the three samples.  Dotted lines in each panel show the models rescaled by the bias values, and the small panels below each show the measurement divided by the model to highlight scale dependencies.  Bias measurements are summarized in the lower half of Table~\ref{tbl:biases}, and shown in the left panel of Figure~\ref{fig:all_bias}.

As seen in the autocorrelation measurements, the lensing cross-correlations are generally scale-independent over the chosen fitting range.  However, beyond this range toward smaller scales (larger $l$), there is a clear increase in cross-correlation power that is particularly strong for the obscured sample, though more prevalent for both samples when \planck\ DR1 is used.  While the measurement is noisy at the largest $l$, with the \planck\ DR2 lensing map the unobscured cross-correlation stays consistent with flat out to the smallest scales ($\sim$0.1$^{\circ}$, consistent with the clustering measurement).  The scale-dependence for the obscured sample is reduced somewhat with DR2, but is still present.  The obscured sample cross-correlation with DR2 also shows a prominent bump around $l \sim 600$, about the scales for which there may be an unknown correlated feature in the DR2 lensing map \citep{PlanckCollaboration:2015tp}. 

The bias values for the unobscured sample are quite consistent across all sample and mask combinations, and with the autocorrelation measurements (Figure~\ref{fig:all_bias})$^{10}$.  Overall the use of \planck\ DR2 reduces the measured bias in both samples (compare for example the measurements for DR1$+$AS and DR2$+$AS).  For the obscured sample, the use of \planck\ DR2 and the AW-selected sample reduces the bias somewhat, until the AS mask is also applied where the bias is raised again.  Overall, the obscured sample tends to have a higher bias, though at a lower significance compared to D15 (at most $\sim$2$\sigma$).

\subsection{\planck\ DR1 vs. DR2 lensing convergence}
Now that we have seen that the updated \planck\ CMB lensing map tends to reduce the measured bias, particularly for the obscured sample, we briefly compare the two maps directly to investigate quantitatively whether one map is superior to the other.  Of course, the fact that the signature of lensing, at least as averaged over the whole sky, is so much stronger in DR2 is already indicative of the superiority of the newest data.  Figure~\ref{fig:kappas} illustrates the improved behavior of the DR2 data as well. Finally, DR2 has a lensing potential power spectrum in better agreement with theoretical predictions \citep[see e.g.\ Figure 6 of][]{PlanckCollaboration:2015tp}.  

\begin{figure}
\centering
\vspace{0.3cm}
\hspace{0cm}
   \includegraphics[width=7.5cm]{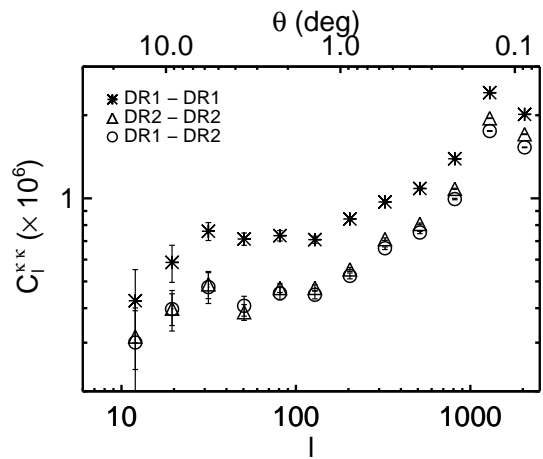}
    \vspace{0cm}
  \caption{Auto- and cross-correlations of the \planck\ DR1 and DR2 $\kappa$ maps. The similar shape but higher power at all scales in the DR1 autocorrelation suggests that there is more correlated noise in the DR1 map.\label{fig:kappa_kappa}}
\vspace{0.2cm}
\end{figure}

We also correlate the DR1 and DR2 $\kappa$ maps with themselves and each other (using the method described in section 3.3.2), to highlight potential differences.  If the maps have similar properties, these auto and cross-correlations should all be quite similar.  The results are shown in Figure~\ref{fig:kappa_kappa}. While the DR2-DR2 auto-correlation and DR2-DR1 cross-correlation are nearly identical, the DR1-DR1 autocorrelation has significantly more power (by $\sim$45\%), while having the same shape at all scales.  This suggests that the DR2 map contains significantly less correlated noise, while still preserving real features in the data.  However, it is not clear how this might affect the obscured sample cross-correlation more than that of the unobscured sample.

\subsection{\wise\ Allsky vs.\ Allwise-selected quasars}
In section 2, many similarities between the AS and AW-selected quasar samples were discussed.  The number densities are similar (three fewer objects per square degree in the AW selected sample), the obscured and unobscured fractions are indistinguishable (Table~\ref{tbl:samples}), the redshift distributions are consistent, and the optical morphological properties are very similar (with a slight increase in unresolved sources in AW).  Despite these similarities, the difference in bias measurements when changing from AS to AW-selected samples (or changing the masks), suggests that there may be some fundamental difference in samples selected from the two catalogues.  To help inform the discussion of the bias measurements, we explore the properties of these samples in more detail here.

\subsubsection{Photometric properties of Allsky and Allwise selected quasars}
\begin{figure*}
\centering
\vspace{0.3cm}
\hspace{0cm}
   \includegraphics[width=4.3cm]{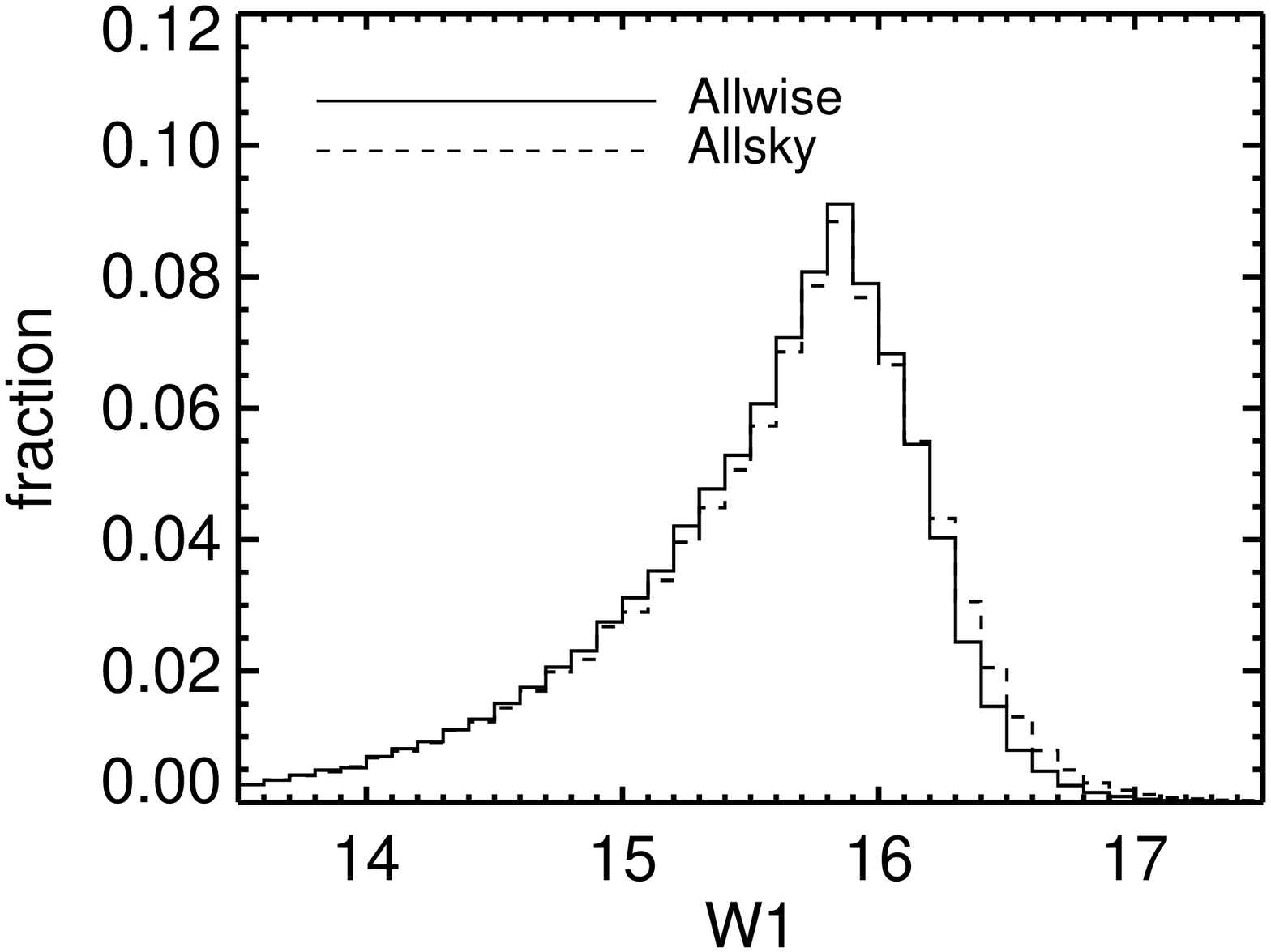}
   \includegraphics[width=4.3cm]{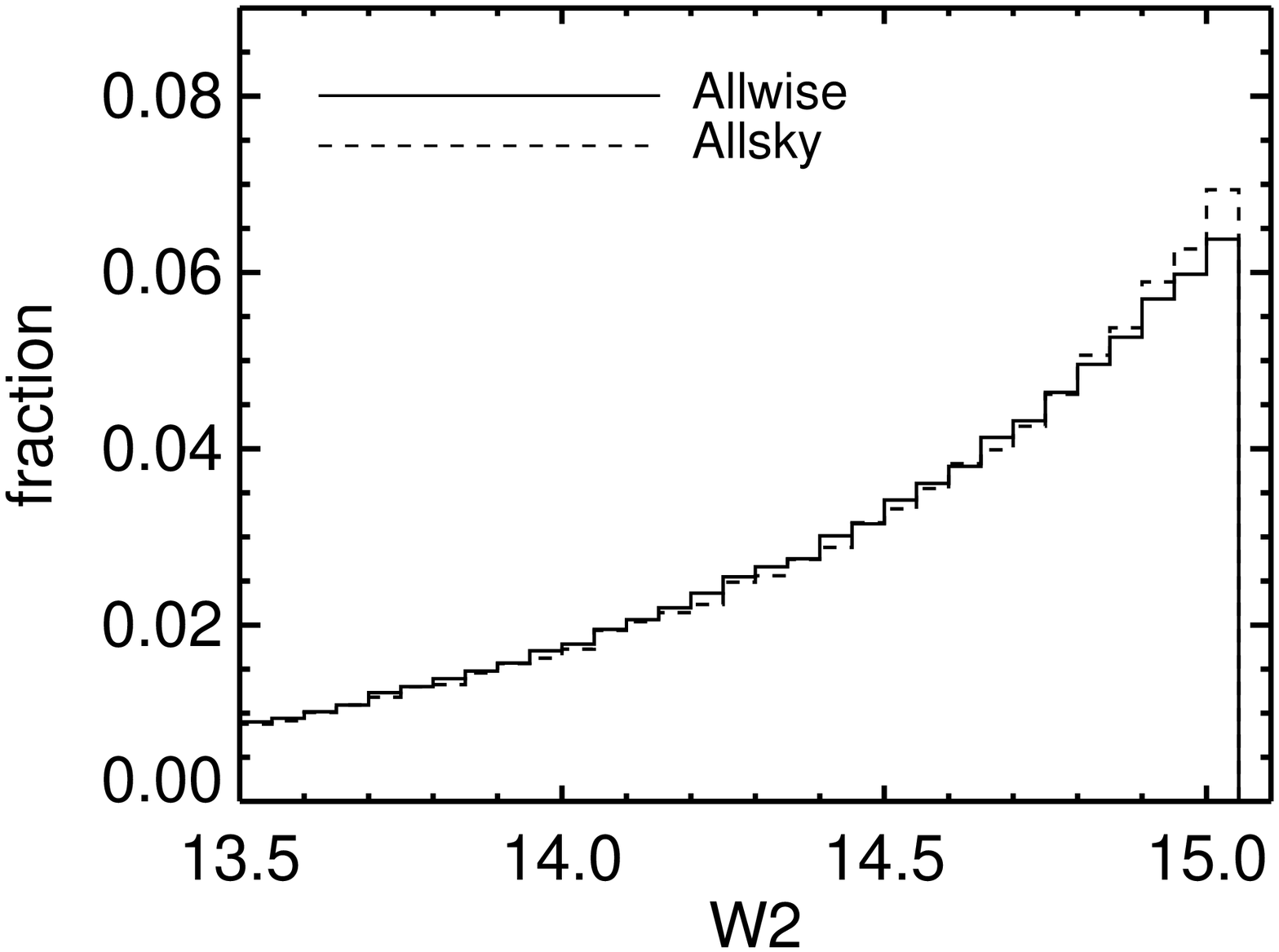}
   \includegraphics[width=4.3cm]{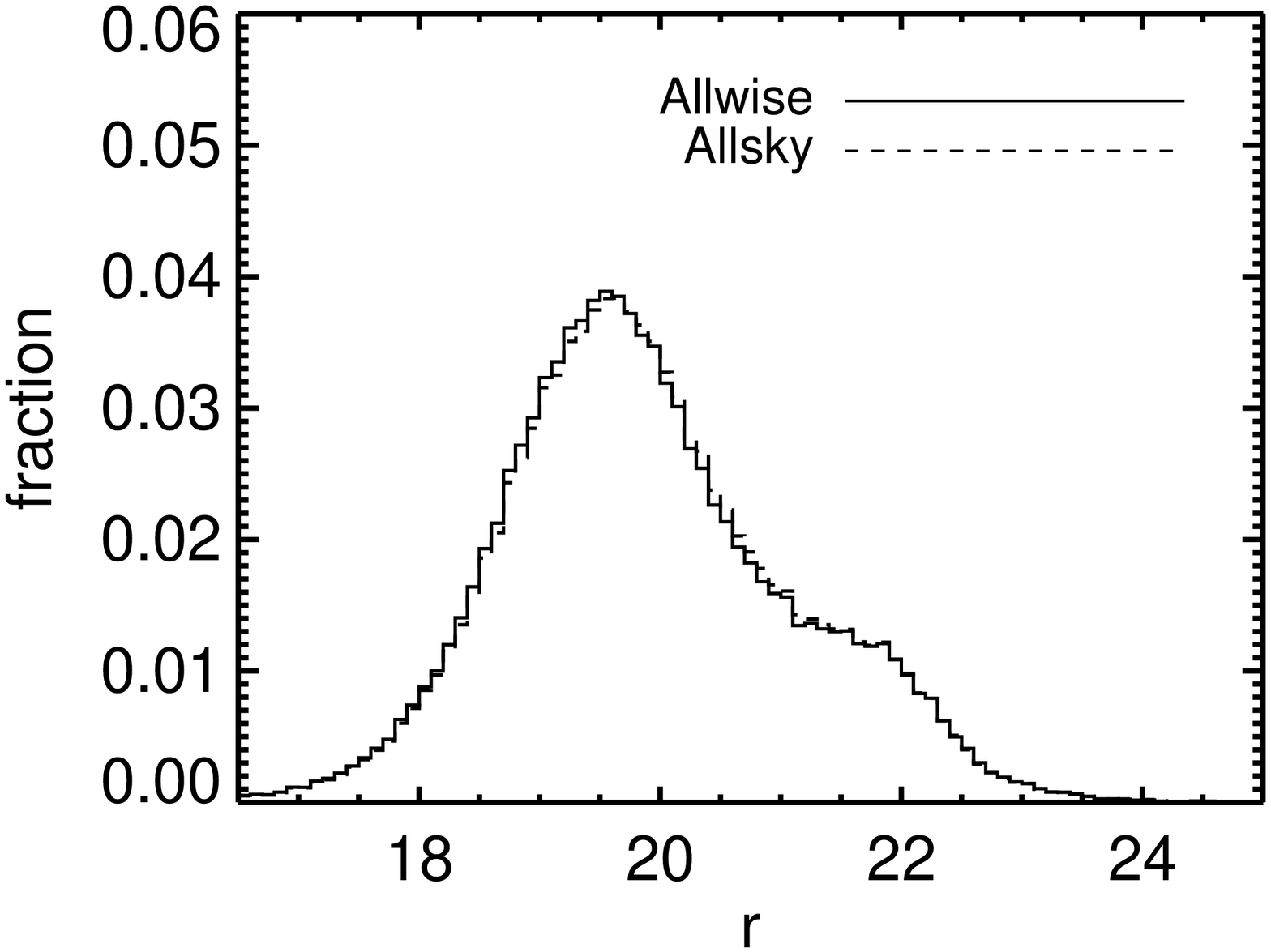}
   \includegraphics[width=4.3cm]{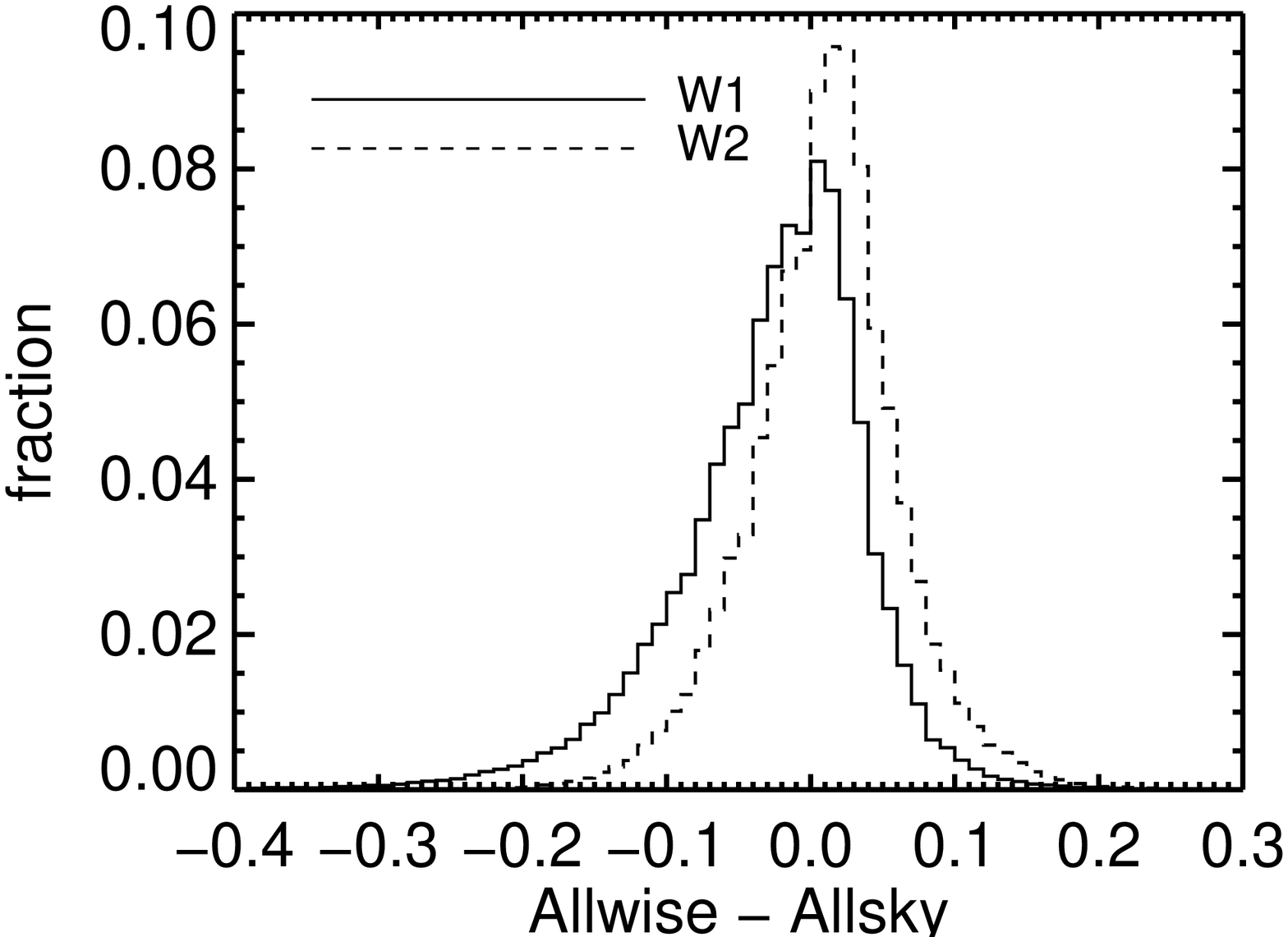}

   \vspace{0.15cm}

   \includegraphics[width=4.3cm]{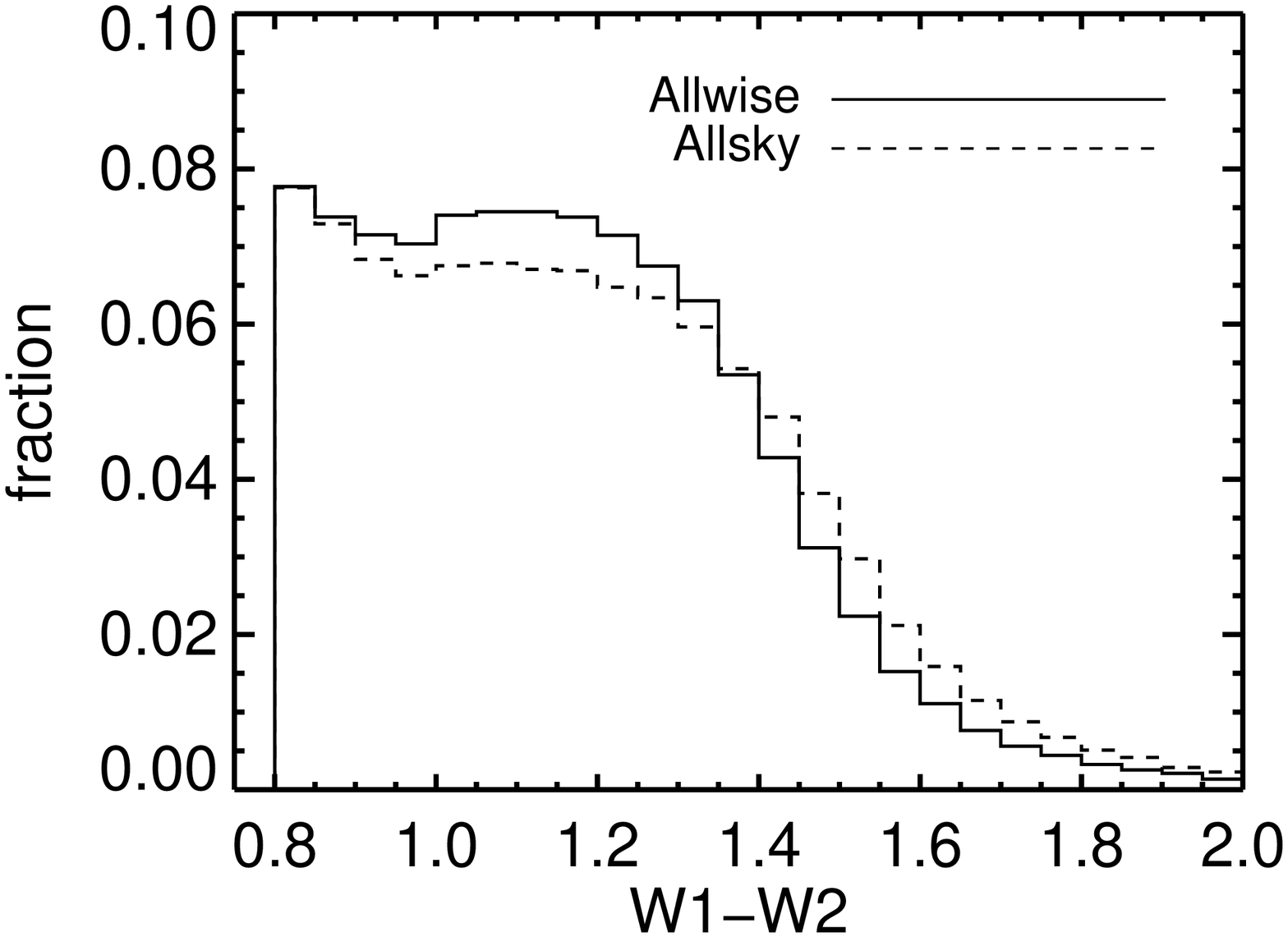}
   \includegraphics[width=4.3cm]{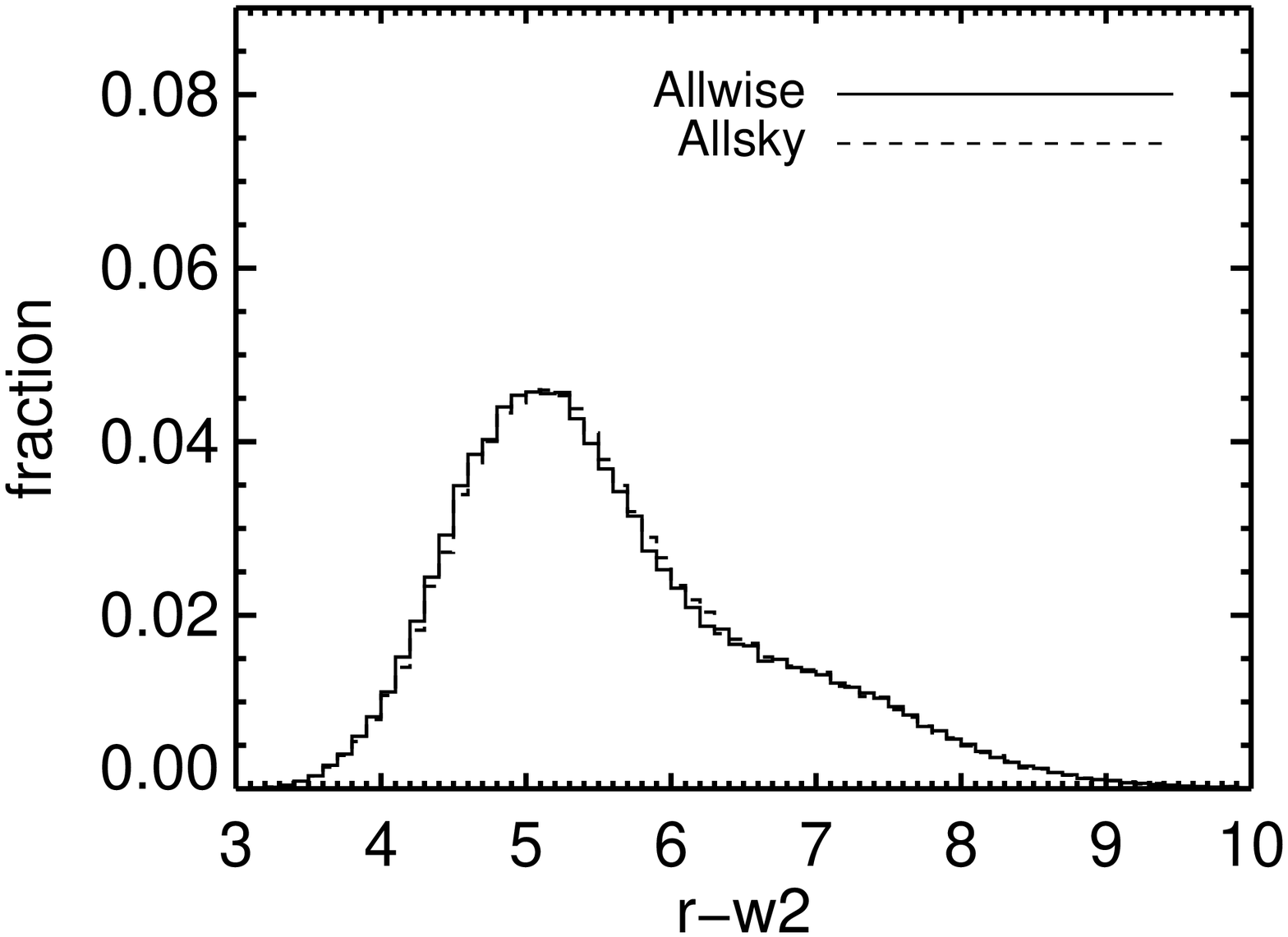}
   \includegraphics[width=4.3cm]{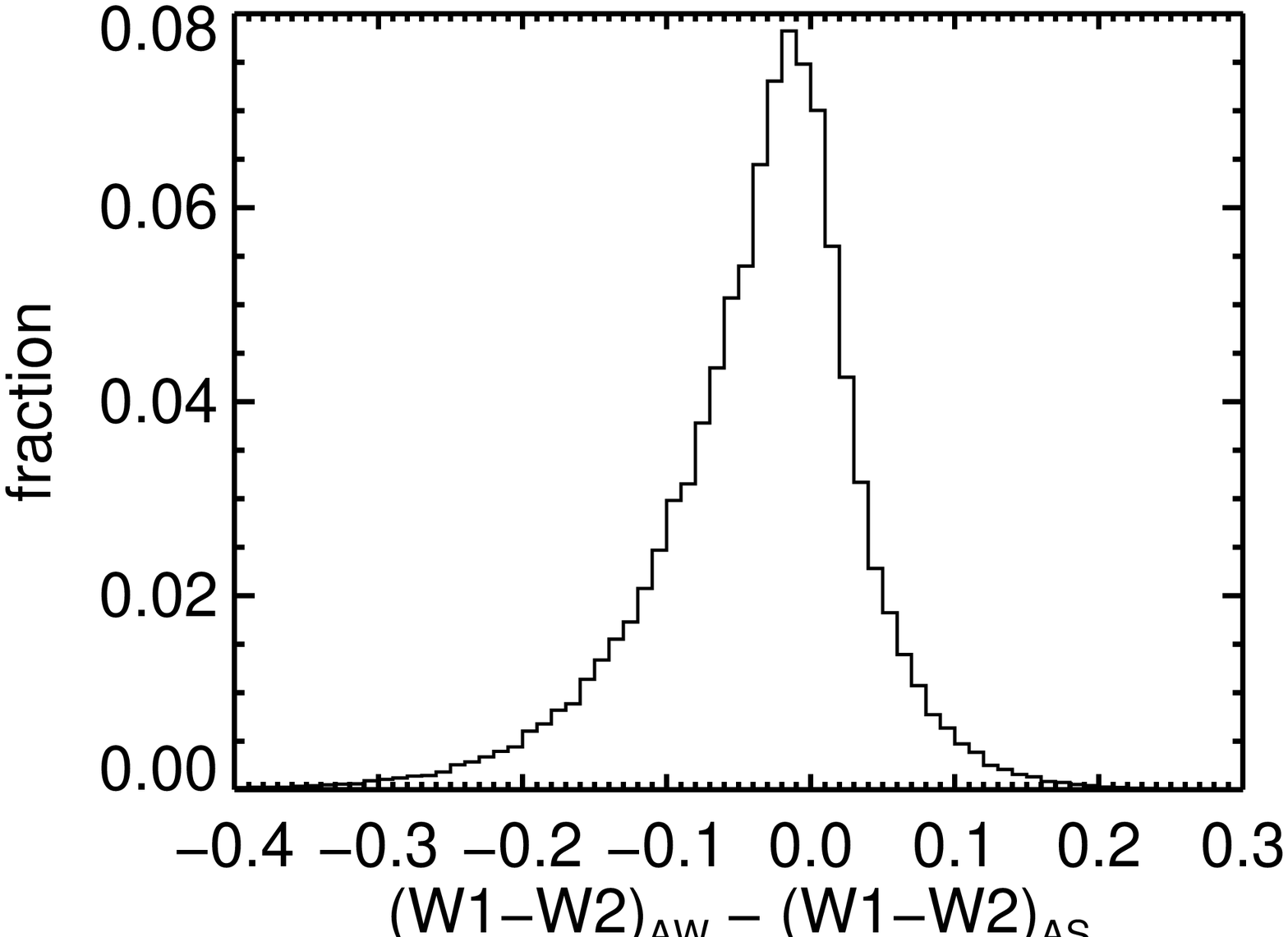}

    \vspace{0cm}
  \caption{{\emph Top row:} Comparison of $W1$, $W2$, and $r$ distributions for the Allwise (solid) and Allsky (dashed) selected samples, after applying both masks to both samples (so differences are not due to masking).  The top-right panel shows the difference between the $W1$ (solid) and $W2$ (dashed) magnitudes for those objects that are common between the samples ($r$-band magnitudes are excluded since the majority of Allwise and Allsky sources are matched to the same optical counterpart, and these differences are null). \emph{Bottom row:} The same, but for colors.\label{fig:aw_as_phot}}
\vspace{0.2cm}
\end{figure*}

The top row of Figure~\ref{fig:aw_as_phot} shows distributions of $W1$, $W2$, and $r$ for the AW and AS selected quasars (the full IR-selected samples; these comparisons are very similar for obscured and unobscured subsamples), as well as the difference between the AW and AS $W1$ and $W2$ magnitudes for objects that are selected from both catalogues.  These comparisons are made after applying both masks to the samples (i.e.\ Allwise* and Allsky*), so any differences are not due to masking (see below).  Of the common objects in AS and AW, the vast majority ($>$99.9\%) are matched to the same optical counterpart so differences between $r$ magnitudes for the common sample are not shown as they are generally null.  

The AW sample shows a subtle shift toward brighter $W1$ and $W2$ fluxes, and as noted the $r$ distributions are nearly identical.  The reason for this is clear in the top right panel showing the difference in the common sample.  The $W1$ difference distribution is strongly asymmetric, with an updated AW flux more likely to be brighter, while the $W2$ difference distribution is much more symmetric about zero and smaller in magnitude .  These effects are noted in the AW explanatory supplement, which states that AW photometry is known to be increasingly brighter at magnitudes fainter than $W1 \sim 14 $ and $W2 \sim 13$ (the majority of our sources) due to correction of a faint source underestimation bias in AS.  These effects are particularly important for color-selected samples such as ours.

The second row of Figure~\ref{fig:aw_as_phot} illustrates how these changes propagate through to our color selection.  We see that the larger brightening in $W1$ compared to $W2$ leads to an IR-redder sample in AS than in AW.  The similarity in the distributions of $r-W2$ illustrates that most of the difference in $W1-W2$ colors is driven by the shift in $W1$.

We also compared the photometric properties of AW-selected quasars that fall within the AS mask (see the middle panel of Figure~\ref{fig:fields}) to the sources outside of the mask.  There are no clear differences that indicate serious problems with the AW objects that fall within the AS mask, and so we do not show them here.  Additionally, the obscured and unobscured fractions of these masked sources are consistent with the overall fractions.  The same is true for AS-selected quasars that are within the AW mask.  These similarities in photometric properties suggest that the fact that the B\"{o}otes field is not strongly affected by differences in masking does not have an important impact on our estimates of the redshift distribution.







\subsubsection{Allsky and Allwise-only sources}
\begin{figure}
\centering
\vspace{0.3cm}
\hspace{0cm}
   \includegraphics[width=4.cm]{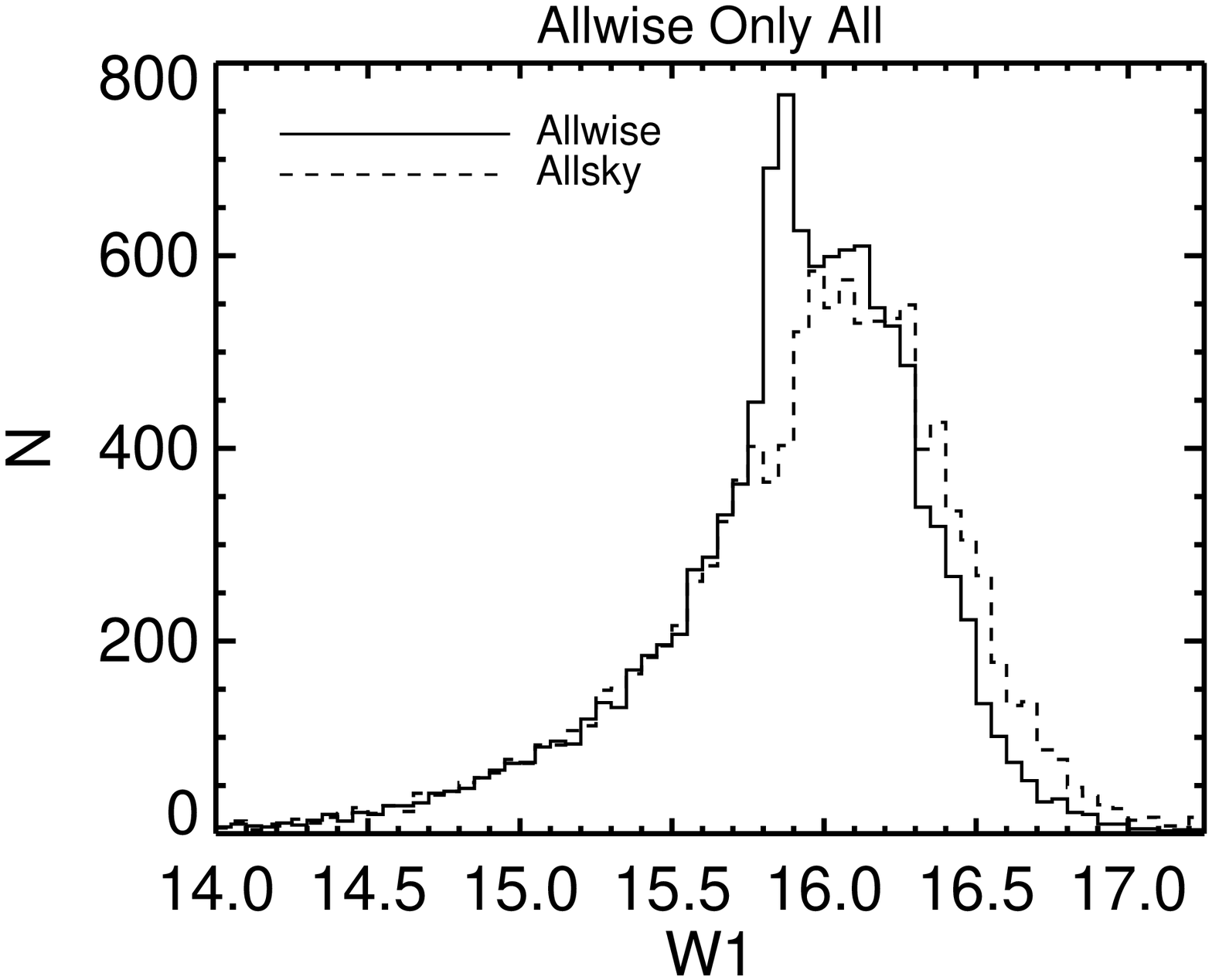} 
   \includegraphics[width=4.cm]{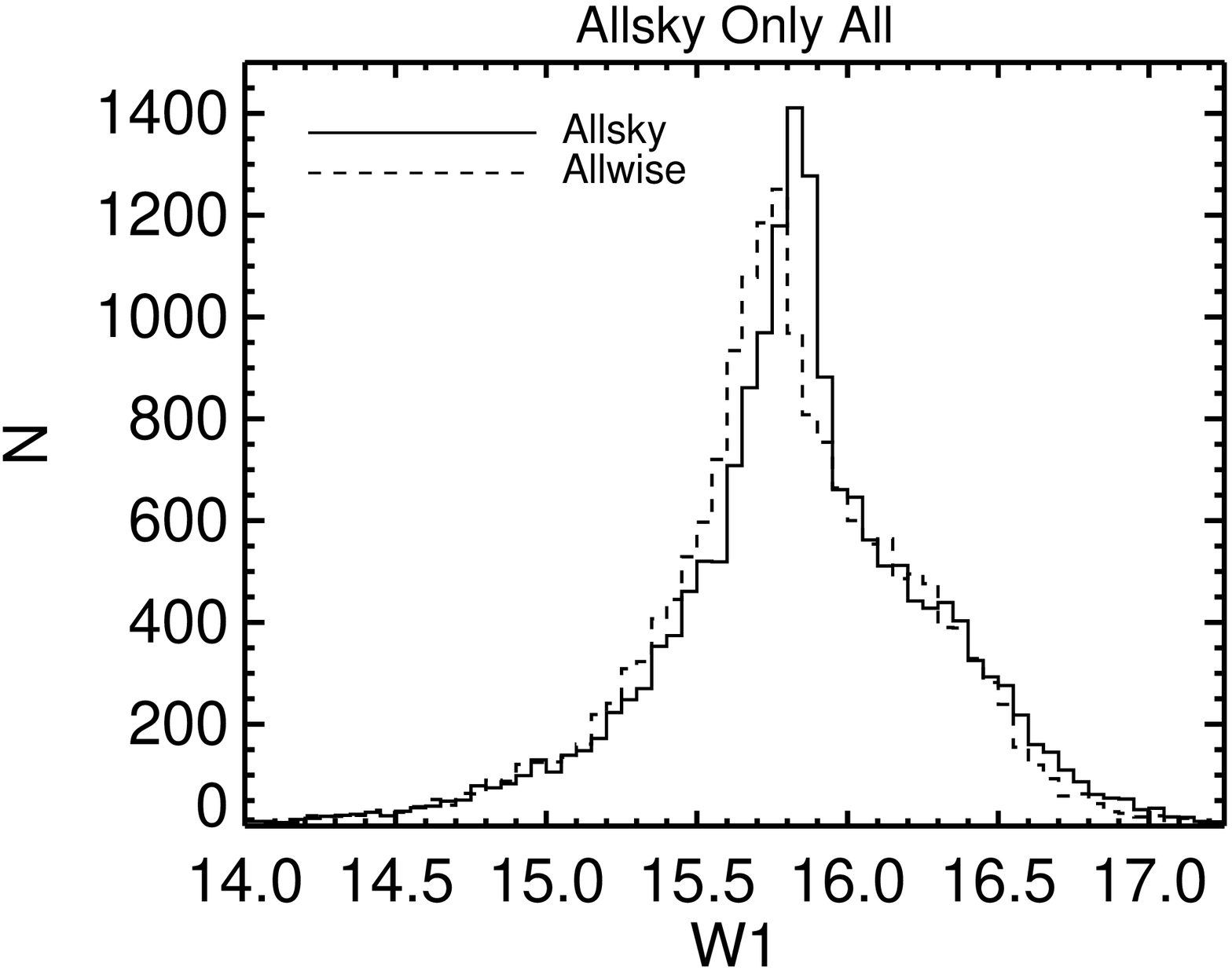}
   
   \vspace{0.15cm}

   \includegraphics[width=4.cm]{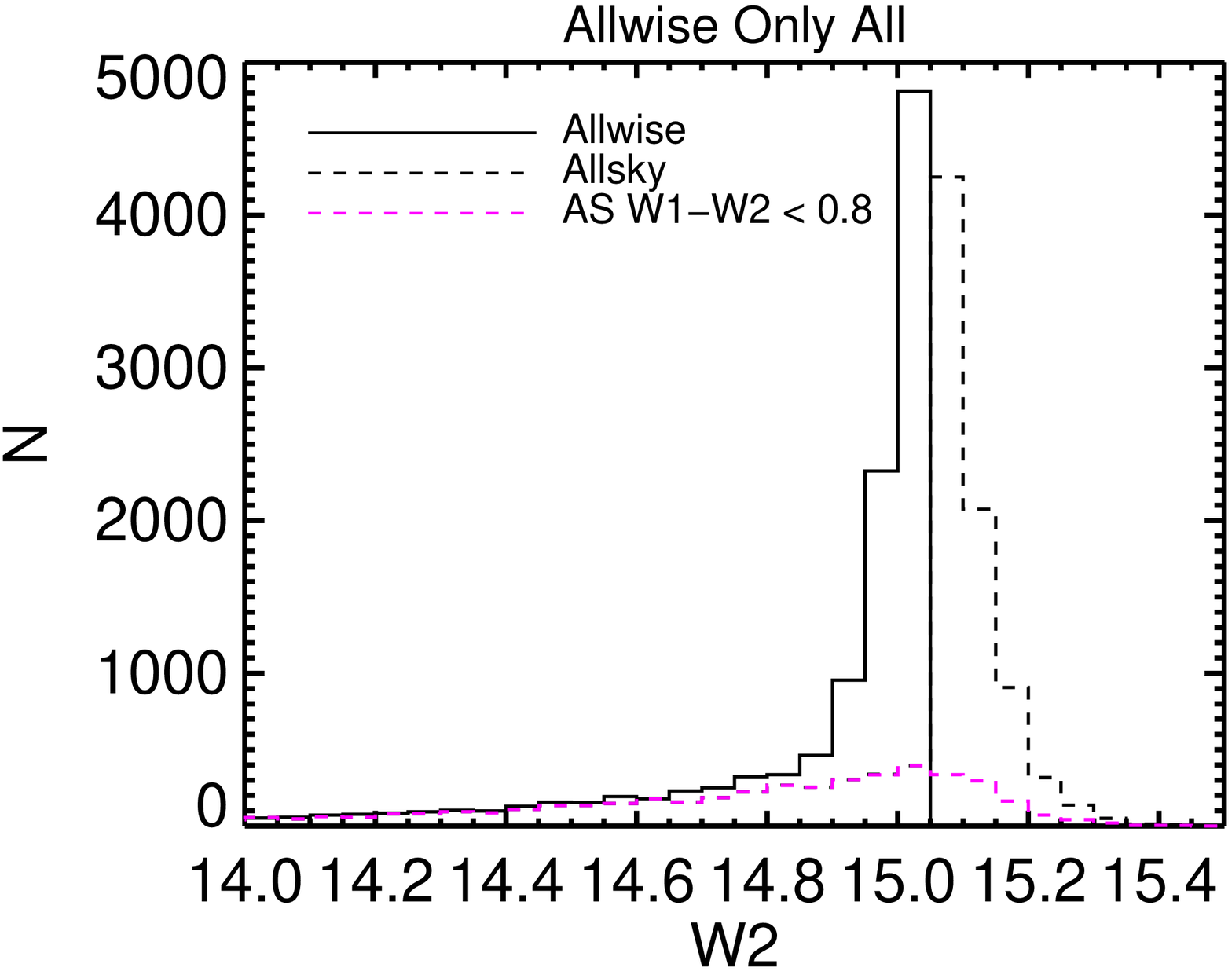}
   \includegraphics[width=4.cm]{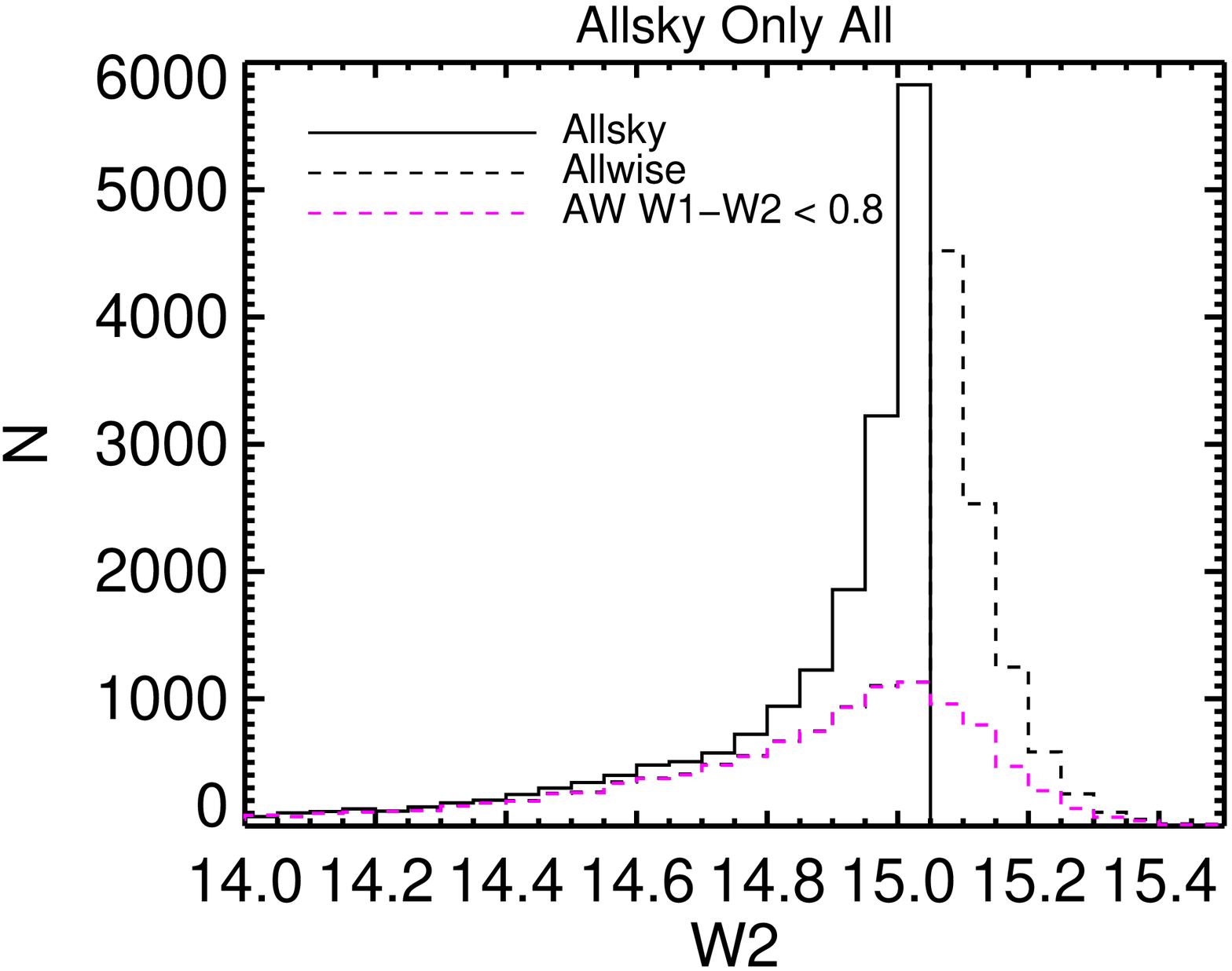} 

   \vspace{0.15cm}

   \includegraphics[width=4.cm]{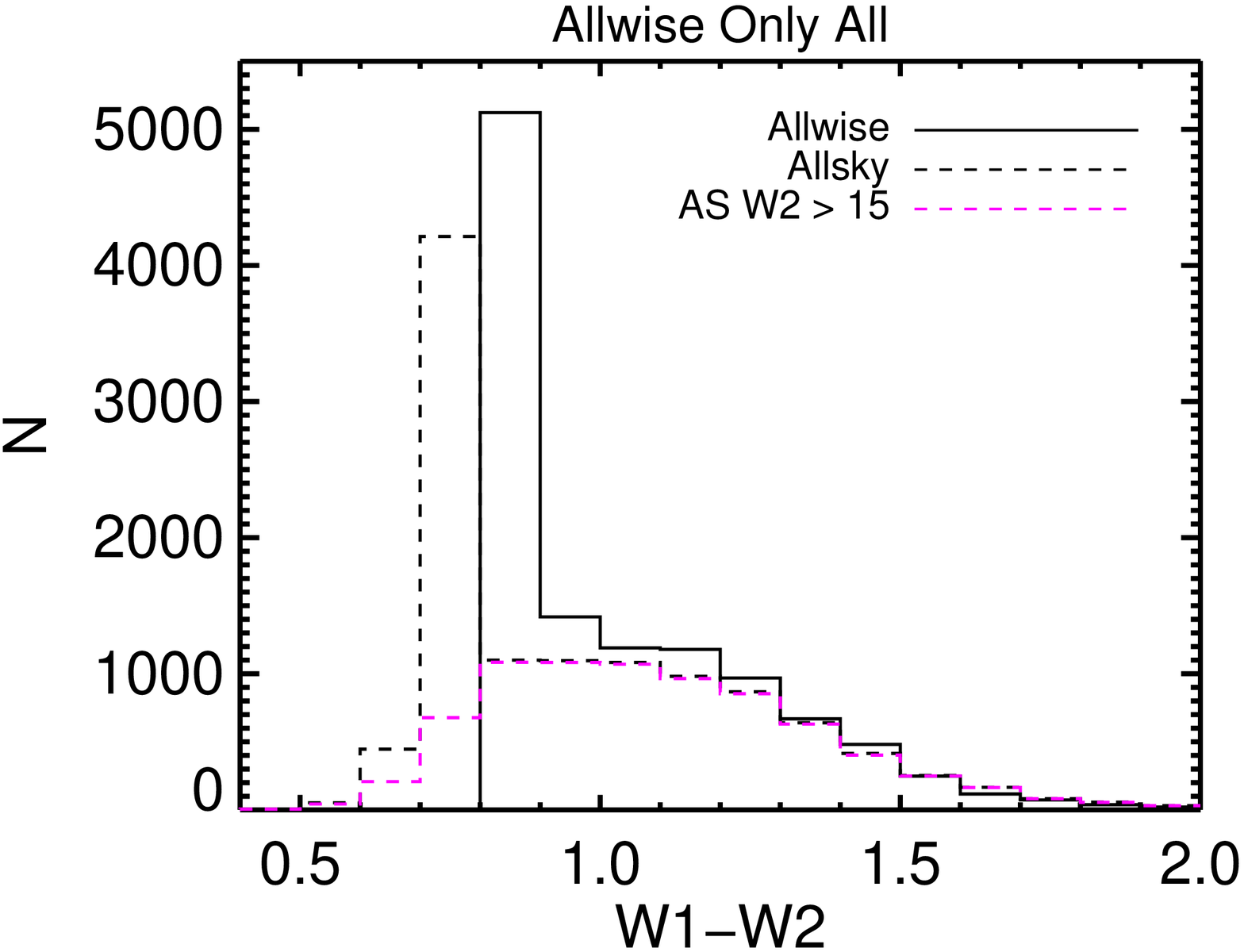}
   \includegraphics[width=4.cm]{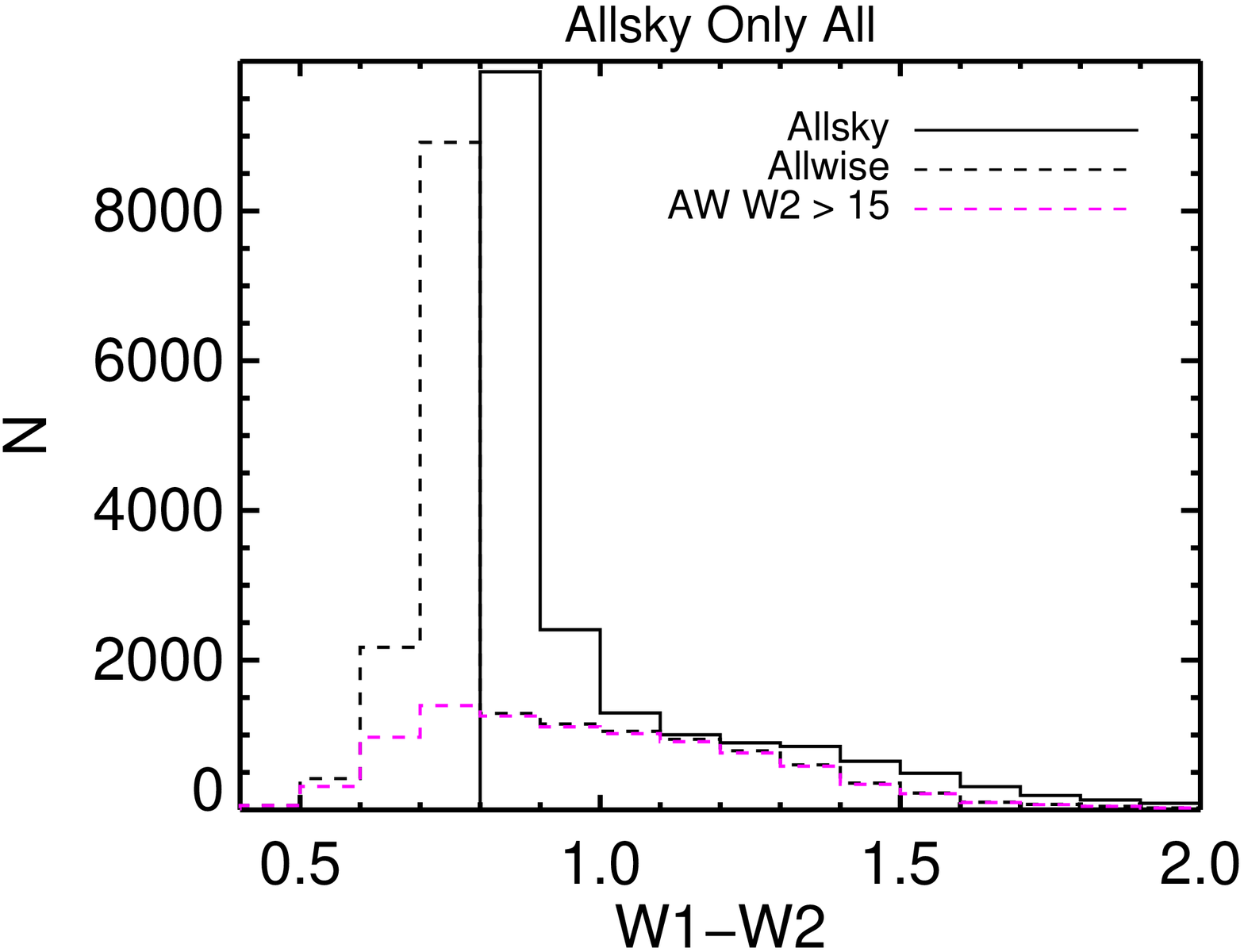}





    \vspace{0cm}
  \caption{A comparison of \wise\ $W1$ (top), $W2$ (middle), and $W1-W2$ (bottom) from the Allwise and Allsky catalogues for objects that satisfy our selection criteria in only one catalogue (right: Allwise only, left: Allsky only).  In the $W2$ and color panels the distributions of objects that would not satisfy the other cut ($W1-W2 < 0.08$ or $W2<15.05$) are shown in magenta.  Clearly, most objects satisfying a cut in one catalogue but not the other are primarily border-line objects in terms of $W2$, which causes them to also appear generally fainter in $W1$.\label{fig:missing_photometry}}
\vspace{0.2cm}
\end{figure}

There are 12,544/5,776/6,632 (total/obscured/unobscured) objects that are only selected by AW and 19,777/9,019/10,523 (total/obscured/unobscured) objects that are only selected in AS.  These sources are not generally missing from the opposite catalogue, but rather their photometry has been updated such that they no longer meet our selection criteria.  Figure~\ref{fig:missing_photometry} compares the AW and AS photometry and colors for sources that are selected as quasars in only one catalogue (again, only the full samples are shown as the obscured and unobscured trends are similar).  In general, objects that do not make the cut in one or the other catalogue are borderline objects, near $W2 \sim 15$ or $W1-W2 \sim 0.8$.  Most of the difference in the two samples is thus due to the small adjustments to $W1$ and $W2$ in AW.  The $W1$ distributions of these objects tend to have a spike around $W1 \sim 15.8$, reflective of the sharp $W2$ cut.  On average however, the AW-only sources are somewhat fainter in $W1$ than the AS-only sources (compare the peaks of either the dashed or solid lines in the top left and top right panels).  This is also true for $W2$, as the tail to brighter magnitudes is smaller in the AW-only sources.

\begin{figure*}
\centering
\vspace{0.3cm}
\hspace{0cm}
   \includegraphics[width=8cm]{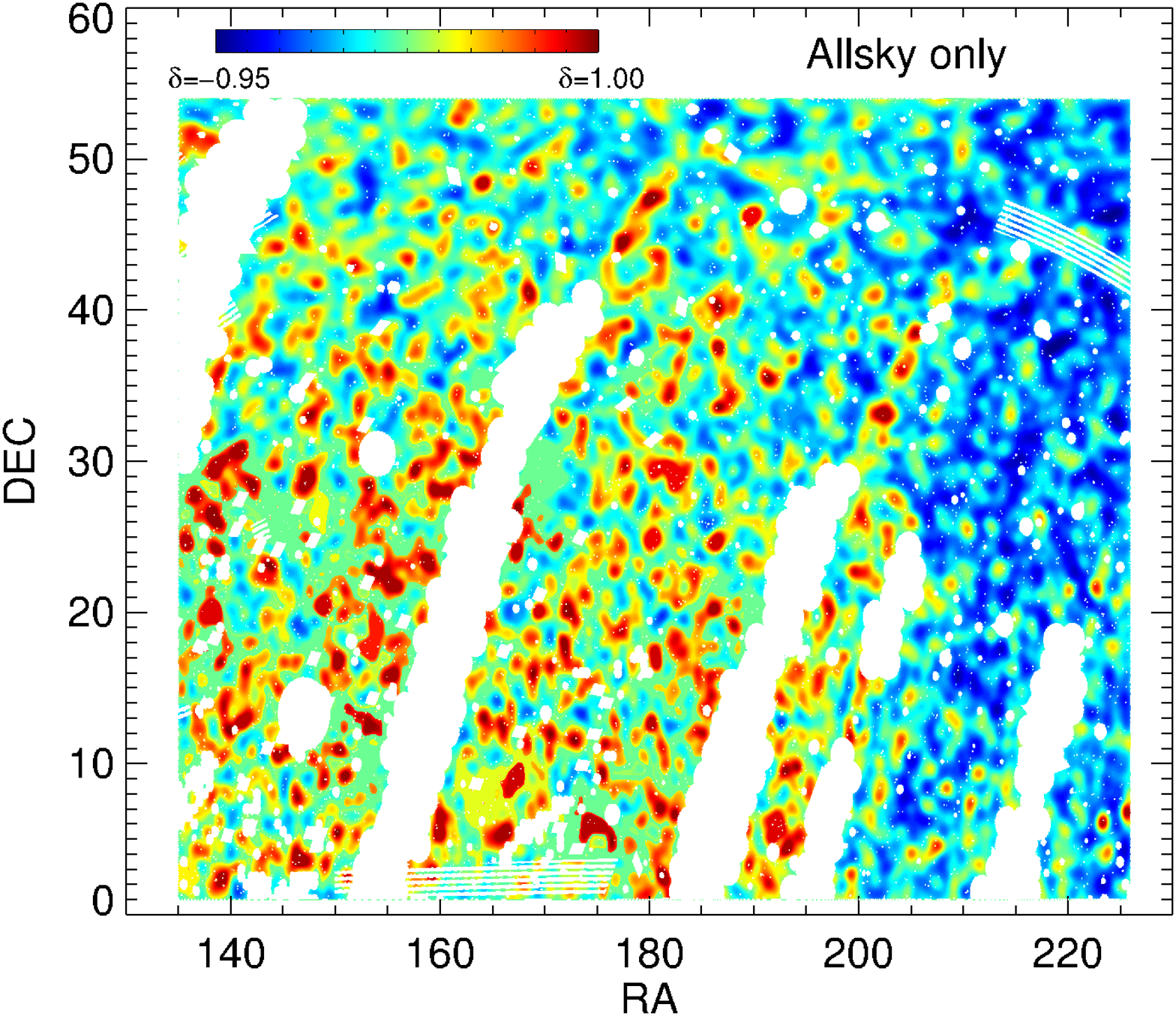}
   \includegraphics[width=8cm]{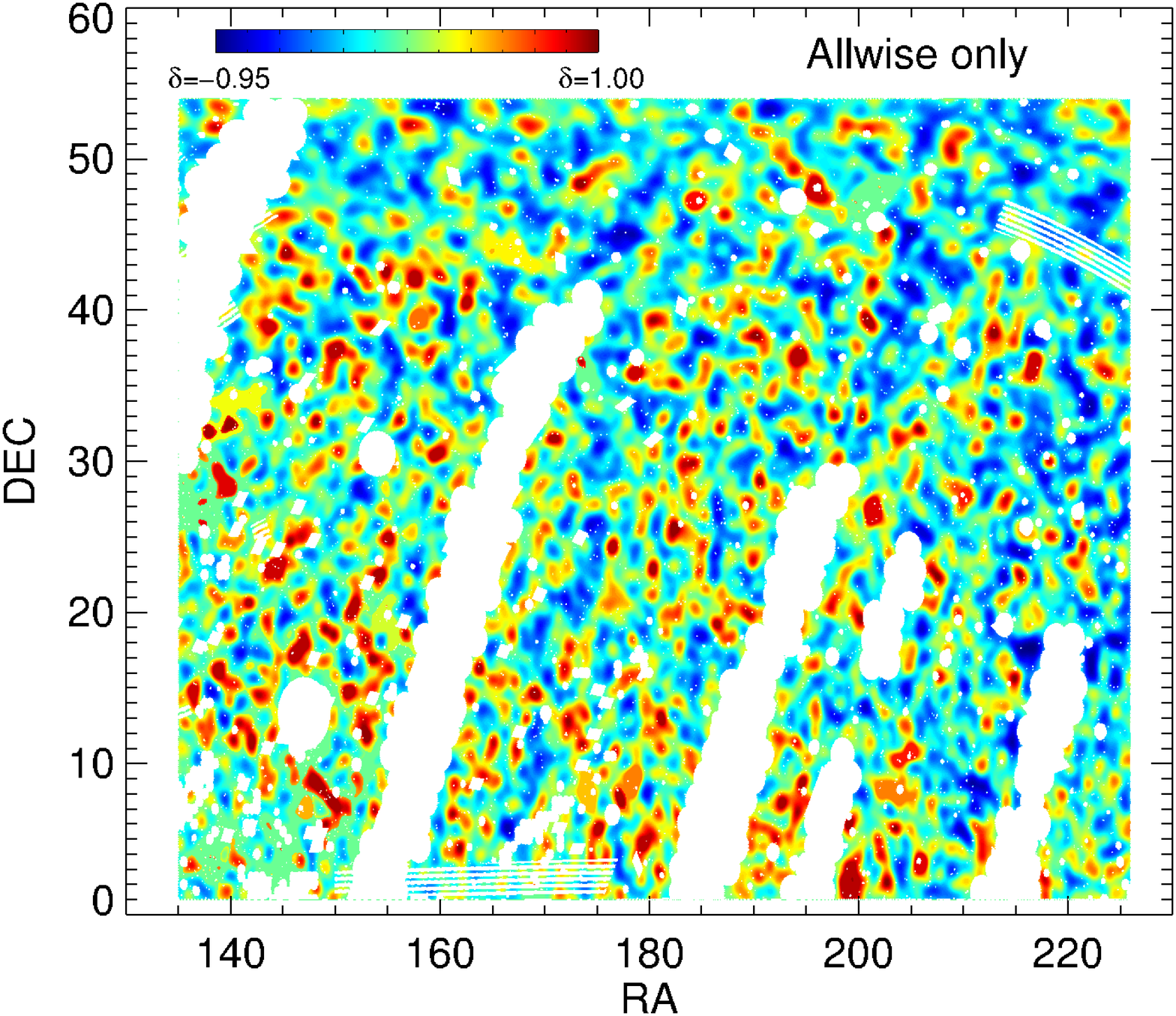}
    \vspace{0cm}
  \caption{The relative densities ($\delta$; see section 3.2) of Allsky-only (right) and Allwise-only (left) selected quasars as a function of position.  Blue indicates an under density relative to the mean, red is over dense.  It is clear that the objects selected only by Allsky are less evenly distributed, suggesting that many are in fact artifacts or that there is a position dependent bias affecting their selection. This is greatly reduced, but still present, in Allwise.\label{fig:rel_densities}}
\vspace{0.2cm}
\end{figure*}

Considering that the sources that only meet our selection in AW or AS tend to have borderline photometric properties, can we argue that one catalogue is truly eliminating more contamination, or is this simply just noise near the cuts?  In Figure~\ref{fig:rel_densities}, we plot the relative densities ($\delta$, see section 3.2) of AS and AW-only sources as a function of position on the sky, using $n_{\textrm{side}}=2048$ \textsc{healpix} pixels smoothed with a 1$^{\circ}$ Gaussian.  We of course expect that intrinsically quasars are uniformly distributed across the field.  However, the AS-only selected sources are heavily biased in position, with their density generally increasing toward the Ecliptic plane (the strips masked for Moon contamination are perpendicular to the Ecliptic).  AW-only sources are more evenly distributed, though the distribution is still not completely uniform.  This suggests that objects selected by AS only may indeed be artifacts, or their selection is biased by a position-dependent factor more-so than in AW (see section 4.4.3).

\begin{figure}
\centering
\vspace{0.3cm}
\hspace{0cm}
   \includegraphics[width=8.5cm]{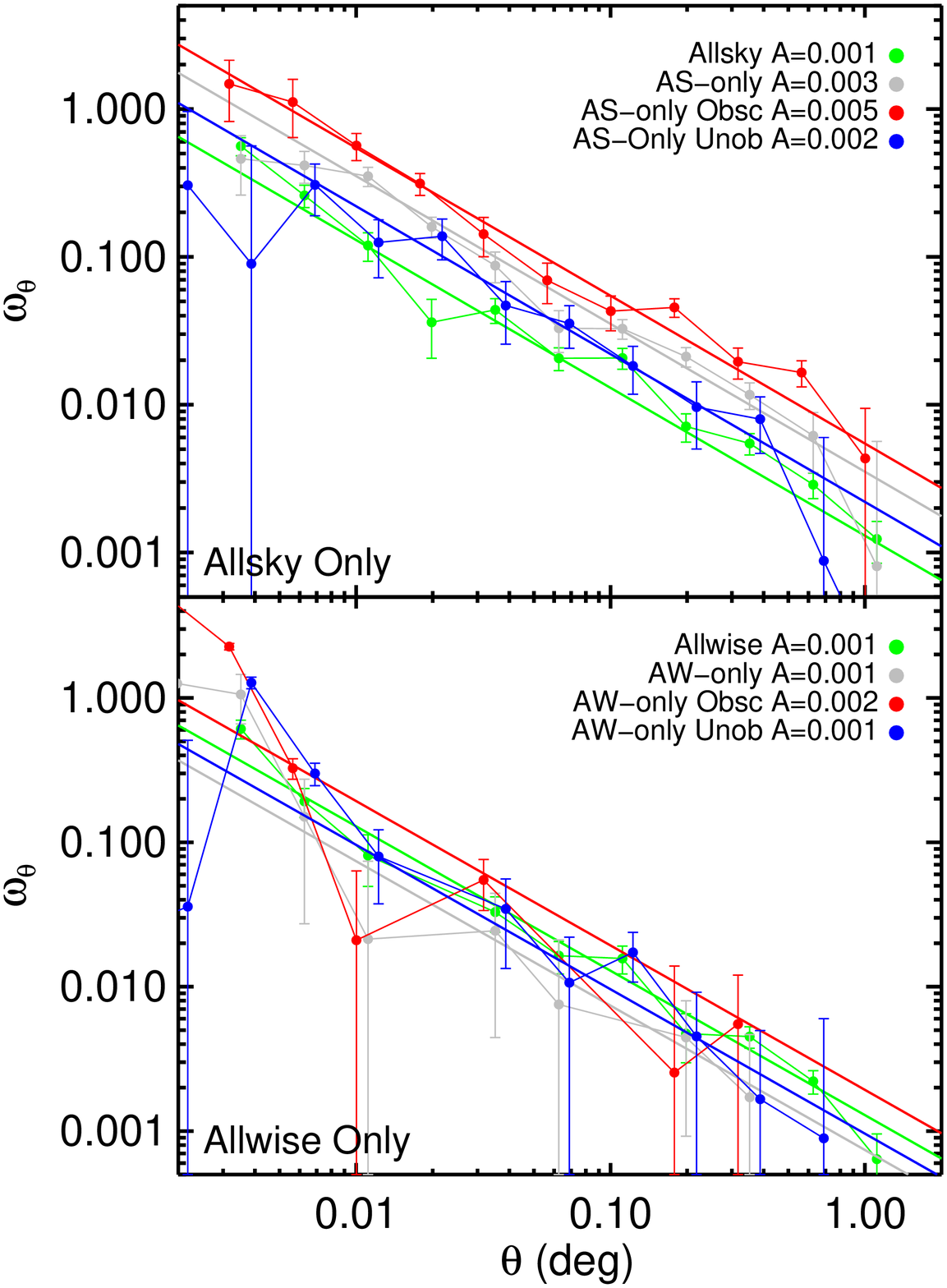}
    \vspace{0cm}
  \caption{The cross-correlation of objects selected by Allsky only (top) and Allwise only (bottom) with their respective full samples, after applying both masks to each sample (differences not due to masking).  In each panel, the green points show the autocorrelation for the complete sample selected from each catalogue, and the amplitudes of power-law fits to the data are listed in the legend (using a fixed slope of $-1$).  The Allsky-only sources clearly cluster more strongly than the full AS sample, and more so for the obscured sample.  This is only weakly seen in the AW-only obscured sample.  Some of this effect might be explained by the faint objects only selected by one catalogue lying at the higher redshift end of the full $z$ distribution, but the fact that the behavior is not the same in both cases suggests additional contamination in the AS sample.\label{fig:in_one_cluster}}
\vspace{0.2cm}
\end{figure}

The clustering and lensing cross-correlation properties of these AS and AW-only samples may shed some light on their nature, but their position-dependent density complicates this because of the need for a random sample that mimics their distribution.  Instead, we measure cross-correlations with the full samples from each catalogue, which are more uniformly distributed and can be normalized with our uniform random sample.  This measurement is done via \citep[e.g.][]{1999MNRAS.305..547C}:
\begin{equation}
\omega(\theta) = \frac{D_{\textrm{F}}D_{\textrm{O}}}{D_{\textrm{F}}R} -1,
\end{equation}
where the `F' and `O' subscripts indicate the full and only-in-one catalogue samples, respectively.  The results are shown in Figure~\ref{fig:in_one_cluster}.  For comparison, the autocorrelation measurements for the full AS and AW samples are shown in green in each panel.   The AS-only sources clearly show a stronger clustering signal relative to the full sample, increasingly so going from unobscured to obscured objects.  There is some indication that the AW-only obscured sources cluster more strongly than the full AW sample, but at much lower significance.  

To quantify this, we fit a simple power-law to the data of the form $\omega_{\textrm{qq}}(\theta) = A\theta^{-1}$ (note that we do not fit our model DM autocorrelation to these measurements, as the redshift distributions for the AS and AW-only samples are not well constrained).  The power-law slope of $-1$ is a typical value for quasar autocorrelations in angular projection \citep[][D14]{2006ApJ...638..622M, 2007AJ....133.2222S, 2009ApJ...697.1656S, 2009ApJ...697.1634R, 2012MNRAS.424..933W}, and fits the full sample results here well.  The amplitudes of these fits for each sample are given in the figure legend, and highlight the qualitative impression discussed above.  

Since objects only selected by one catalogue tend to be faint (see section 4.4.1) it is likely that at least some of these sources represent the higher redshift end of the distribution, on average, partially explaining their larger clustering signal.  However, it isn't clear why this would affect one sample more than the other, which suggests that there is some additional signal from contamination present in the AS-only sources that isn't present in the AW-only objects.  The fact that this is stronger in the obscured subsample is also reflective of why the bias of the obscured sample is affected more significantly by changing samples and masks. 


\begin{figure}
\centering
\vspace{0.3cm}
\hspace{0cm}
   \includegraphics[width=8.5cm]{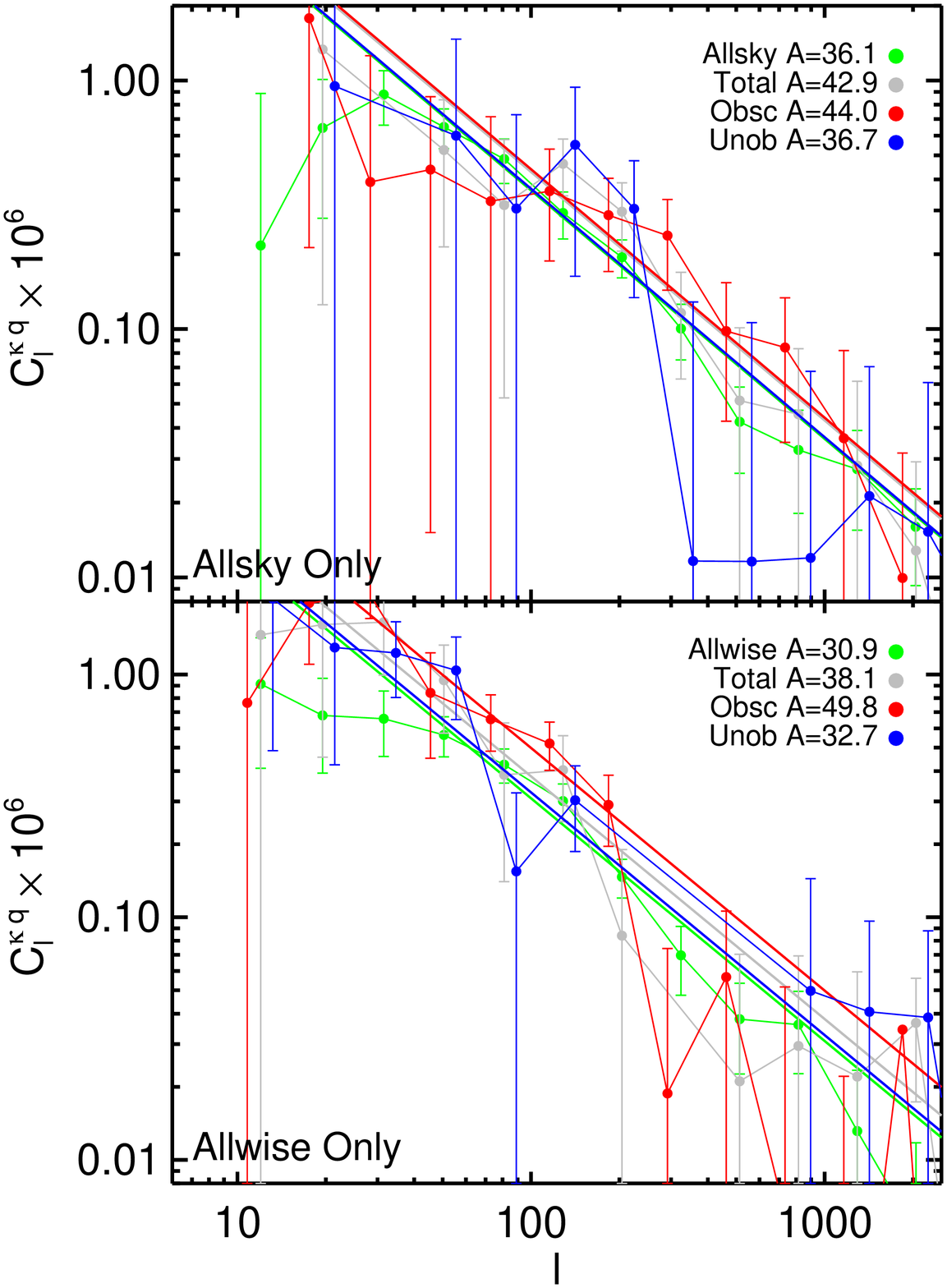}
    \vspace{0cm}
  \caption{A comparison of the CMB lensing cross-correlation for objects selected by Allsky only (top) and Allwise only (bottom), after applying both masks to each sample (differences not due to masking).  In each panel, the green points show the results for the complete sample selected from each catalogue.  There is significant noise in all cases due to the small sample sizes, but there is a signal present indicating that these objects are not just artifacts and are indeed extragalactic.  However, there is no evidence that the cross-correlation signal is different from the full samples.  Without understanding the $dndz$ of these samples, these results are difficult to interpret.\label{fig:in_one_lensing}}
\vspace{0.2cm}
\end{figure}

We also cross-correlate the AS and AW-only samples with the CMB lensing maps, and the results are shown in Figure~\ref{fig:in_one_lensing}.  The green points in each panel show the result from the full AS and AW samples.  To zeroth order, the fact that there is any cross-power here confirms what we found above, that many of these objects are indeed extragalactic, probably at the high redshift end of our sample.  We again fit fixed-slope-power-laws to these results for more direct comparisons, and the amplitudes are given in the legend. Given the amount of noise in these measurements, it is difficult to draw stronger conclusions other than the AS and AW-only samples have a similar cross-power as the full samples.

\subsubsection{Cross-correlations with systematics}
\begin{figure*}
\centering
\vspace{0.3cm}
\hspace{0cm}

   \includegraphics[width=5.5cm]{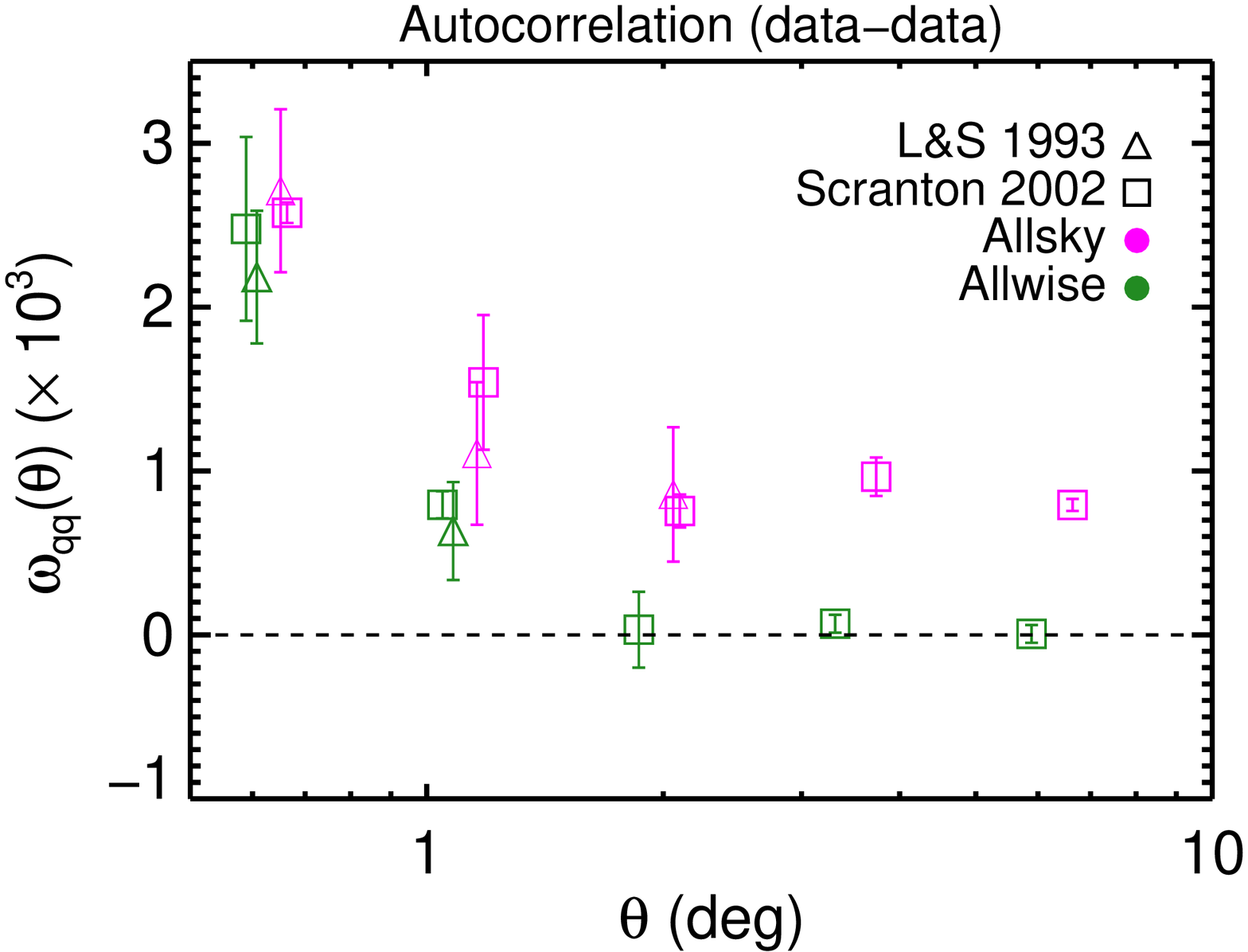}
   \includegraphics[width=5.5cm]{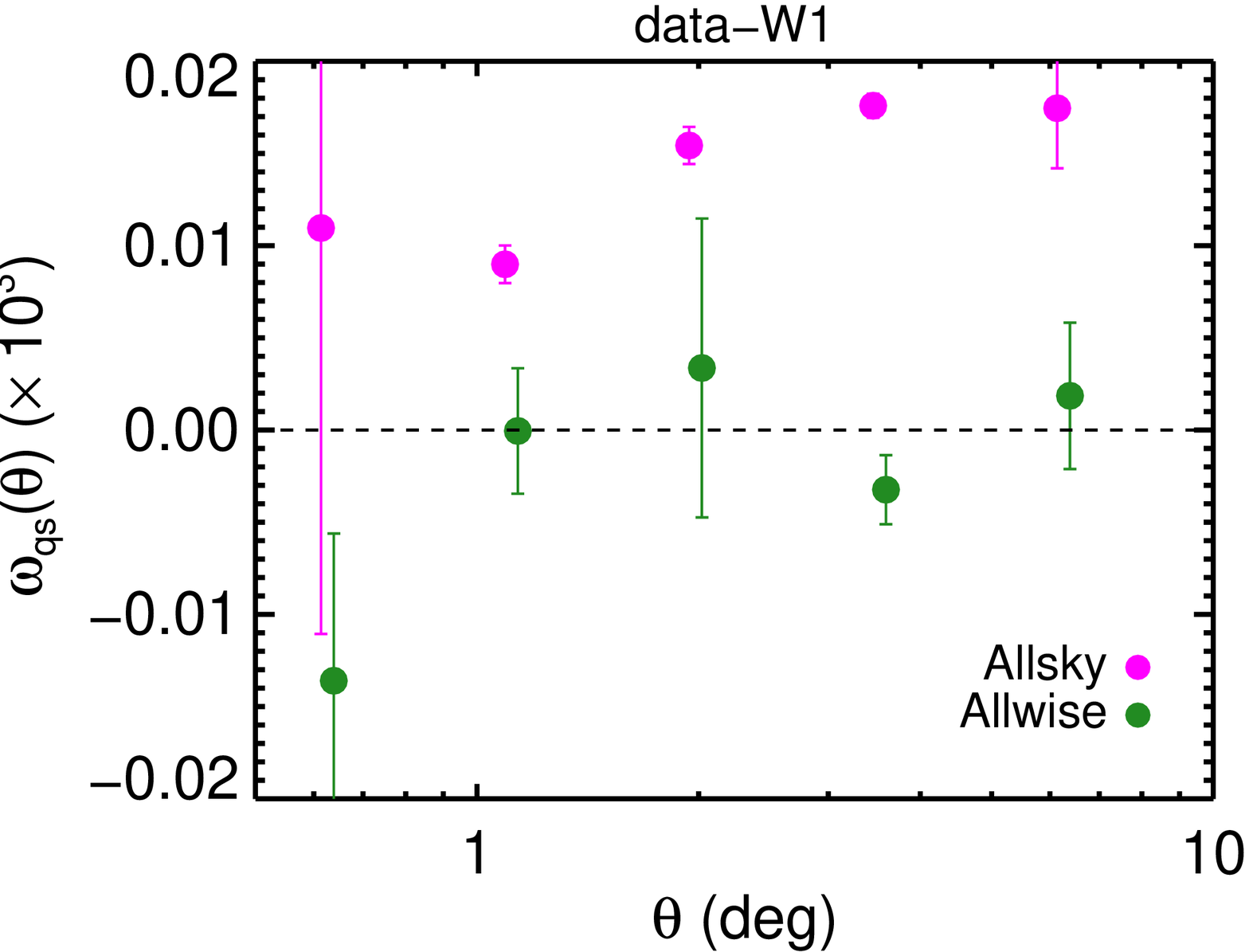}
   \includegraphics[width=5.5cm]{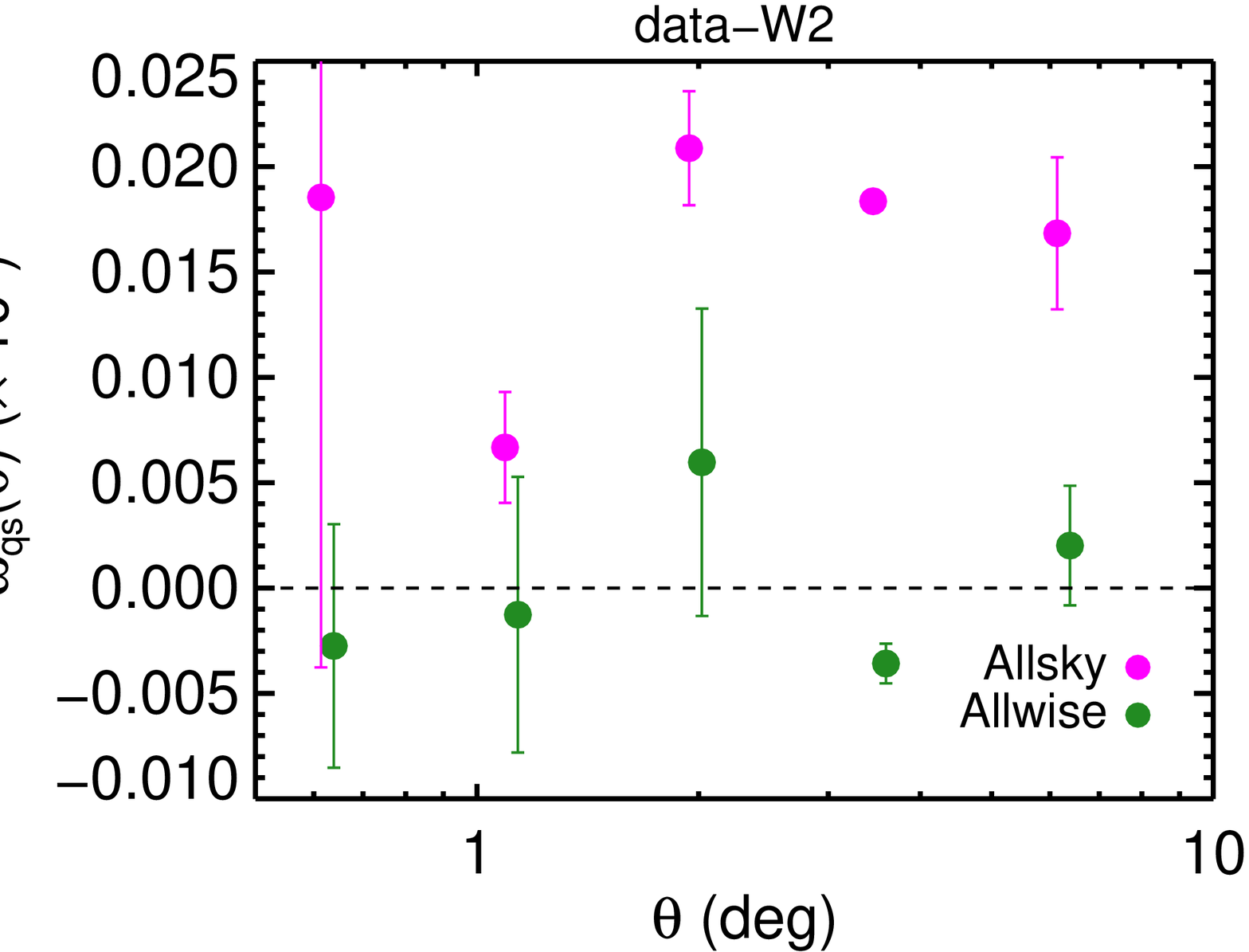}

   \vspace{0.4cm}
   
   \includegraphics[width=5.5cm]{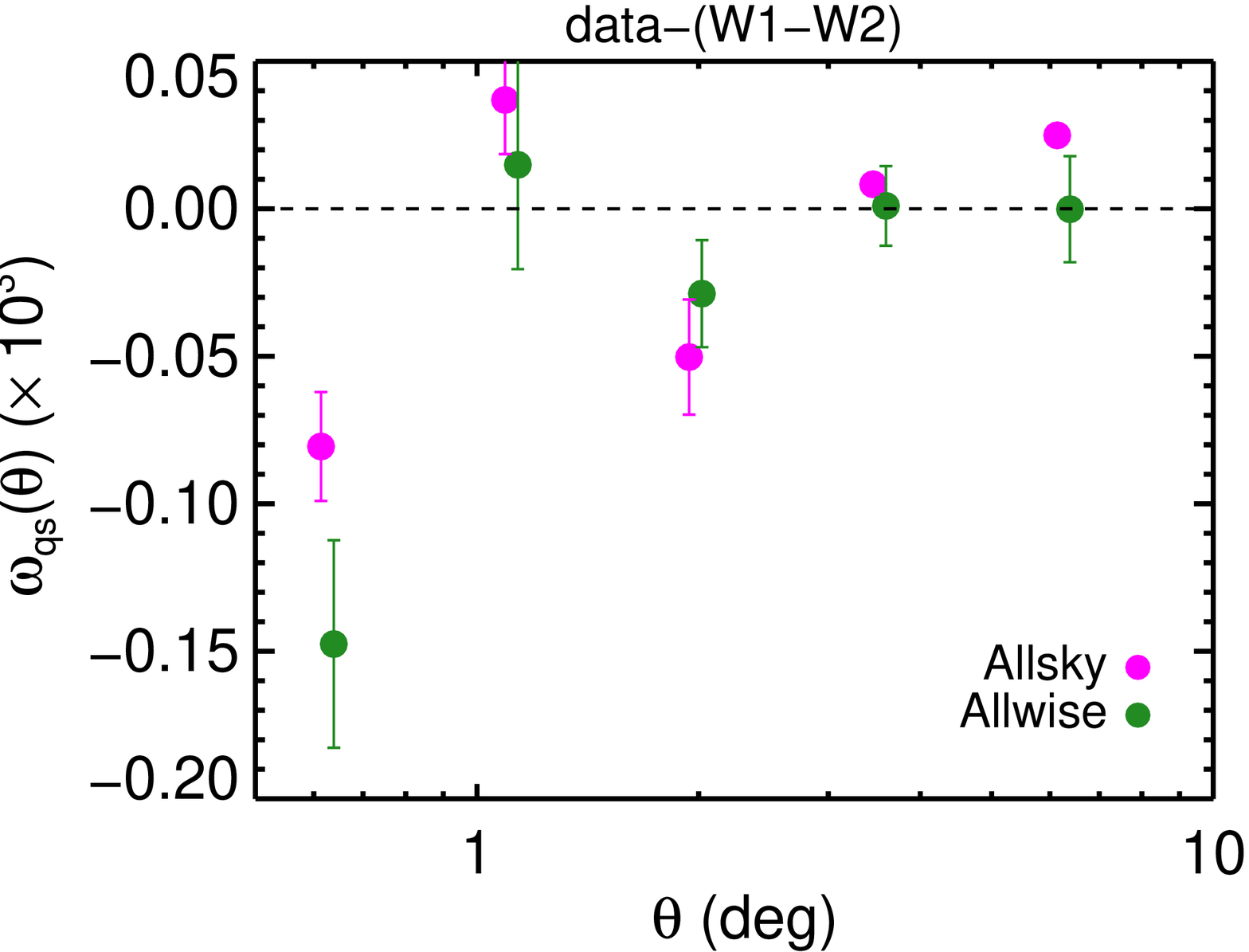}
   \includegraphics[width=5.5cm]{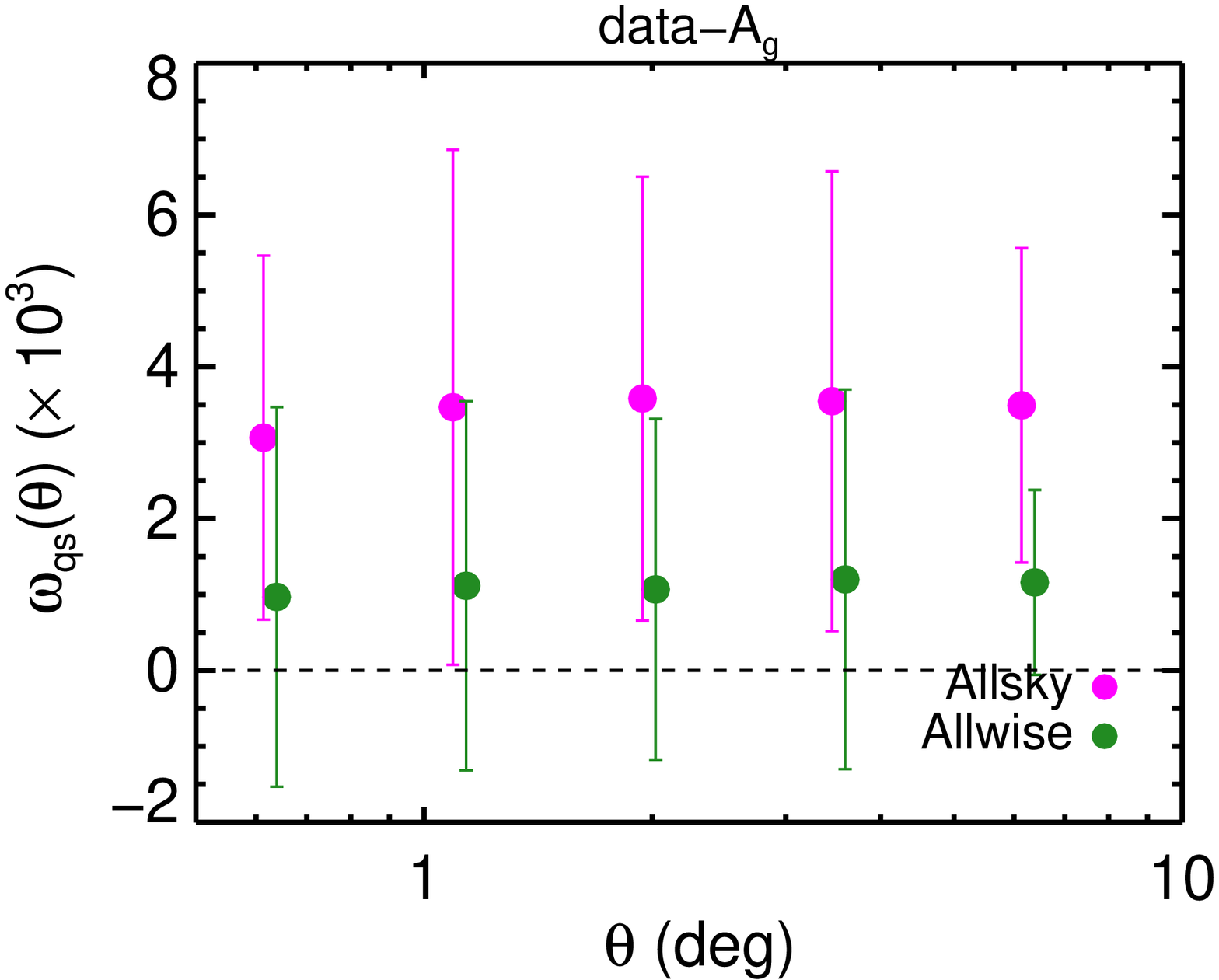}
   \includegraphics[width=5.5cm]{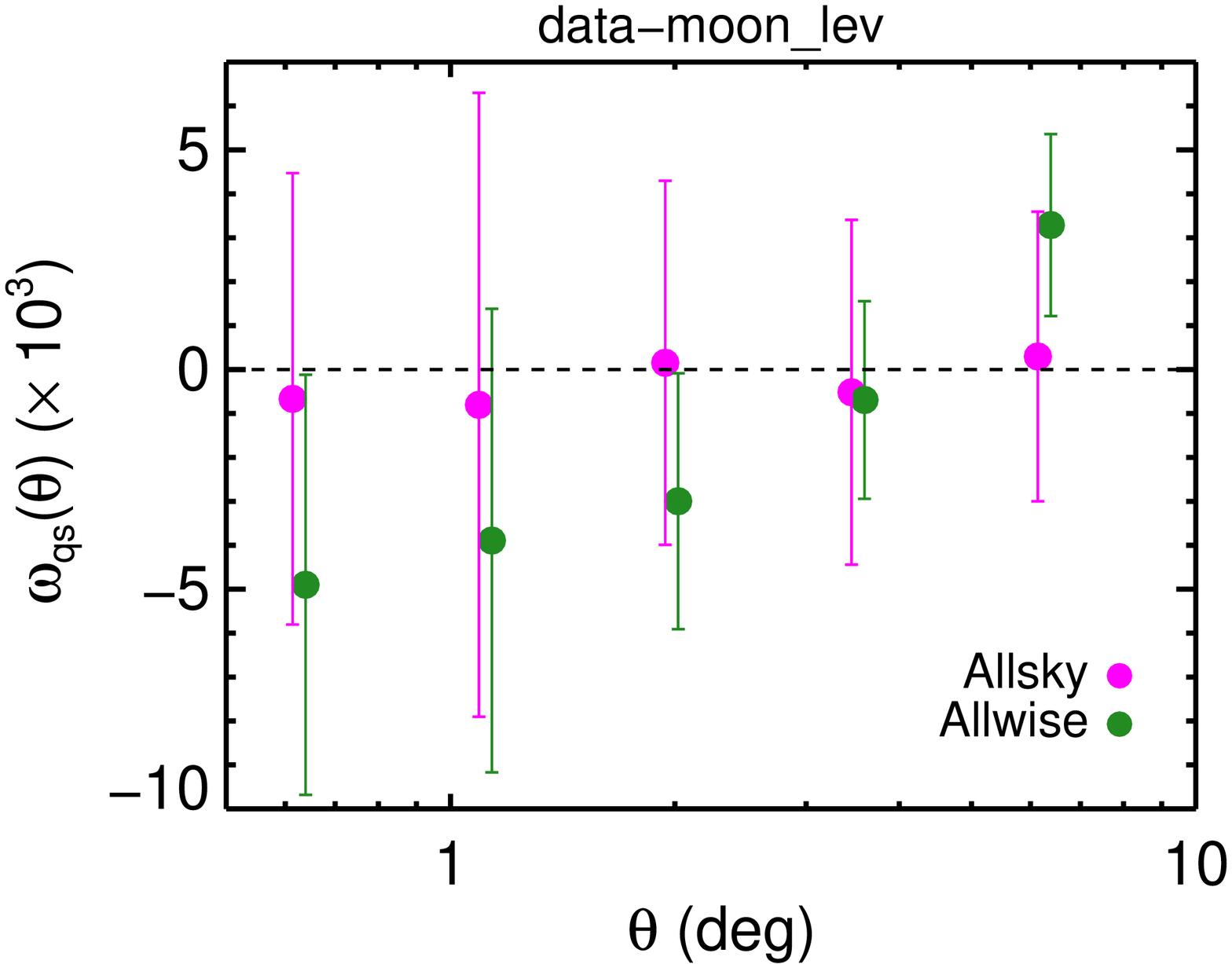}   

    \vspace{0cm}
  \caption{All panels include a dashed line marking zero correlation as a guide. \emph{Top left:} The autocorrelation function of the total Allsky and Allwise samples on large scales, illustrating both the agreement between the two methods used for the calculation (see sections 3.1 and 3.2) and the fact that the Allwise data has less excess signal on large scales. \emph{Clockwise from top-center:} Cross-correlations of the Allsky and Allwise data with various systematics --- $W1$ magnitude, $W2$ magnitude, $W1-W2$ color, extinction in the $g$-band ($A_g$), and the Moon level in the \wise\ image tiles.  Allwise cross-correlations are consistent with zero in most cases, while those with Allsky are not, with the exception of the Moon level.\label{fig:data_sys}}
\vspace{0.2cm}
\end{figure*}

To further compare the samples selected from AS and AW, we cross-correlate the quasar density with several systematics, using the method described in section 3.2.  We do this first for the full AW and AS samples, with both masks applied, cross-correlating with $W1$ and $W2$ magnitude, $W1-W2$, Galactic reddening $A_g$, and Moon level.  The results are shown in Figure~\ref{fig:data_sys}.  

The first panel shows the quasar autocorrelation using the pixelization method (Equation~\ref{eq:pix_cross}) as compared to the estimator in Equation~\ref{eq:LS}.  They agree quite well on scales where they overlap, and the pixelization method allows us to probe larger scales (above $\sim$2$^{\circ}$).   On these scales ($>$50 Mpc/$h$ at $z=1$), the quasar autocorrelation should approach zero \citep[e.g.][]{2006ApJ...638..622M, Krumpe:2013tv}.  This is true of the AW-selected sample, but the AS sample retains a significant signal even out to $\sim$7$^{\circ}$.

In the absence of systematic observational effects, $W1$ and $W2$ should be uncorrelated with the quasars as a function of scale.  The next two panels of Figure~\ref{fig:data_sys} show that this is generally true of AW, within errors, but not for AS.   This is also true, though to a lesser degree, for $W1-W2$.

The next panel shows the cross-correlation of the quasar density with Galactic reddening in the $g$-band, $A_g$.  Again, this is consistent with zero for AW (though there is a slight systematic shift from zero), but not AS.  Note that the reddening component of the mask did not change from AS to AW, and so this reflects a change in the data itself.  It is also interesting to note the flatness of these cross-correlations, which is most likely a consequence of the fact that the Galactic dust density varies slowly with scale, on average.

Finally, the last panel shows the cross-correlation with the Moon level.  Recall that regions with {\tt moon\_lev} $>1$ in $W4$ were masked, which does leave some minor Moon contamination possible.  In this case, the results are reversed --- the AS data do not correlate with the Moon level at all, but AW is anti-correlated.  This seems to imply that there may be problems with \wise\ data calibration due to the Moon present in AW that were not present in AS.

\begin{figure*}
\centering
\vspace{0.3cm}
\hspace{0cm}

   \includegraphics[width=5.5cm]{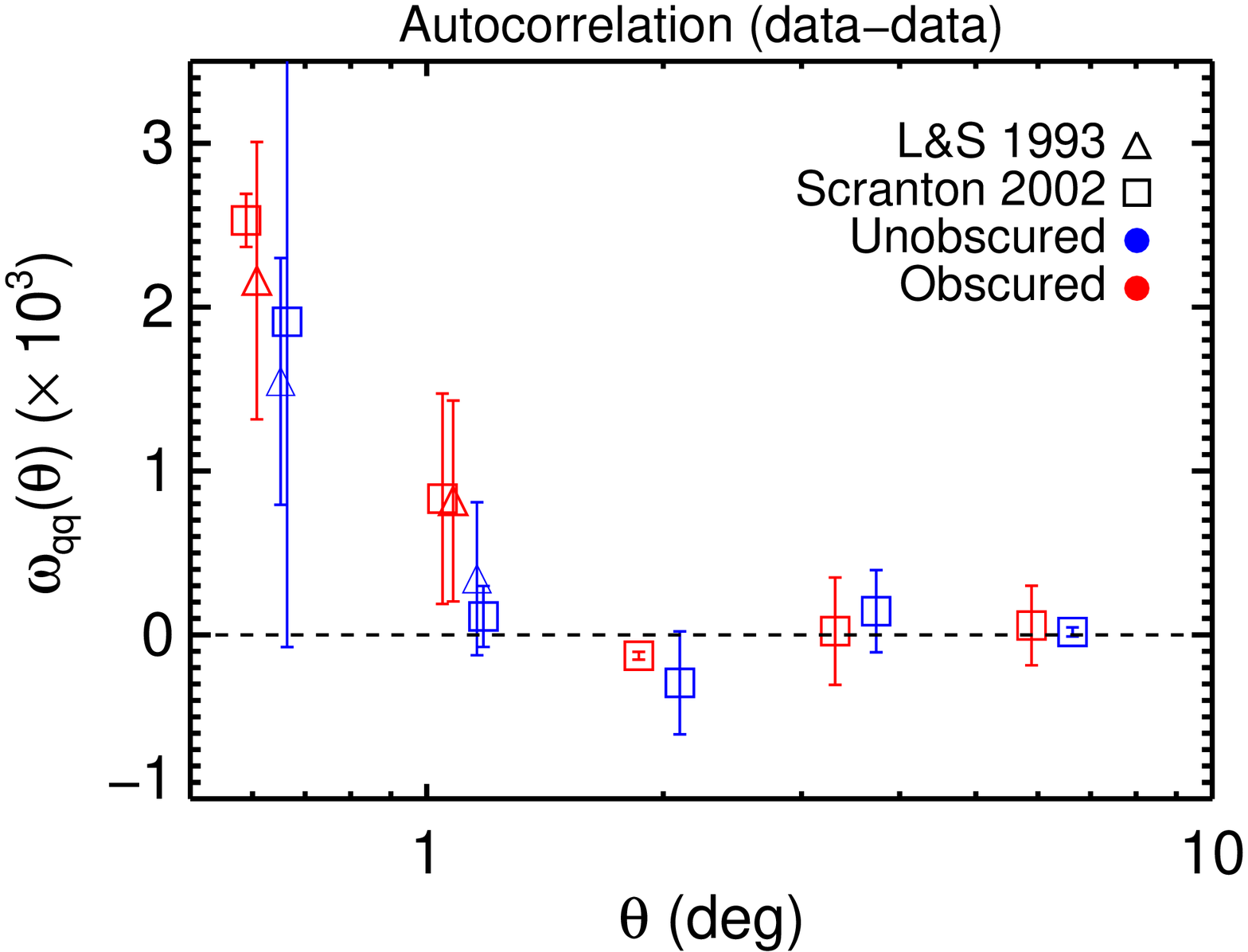}
   \includegraphics[width=5.5cm]{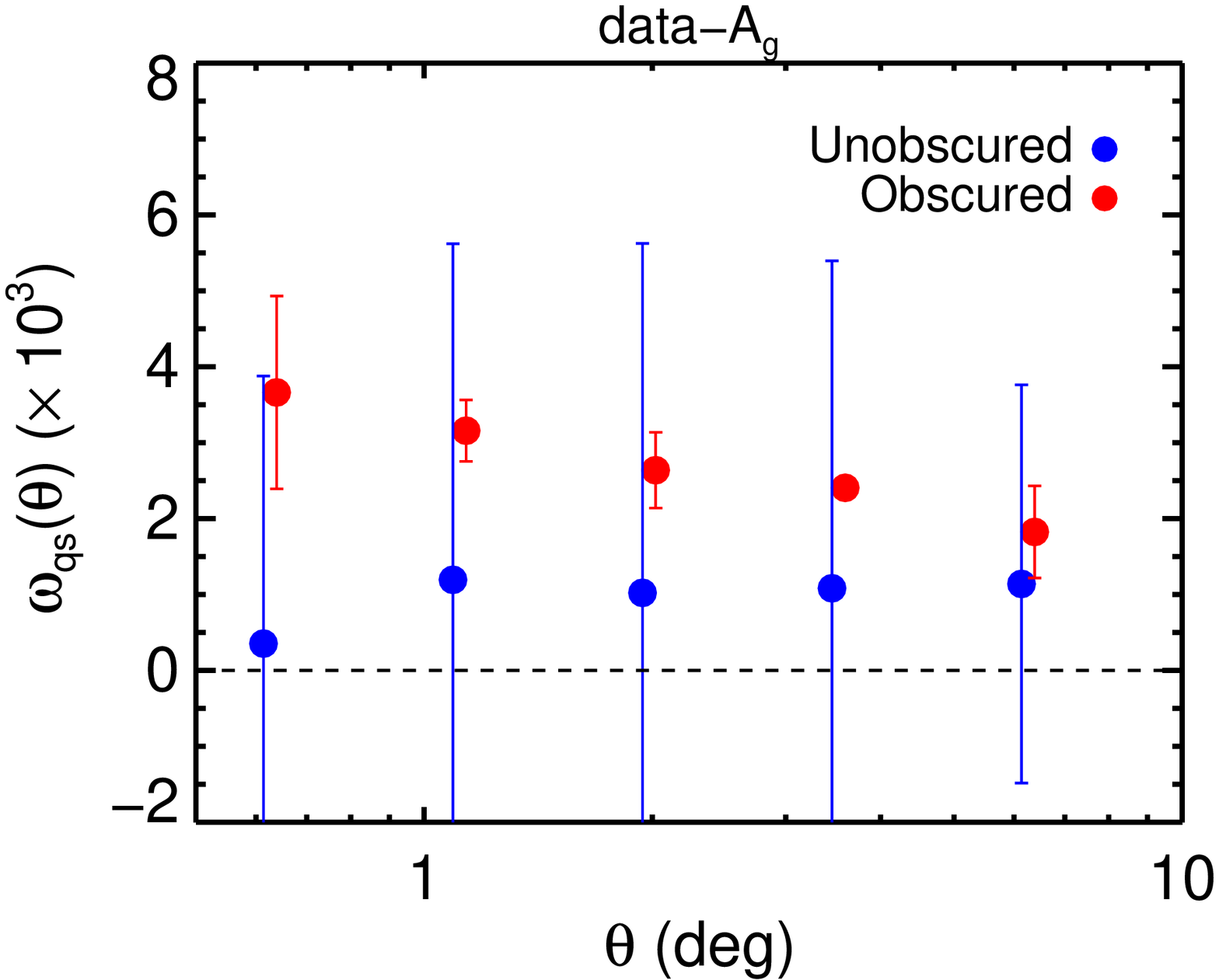}
   \includegraphics[width=5.5cm]{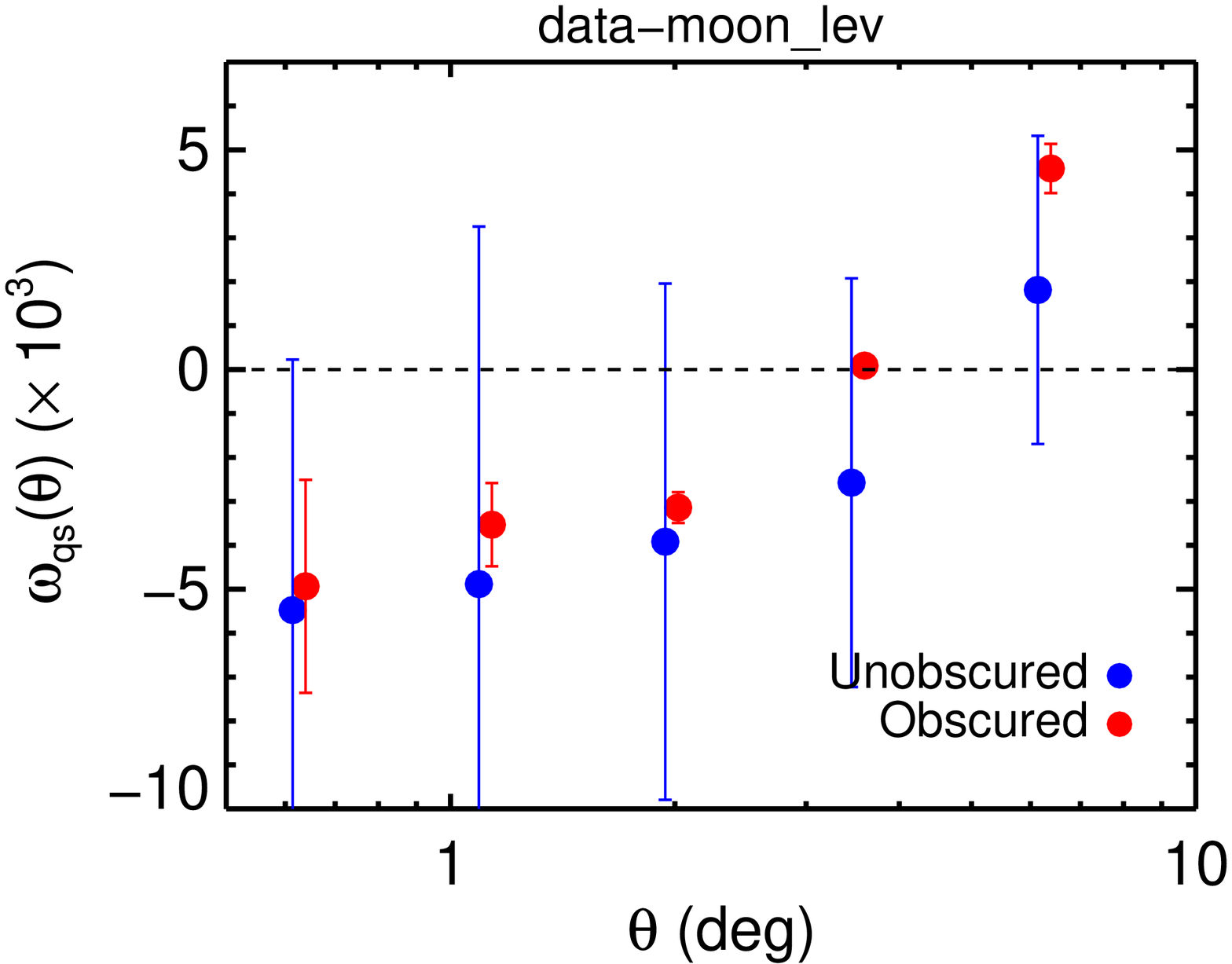}   

    \vspace{0cm}
  \caption{All panels contain a dashed line at zero correlation as a guide.  \emph{Left:} The autocorrelation function of Allwise obscured and unobscured quasars, illustrating both the agreement between the two methods (see sections 3.1 and 3.2) and the lack of signal on large scales for both samples. \emph{Center, right:} Cross-correlation of the samples with Galactic extinction in the $g$-band ($A_g$; center), and the Moon level in the \wise\ imaging tiles (right).  The correlation of the data with $A_g$ appears to be primarily driven by the obscured sample, while the correlation with the Moon level is similar for both.\label{fig:data_sys_type}}
\vspace{0.2cm}
\end{figure*}

Because the change from AS to AW seems to affect the obscured sample more than the unobscured, we focus in Figure~\ref{fig:data_sys_type} on cross-correlations of these subsamples of the AW-selected quasars with systematics.  The first panel shows agreement with the pair-counting method for calculating the autocorrelation, as well as the fact that both autocorrelations approach zero on larger scales.  We omit the cross-correlations with $W1$, $W2$, and $W1-W2$ as these are null for both subsets of data.  However, as seen in the center panel, obscured sources do correlate with reddening in this sample, while unobscured sources do not.  In the final panel, we see that both obscured and unobscured sources correlate in a similar way with the Moon level.

\section{DISCUSSION}
\subsection{Which measurement is the most reliable?}
Based on the analysis of \planck\ DR1/DR2 and the quasars selected from the \wise\ Allsky and Allwise catalogues, it is clear that the latter in both cases are superior and more reliable data sets.  The apparent reduction in correlated noise in \planck\ DR2, along with the increased lensing signal are obvious improvements.  For the \wise\ selected quasars, the lack of correlations with $W1$ and $W2$ magnitudes and $A_g$, as well as the more uniform distribution of AW-only sources, indicate that this catalogue has certainly improved systematic effects that impact selection and contamination by artifacts. However, Figure~\ref{fig:rel_densities} illustrates that this is not completely eliminated with AW, and there is still a slight correlation of obscured quasars with Galactic reddening in AW.

Despite these overall improvements, the correlation of AW quasars with the Moon level is concerning.  In order to address this, we perform one final check by repeating our clustering and CMB lensing cross-correlation analysis with any region with {\tt moon\_lev} $>0$, in any \wise\ band, masked out.   This removes an additional $\sim$1000 deg$^2$ of area, and the reduction in sample sizes inflates the uncertainties accordingly.  However, we find that the results do not change significantly, though the bias is slightly reduced for both samples (by a few per cent).  While it appears that the Moon may have an adverse effect on our selection, we are unable to remove it from our measurement with the current information from \wise.


Given the lack of understanding on how the Moon level is affecting the AW-selected quasars, and the fact that the objects selected only in AW still show a position-dependent relative density, our most conservative approach currently is to use objects selected as quasars by both AW and AS.  This naturally includes the mask components of both data sets as well.  We adopt the bias and halo masses from this sample as our best current constraints, and these values are summarized in Table~\ref{tbl:results} for convenience.

\begin{table}
  \caption{Adopted bias and halo mass measurements}
  \label{tbl:results}
  \begin{tabular}{lcccc}
  \hline
                                 &  &  $\langle z \rangle$  & $b_q$             & $\log(M_h/M_{\odot} \ h^{-1})$   \\
  \hline
                                 &  &                            \multicolumn{3}{c}{Quasar Autocorrelation}                \\          
                                                                                      \cline{3-5}                      \\
  All IR            &  & 1.02                         & 1.92$\pm$0.12 & $12.79_{-0.11}^{+0.10}$    \\
  Obscured     &  & 0.98                         & 2.13$\pm$0.21 & $13.00_{-0.16}^{+0.14}$     \\
  Unobscured &  & 1.05                         & 1.88$\pm$0.15 & $12.72_{-0.15}^{+0.13}$    \\
\\
                                 &  &                            \multicolumn{3}{c}{CMB Lensing cross-correlation}                \\          
                                                                                      \cline{3-5}                      \\
   All IR            &  & 1.02                         &  1.87$\pm$0.14 & $12.74_{-0.14}^{+0.12}$   \\
  Obscured     &  & 0.98                         &  2.06$\pm$0.22 & $12.94_{-0.18}^{+0.15}$   \\
  Unobscured &  & 1.05                         &  1.72$\pm$0.18 & $12.56_{-0.21}^{+0.17}$   \\ 
\hline
   \end{tabular}
   \\  
{
\raggedright    
 The adopted bias and halo masses from the angular autocorrelation (top) and CMB lensing cross-correlations (bottom), using the most conservative sample --- those objects satisfying the selection criteria in both Allsky and Allwise catalogues.\\
 }
\end{table}

\subsection{The redshift distribution and evolution of the bias}
There are two approaches we have adopted that can impact the models from Equations~\ref{eq:omega_mod} and \ref{eq:lens_model}, and thus the final inferred bias values.  The first is in the way we handle the empirical redshift distributions. While the spline interpolation of $dN/dz$ preserves the features seen in Figure~\ref{fig:z_dist}, it is possible that some of these are artificial and do not fully reflect the true intrinsic distribution (though the application of our selection criteria to the B\"{o}otes field should mitigate some of these issues).  Therefore, we also test fitting a smoother function to these distributions --- a Gaussian with an exponential tail --- which results in a smoother overall $dN/dz$.  Propagating these fits through to the models, we find that they shift by at most a few percent over the scales of interest.  The effect is most dramatic for the unobscured sample, due to the spike at low-$z$, but still only shifts the model by $\sim$4\%, and only at larger scales.  Re-fitting the effective bias, we find that the results shift by $\sim$1\%.  Our choice of model for $dN/dz$ does not impact our results significantly.

We have also adopted a simple bias model that does not account for evolution with redshift, and we are really measuring the effective bias, or the bias averaged over our redshift range.  Since we lack individual source redshifts, and our error bars are still sufficiently large, we are unable to accurately generate our own empirical $b(z)$ model directly from our data.  However, we explore the role of such evolution in our fits, by including in our model calculations the $b(z)$ of \citet{2005MNRAS.356..415C}: $b(z) = 0.53 + 0.289(1+z)^2$.  We then fit these new models, with an additional scale factor $b_0$, to our data.  

In general, a model of the bias including redshift evolution does not significantly change the quality our fits (based on the values of $\chi^2$).  In the case of the angular autocorrelation, both obscured and unobscured populations require an additional rescaling, with $b_{0,\textrm{obsc}} = 1.34$ and $b_{0,\textrm{unob}} = 1.26$.  Including these factors and inserting the mean redshifts of each sample into the model (i.e. $b = b_0 [0.53 + 0.289(1+\langle z \rangle)^2$]), we derive bias values of $b_{\textrm{obsc}} = 2.23 \pm 0.18$ and $b_{\textrm{unob}} = 2.21 \pm 0.13$.  These are both increased but roughly consistent with our simpler effective bias measurement, and the values for the two populations are consistent.  For the CMB lensing cross-correlations, we measure $b_{0,\textrm{obsc}} = 1.28$ and $b_{0,\textrm{unob}} = 1.04$.  In this case, the unobscured sample is consistent with the fiducial unobscured \citet{2005MNRAS.356..415C} model, while the obscured sample requires additional rescaling.  With this model we find $b_{\textrm{obsc}} = 2.13 \pm 0.14$ and $b_{\textrm{unob}} = 1.82 \pm 0.11$, fully consistent with our values assuming a constant bias.

Given that we are unable to currently determine $b(z)$ for these samples directly, the fact that such a model does not improve our fits, and that overall the results are consistent between a constant and evolving bias model, we prefer to adopt the constant bias measurements here as they involve fewer assumptions.  Future work will revisit this issue in more detail.

\subsection{Comparison with previous results}
There have been several measurements of the obscured and unobscured quasar bias and halo masses in the last several years, as it is only recently that samples have grown large enough to do so precisely.  There is yet to be convergence on a result.

Our values here are lower than those of D14 and D15 due both to the subtle differences in the samples selected as discussed above, as well as some updates in our procedures.  First, the use of CAMB with its most recent default parameters to produce the model power spectra in a consistent fashion for both the clustering and lensing cross-correlation measurements results in models with slightly larger amplitudes, which naturally decreases the inferred bias.  This is seen in the ``DR1/Allsky'' points of Figure~\ref{fig:all_bias}, as compared to e.g. Figure 7 of D15.  This is a systematic effect that impacts both the obscured and unobscured samples, but not their relative values.  However, in Figure~\ref{fig:all_bias} we also illustrate that using the most recent data results in a decrease of the bias as well, for both unobscured and obscured quasars, and more significantly for the clustering measurement.  As we argued in the previous section, the new data are quantifiably superior, and these new values should be considered more reliable.  

It is difficult however to pinpoint the exact reason for this reduction in the bias.  The fact that it is present in both the clustering and CMB lensing cross-correlations, which depend on different systematics, indicates that it is a real effect --- AW is picking out objects with lower clustering and lensing cross-correlation amplitudes, especially for the obscured sample.  Since the AW $W1-W2$ color distribution is slightly bluer there may be a relationship between IR color and bias (as opposed to just optical-IR color, or obscured and unobscured, and bias).  However, exploring such a trend will require even larger samples, over extended areas, in order to have large enough sub-samples to explore with sufficiently small statistical errors.  We will revisit this in future work.  The systematic shift in color in AW may also point to the need for updates to the various \wise\ color selection techniques for quasars.

While D14 showed that the factor of 10 larger halo masses for obscured quasars of \citet{2014ApJ...789...44D} was due to insufficient masking of the \wise\ data, and found instead a factor of $\sim$3 difference, that is further reduced here to roughly a factor of two.  However, the significance of this difference is now only $\sim$2$\sigma$.  

\begin{figure*}
\centering
\vspace{0.3cm}
\hspace{0cm}
   \includegraphics[width=17cm]{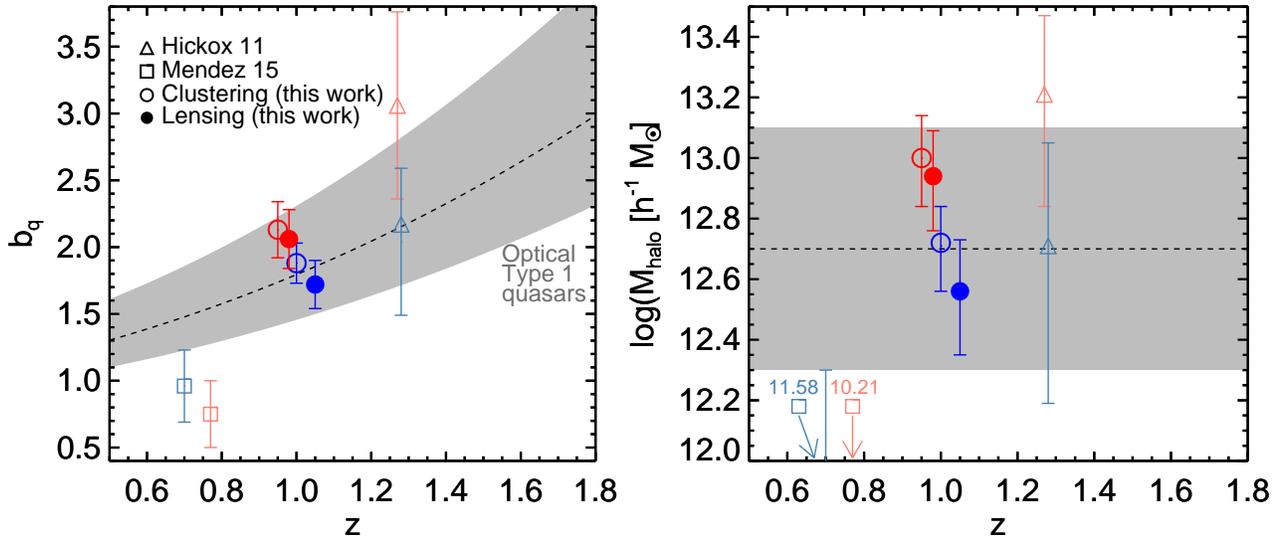}
    \vspace{0cm}
  \caption{The adopted bias (left) and halo mass (right) measurements from Table~\ref{tbl:results} compared to other recent results (comparison with the D14/D15 results can be seen in Figure~\ref{fig:all_bias}).  Points have been shifted slightly in redshift where necessary for clarity.  The grey band represents the range of results typical for optically-selected unobscured quasars, largely from the SDSS.  The \citet{2015arXiv150406284M} results utilize individual redshifts for a projected clustering measurement --- note that their halo mass measurements fall outside of the plot range (though the obscured halo mass error bar can be seen), so the actual values are listed.  While the \citet{2011ApJ...731..117H} results agree well with what we find here, the \citet{2015arXiv150406284M} estimates are significantly lower.  Obscured quasars have halo masses $\sim$2 times larger than unobscured quasars, with a significance of $\sim$2$\sigma$.\label{fig:bias_comp}}
\vspace{0.2cm}
\end{figure*}

In Figure~\ref{fig:bias_comp} we compare our updated measurements with several from the recent literature.  Note that all halo masses in the right panel are calculated using our prescription based on the bias values taken from the respective studies shown on the left (i.e. the halo masses shown may not be the same as those calculated in the original reference).  The grey band gives the approximate range of values for optically-selected unobscured quasars, primarily from the works referenced in section 1.  We first note that both obscured and unobscured quasars are now roughly consistent with this range, despite the difference in the samples here.  While the measurements of \citet{2011ApJ...731..117H} are at slightly higher redshift, our results remain consistent with theirs.  Though our error bars are significantly smaller due to the dramatic increase in area and sample size, the significance of the difference in bias/halo masses is roughly the same once the reduction in the magnitude of the difference is considered.

We also show the recent results of \citet{2015arXiv150406284M}, at slightly lower redshift.  Their sample used the IR selection technique of \citet{2013ApJ...772...26A}, which extends deeper to $W2 < 17.11$ with a magnitude-dependent color criteria.  They also used samples with individual redshifts from several fields, and we show their results with the COSMOS field excluded, due to this region containing structures that are particularly over dense as compared to the cosmic average \citep[e.g.][]{2007ApJS..172...70L}.  The deeper $W2$ limit likely results in a slightly lower average luminosity, though the requirement of spectroscopic redshifts somewhat  offsets this effect.  Note that the \citet{2015arXiv150406284M} halo masses are quite low (the lowest of all the samples they considered) compared to ours and those of other groups --- more than two orders of magnitude in the obscured quasar case.  Aside from the small halo masses in general, the sense of the difference between obscured and unobscured samples is the opposite of what we find here and was found in \citet{2011ApJ...731..117H}.  

The individual redshifts of \citet{2015arXiv150406284M} should reduce systematic errors due to differences in the redshift distributions that could be present in our measurements, despite the improved statistical errors we achieve with a larger sample.  However, our bias values for the obscured and unobscured samples differ by $\sim$10\%.  As shown in Figure 8 of D15, our effective redshift estimates would need to be systematically offset by $\sim$0.25 to fully account for this difference.  It is unlikely that the B\"{o}otes field is this poorly representative of the full population.  On the other hand, considering that the bias is not weighted equally at each redshift, the presence of a high-$z$ tail in the obscured sample could skew the bias significantly.  Such a tail would most likely be present in the sources lacking SDSS counterparts.  As a check we repeat our clustering and CMB lensing cross-correlation for only the obscured sources with $r$-band detections.  These objects do not have a significantly different redshift distribution in B\"{o}otes, and as in D14/D15, the measurement is completely consistent with that of the entire obscured sample.  While incorporating full redshift information into our measurements would certainly be an improvement, it is difficult to see how redshift distribution errors would dramatically impact our findings.

\subsection{Interpretation and future work}
While the magnitude and significance of the difference in bias and halo mass between obscured and unobscured quasars is reduced in this work, it is still present.  The simplest interpretation of this is that obscured quasars reside in higher mass haloes, and are not simple analogues of unobscured quasars seen through a dusty torus.  They could instead represent different phases common to the quasar phase of black holes.  

It is also possible that our obscured sample is a mix of genuinely obscured objects and lower luminosity unobscured AGN.  This ``contamination'' could imply either that lower luminosity AGN are more biased \citep[which is unexpected, and previous results regarding trends of bias with luminosity have been weak at best;][]{2009ApJ...697.1656S, 2013ApJ...778...98S, Krolewski:2015wb, 2015arXiv150708380E}, or that the signal from the true obscured population is being diluted.  In addition, it is likely that some fraction of quasars are in fact obscured only due to our particular line of sight, and are intrinsically the same as the unobscured population if seen from a different angle.  This would also serve to dilute the bias measurement of sources obscured by other means.  Both of these factors imply that the difference we are finding between obscured and unobscured quasars may be a lower limit.

We again use abundance matching techniques \citep[e.g.][]{1999ApJ...523...32C, 1999ApJ...520..437K, 2004MNRAS.353..189V, 2006ApJ...643...14S, 2010MNRAS.404.1111G} to estimate the implied lifetimes of obscured and unobscured phases, assuming there is an evolutionary trend.  The median $L_{\textrm{bol}} \sim 10^{46}$ ergs/s \citep{2011ApJ...731..117H, 2014ApJ...795..124H} implies an overall space density at $z =1$ of $\sim$$2 \times 10^{-5}$ Mpc$^{-3}$ \citep{2007ApJ...654..731H}.  This is split by roughly 60\% and 40\% unobscured and obscured, respectively.  For the halo masses measured here (we adopt the results from CMB lensing cross-correlations), and the halo mass function of \citet{2010ApJ...724..878T}, we find host halo space densities of $dn/d\log_{10}(M) = 8.1_{-5.3}^{+2.8} \times 10^{-4}$ Mpc$^{-3}$ (unobscured) and $3.0_{-1.8}^{+1.0} \times 10^{-4}$ Mpc$^{-4}$ (obscured).  Over $0.5 < z < 1.5$, or roughly 4 Gyr of cosmic time, this implies lifetimes of $70_{-37}^{+27}$ Myr for the unobscured phase and $100_{-52}^{+38}$ Myr for the obscured phase.  Each is on the order of 1\% of the Hubble time, with the obscured phase lasting about 1.5 times longer than the unobscured phase.

Given the reduced significance of the difference in halo masses found here, it is possible that the large-scale bias of obscured and unobscured quasars is consistent with a pure orientation model.  However, there are other indications that even if these sources occupy similar mass haloes overall, there are other differences.  For example, the smaller scale signal (below 1 Mpc/$h$, or $0.04^{\circ}$ at our redshifts) appears to differ between the samples in both the clustering and CMB-lensing cross correlations.  This could imply a difference in the HODs and in particular the satellite fractions of each population \citep{2005ApJ...633..791Z, 2007ApJ...659....1Z, 2012MNRAS.424..933W, 2013ApJ...779..147C}.  However, studying the HOD of obscured quasars in detail is difficult without full redshift information, but such additional information could provide higher-order correlation functions or direct measurement of the mean occupation function of obscured quasars \citep{2013ApJ...779..147C}.

Our next steps are to leverage the all sky nature of \wise\, along with the full footprint of SDSS and other optical surveys (e.g. The Dark Energy Survey) to build the largest IR-selected obscured and unobscured quasar samples possible.  This will provide dramatically improved statistical error bars on these bias measurements, to further constrain the potential difference in halo masses.  We will also use photometric redshift estimations from multi-wavelength photometry and SED fitting \citep[e.g.][]{2014ApJ...795..124H, 2015MNRAS.452.3124D, Carroll:2015vp} to explore the redshift evolution of obscured quasars.

\section{Summary}
Using the most recent Allwise source catalogue from \wise, along with new products from \planck, we present updated measurements of the obscured and unobscured quasar bias via angular autocorrelations and cross-correlation with CMB lensing.  We find, as with our previous work, that obscured quasars have a larger bias and therefore halo mass and longer lifetime.  However, this difference is reduced with respect to our results using previous data products from \wise\ and \planck.  The inferred typical halo mass of obscured quasars is roughly a factor of two larger than that of unobscured quasars, at a significance of $\sim$1-2$\sigma$.  

In order to explain this change, we have carefully compared the properties of quasars selected from both \wise\ catalogues using standard color-cuts.  The general properties of these sources --- morphology, redshift distribution, unobscured/obscured fractions --- are indistinguishable.  However, the photometric properties do differ slightly.  In particular, the more recent \wise\ catalogue tends to select brighter and bluer quasars, which could contribute to the change in our results.

We have also carefully explored systematic effects in \wise-selected quasars from both catalogues, and find that the updated version is clearly superior.  In particular, the quasar selection in the Allwise catalogue appears less biased with respect to flux measurements in $W1$ and $W2$, as well as with Galactic extinction.  However, there may be some complications with respect to Moon contamination in Allwise, which leads us to rely on the sample that is selected from both catalogues for our current measurement.  

Finally, we make available with this paper several sets of codes for making these measurements and generating consistent models.


\section*{Acknowledgements}
MAD, RCH, and ADM were partially supported by NASA through ADAP award NNX12AE38G and by the National Science Foundation through grant number 1211096.  MAD and ADM were also supported by NSF grant numbers 1211112 and 1515404, and MAD and RHC were also supported by NSF grant number 1515364.  RCH also acknowledges support from an Alfred P. Sloan Research Fellowship, and a Dartmouth Class of 1962 Faculty Fellowship.

\clearpage

\bibliography{/Users/Mike/Research/bibliographies/full_library}

\label{lastpage}

\end{document}

%% file: commands_mnras.tex
\newcommand\ion[2]{#1$\;${\small\rmfamily\@Roman{#2}}\relax}%

\def\lsim{\lower0.3em\hbox{$\,\buildrel <\over\sim\,$}}
\def\gsim{\lower0.3em\hbox{$\,\buildrel >\over\sim\,$}}

%% file: bias_allwise.bbl
\begin{thebibliography}{127}
\expandafter\ifx\csname natexlab\endcsname\relax\def\natexlab#1{#1}\fi

\bibitem[{Abazajian {et~al}\mbox{.}(2009)Abazajian, Adelman-McCarthy,
  Ag{\"u}eros, Allam, Allende~Prieto, An, Anderson, Anderson, Annis, Bahcall,
  Bailer-Jones, Barentine, Bassett, Becker, Beers, Bell, Belokurov, Berlind,
  Berman, Bernardi, Bickerton, Bizyaev, Blakeslee, Blanton, Bochanski, Boroski,
  Brewington, Brinchmann, Brinkmann, Brunner, Budav{\'a}ri, Carey, Carliles,
  Carr, Castander, Cinabro, Connolly, Csabai, Cunha, Czarapata, Davenport,
  de~Haas, Dilday, Doi, Eisenstein, Evans, Evans, Fan, Friedman, Frieman,
  Fukugita, G{\"a}nsicke, Gates, Gillespie, Gilmore, Gonzalez, Gonzalez,
  Grebel, Gunn, Gy{\"o}ry, Hall, Harding, Harris, Harvanek, Hawley, Hayes,
  Heckman, Hendry, Hennessy, Hindsley, Hoblitt, Hogan, Hogg, Holtzman, Hyde,
  Ichikawa, Ichikawa, Im, Ivezi{\'c}, Jester, Jiang, Johnson, Jorgensen,
  Juri{\'c}, Kent, Kessler, Kleinman, Knapp, Konishi, Kron, Krzesinski,
  Kuropatkin, Lampeitl, Lebedeva, Lee, Lee, French~Leger, L{\'e}pine, Li, Lima,
  Lin, Long, Loomis, Loveday, Lupton, Magnier, Malanushenko, Malanushenko,
  Mandelbaum, Margon, Marriner, Mart{\'\i}nez-Delgado, Matsubara, McGehee,
  McKay, Meiksin, Morrison, Mullally, Munn, Murphy, Nash, Nebot, Neilsen,
  Newberg, Newman, Nichol, Nicinski, Nieto-Santisteban, Nitta, Okamura,
  Oravetz, Ostriker, Owen, Padmanabhan, Pan, Park, Pauls, Peoples, Percival,
  Pier, Pope, Pourbaix, Price, Purger, Quinn, Raddick, Re~Fiorentin, Richards,
  Richmond, Riess, Rix, Rockosi, Sako, Schlegel, Schneider, Scholz, Schreiber,
  Schwope, Seljak, Sesar, Sheldon, Shimasaku, Sibley, Simmons, Sivarani,
  Allyn~Smith, Smith, Smolcic, Snedden, Stebbins, Steinmetz, Stoughton,
  Strauss, SubbaRao, Suto, Szalay, Szapudi, Szkody, Tanaka, Tegmark, Teodoro,
  Thakar, Tremonti, Tucker, Uomoto, Vanden~Berk, Vandenberg, Vidrih, Vogeley,
  Voges, Vogt, Wadadekar, Watters, Weinberg, West, White, Wilhite, Wonders,
  Yanny, Yocum, York, Zehavi, Zibetti, \& Zucker}]{2009ApJS..182..543A}
Abazajian K.~N. {et~al.}, 2009, ApJS, 182, 543

\bibitem[{Agarwal, Ho \& Shandera(2014)Agarwal, Ho, \&
  Shandera}]{2014JCAP...02..038A}
Agarwal N., Ho S., Shandera S., 2014, JCAP, 02, 038

\bibitem[{Alexander \& Hickox(2012)}]{2012NewAR..56...93A}
Alexander D.~M., Hickox R.~C., 2012, NewAR, 56, 93

\bibitem[{Antonucci(1993)}]{1993ARA&A..31..473A}
Antonucci R., 1993, ARA{\&}A, 31, 473

\bibitem[{Assef {et~al}\mbox{.}(2015)Assef, Eisenhardt, Stern, Tsai, Wu,
  Wylezalek, Blain, Bridge, Donoso, Gonzales, Griffith, \&
  Jarrett}]{2015ApJ...804...27A}
Assef R.~J. {et~al.}, 2015, ApJ, 804, 27

\bibitem[{Assef {et~al}\mbox{.}(2013)Assef, Stern, Kochanek, Blain, Brodwin,
  Brown, Donoso, Eisenhardt, Jannuzi, Jarrett, Stanford, Tsai, Wu, \&
  Yan}]{2013ApJ...772...26A}
Assef R.~J. {et~al.}, 2013, ApJ, 772, 26

\bibitem[{Bardeen {et~al}\mbox{.}(1986)Bardeen, Bond, Kaiser, \&
  Szalay}]{1986ApJ...304...15B}
Bardeen J.~M., Bond J.~R., Kaiser N., Szalay A.~S., 1986, ApJ, 304, 15

\bibitem[{Becker, White \& Helfand(1995)Becker, White, \&
  Helfand}]{Becker:1995p345}
Becker R.~H., White R.~L., Helfand D.~J., 1995, Astrophysical Journal v.450,
  450, 559

\bibitem[{Berlind \& Weinberg(2002)}]{2002ApJ...575..587B}
Berlind A.~A., Weinberg D.~H., 2002, ApJ, 575, 587

\bibitem[{Berlind {et~al}\mbox{.}(2003)Berlind, Weinberg, Benson, Baugh, Cole,
  Dav{\'e}, Frenk, Jenkins, Katz, \& Lacey}]{2003ApJ...593....1B}
Berlind A.~A. {et~al.}, 2003, ApJ, 593, 1

\bibitem[{Blanton \& Roweis(2007)}]{2007AJ....133..734B}
Blanton M.~R., Roweis S., 2007, AJ, 133, 734

\bibitem[{Bleem {et~al}\mbox{.}(2012)Bleem, van Engelen, Holder, Aird,
  Armstrong, Ashby, Becker, Benson, Biesiadzinski, Brodwin, Busha, Carlstrom,
  Chang, Cho, Crawford, Crites, de~Haan, Desai, Dobbs, Dor{\'e}, Dudley, Geach,
  George, Gladders, Gonzalez, Halverson, Harrington, High, Holden, Holzapfel,
  Hoover, Hrubes, Joy, Keisler, Knox, Lee, Leitch, Lueker, Luong-Van, Marrone,
  Martinez-Manso, McMahon, Mehl, Meyer, Mohr, Montroy, Natoli, Padin, Plagge,
  Pryke, Reichardt, Rest, Ruhl, Saliwanchik, Sayre, Schaffer, Shaw, Shirokoff,
  Spieler, Stalder, Stanford, Staniszewski, Stark, Stern, Story, Vallinotto,
  Vanderlinde, Vieira, Wechsler, Williamson, \& Zahn}]{2012ApJ...753L...9B}
Bleem L.~E. {et~al.}, 2012, ApJL, 753, L9

\bibitem[{Booth \& Schaye(2010)}]{2010MNRAS.405L...1B}
Booth C.~M., Schaye J., 2010, MNRAS, 405, L1

\bibitem[{Brodwin {et~al}\mbox{.}(2006)Brodwin, Brown, Ashby, Bian, Brand, Dey,
  Eisenhardt, Eisenstein, Gonzalez, Huang, Jannuzi, Kochanek, McKenzie, Murray,
  Pahre, Smith, Soifer, Stanford, Stern, \& Elston}]{2006ApJ...651..791B}
Brodwin M. {et~al.}, 2006, ApJ, 651, 791

\bibitem[{Carroll \& Hickox(2015)}]{Carroll:2015vp}
Carroll C.~M., Hickox R.~C., 2015, in AGN versus star formation: the fate of
  the gas in galaxies, Durham University, p.~A2

\bibitem[{Chatterjee {et~al}\mbox{.}(2013)Chatterjee, Nguyen, Myers, \&
  Zheng}]{2013ApJ...779..147C}
Chatterjee S., Nguyen M.~L., Myers A.~D., Zheng Z., 2013, ApJ, 779, 147

\bibitem[{Coil {et~al}\mbox{.}(2007)Coil, Hennawi, Newman, Cooper, \&
  Davis}]{2007ApJ...654..115C}
Coil A.~L., Hennawi J.~F., Newman J.~A., Cooper M.~C., Davis M., 2007, ApJ,
  654, 115

\bibitem[{Col{\'\i}n {et~al}\mbox{.}(1999)Col{\'\i}n, Klypin, Kravtsov, \&
  Khokhlov}]{1999ApJ...523...32C}
Col{\'\i}n P., Klypin A.~A., Kravtsov A.~V., Khokhlov A.~M., 1999, ApJ, 523, 32

\bibitem[{Comastri {et~al}\mbox{.}(1995)Comastri, Setti, Zamorani, \&
  Hasinger}]{1995A&A...296....1C}
Comastri A., Setti G., Zamorani G., Hasinger G., 1995, A{\&}A, 296, 1

\bibitem[{Cooray \& Hu(2000)}]{2000ApJ...534..533C}
Cooray A., Hu W., 2000, ApJ, 534, 533

\bibitem[{Croft, Dalton \& Efstathiou(1999)Croft, Dalton, \&
  Efstathiou}]{1999MNRAS.305..547C}
Croft R. A.~C., Dalton G.~B., Efstathiou G., 1999, MNRAS, 305, 547

\bibitem[{Croom {et~al}\mbox{.}(2005)Croom, Boyle, Shanks, Smith, Miller,
  Outram, Loaring, Hoyle, \& da~{\^A}ngela}]{2005MNRAS.356..415C}
Croom S.~M. {et~al.}, 2005, MNRAS, 356, 415

\bibitem[{Croom {et~al}\mbox{.}(2004)Croom, Smith, Boyle, Shanks, Miller,
  Outram, \& Loaring}]{2004MNRAS.349.1397C}
Croom S.~M., Smith R.~J., Boyle B.~J., Shanks T., Miller L., Outram P.~J.,
  Loaring N.~S., 2004, MNRAS, 349, 1397

\bibitem[{Croton(2009)}]{2009MNRAS.394.1109C}
Croton D.~J., 2009, MNRAS, 394, 1109

\bibitem[{da~{\^A}ngela {et~al}\mbox{.}(2008)da~{\^A}ngela, Shanks, Croom,
  Weilbacher, Brunner, Couch, Miller, Myers, Nichol, Pimbblet, de~Propris,
  Richards, Ross, Schneider, \& Wake}]{2008MNRAS.383..565D}
da~{\^A}ngela J. {et~al.}, 2008, MNRAS, 383, 565

\bibitem[{Das {et~al}\mbox{.}(2011)Das, Sherwin, Aguirre, Appel, Bond,
  Carvalho, Devlin, Dunkley, D{\"u}nner, Essinger-Hileman, Fowler, Hajian,
  Halpern, Hasselfield, Hincks, Hlozek, Huffenberger, Hughes, Irwin, Klein,
  Kosowsky, Lupton, Marriage, Marsden, Menanteau, Moodley, Niemack, Nolta,
  Page, Parker, Reese, Schmitt, Sehgal, Sievers, Spergel, Staggs, Swetz,
  Switzer, Thornton, Visnjic, \& Wollack}]{2011PhRvL.107b1301D}
Das S. {et~al.}, 2011, Physical Review Letters, 107, 21301

\bibitem[{DiPompeo {et~al}\mbox{.}(2015{\natexlab{a}})DiPompeo, Bovy, Myers, \&
  Lang}]{2015MNRAS.452.3124D}
DiPompeo M.~A., Bovy J., Myers A.~D., Lang D., 2015{\natexlab{a}}, MNRAS, 452,
  3124

\bibitem[{DiPompeo {et~al}\mbox{.}(2014)DiPompeo, Myers, Hickox, Geach, \&
  Hainline}]{2014MNRAS.442.3443D}
DiPompeo M.~A., Myers A.~D., Hickox R.~C., Geach J.~E., Hainline K.~N., 2014,
  MNRAS, 442, 3443

\bibitem[{DiPompeo {et~al}\mbox{.}(2015{\natexlab{b}})DiPompeo, Myers, Hickox,
  Geach, Holder, Hainline, \& Hall}]{2015MNRAS.446.3492D}
DiPompeo M.~A., Myers A.~D., Hickox R.~C., Geach J.~E., Holder G., Hainline
  K.~N., Hall S.~W., 2015{\natexlab{b}}, MNRAS, 446, 3492

\bibitem[{Donley {et~al}\mbox{.}(2012)Donley, Koekemoer, Brusa, Capak,
  Cardamone, Civano, Ilbert, Impey, Kartaltepe, Miyaji, Salvato, Sanders,
  Trump, \& Zamorani}]{2012ApJ...748..142D}
Donley J.~L. {et~al.}, 2012, ApJ, 748, 142

\bibitem[{Donley {et~al}\mbox{.}(2007)Donley, Rieke, P{\'e}rez-Gonz{\'a}lez,
  Rigby, \& Alonso-Herrero}]{2007ApJ...660..167D}
Donley J.~L., Rieke G.~H., P{\'e}rez-Gonz{\'a}lez P.~G., Rigby J.~R.,
  Alonso-Herrero A., 2007, ApJ, 660, 167

\bibitem[{Donoso {et~al}\mbox{.}(2014)Donoso, Yan, Stern, \&
  Assef}]{2014ApJ...789...44D}
Donoso E., Yan L., Stern D., Assef R.~J., 2014, ApJ, 789, 44

\bibitem[{Eftekharzadeh {et~al}\mbox{.}(2015)Eftekharzadeh, Myers, White,
  Weinberg, Schneider, Shen, Font-Ribera, Ross, P{\^a}ris, \&
  Streblyanska}]{2015arXiv150708380E}
Eftekharzadeh S. {et~al.}, 2015, arXiv, 8380

\bibitem[{Ellison, Patton \& Hickox(2015)Ellison, Patton, \&
  Hickox}]{2015MNRAS.451L..35E}
Ellison S.~L., Patton D.~R., Hickox R.~C., 2015, MNRAS, 451, L35

\bibitem[{Elvis {et~al}\mbox{.}(1994)Elvis, Wilkes, McDowell, Green, Bechtold,
  Willner, Oey, Polomski, \& Cutri}]{1994ApJS...95....1E}
Elvis M. {et~al.}, 1994, ApJS, 95, 1

\bibitem[{Fan {et~al}\mbox{.}(2006)Fan, Strauss, Richards, Hennawi, Becker,
  White, Diamond-Stanic, Donley, Jiang, Kim, Vestergaard, Young, Gunn, Lupton,
  Knapp, Schneider, Brandt, Bahcall, Barentine, Brinkmann, Brewington,
  Fukugita, Harvanek, Kleinman, Krzesinski, Long, Neilsen, Nitta, Snedden, \&
  Voges}]{2006AJ....131.1203F}
Fan X. {et~al.}, 2006, AJ, 131, 1203

\bibitem[{Fitzpatrick \& Massa(2009)}]{2009ApJ...699.1209F}
Fitzpatrick E.~L., Massa D., 2009, ApJ, 699, 1209

\bibitem[{Geach {et~al}\mbox{.}(2013)Geach, Hickox, Bleem, Brodwin, Holder,
  Aird, Benson, Bhattacharya, Carlstrom, Chang, Cho, Crawford, Crites, de~Haan,
  Dobbs, Dudley, George, Hainline, Halverson, Holzapfel, Hoover, Hou, Hrubes,
  Keisler, Knox, Lee, Leitch, Lueker, Luong-Van, Marrone, McMahon, Mehl, Meyer,
  Millea, Mohr, Montroy, Myers, Padin, Plagge, Pryke, Reichardt, Ruhl, Sayre,
  Schaffer, Shaw, Shirokoff, Spieler, Staniszewski, Stark, Story, van Engelen,
  Vanderlinde, Vieira, Williamson, \& Zahn}]{2013ApJ...776L..41G}
Geach J.~E. {et~al.}, 2013, ApJL, 776, L41

\bibitem[{G{\'o}rski {et~al}\mbox{.}(2005)G{\'o}rski, Hivon, Banday, Wandelt,
  Hansen, Reinecke, \& Bartelmann}]{2005ApJ...622..759G}
G{\'o}rski K.~M., Hivon E., Banday A.~J., Wandelt B.~D., Hansen F.~K., Reinecke
  M., Bartelmann M., 2005, ApJ, 622, 759

\bibitem[{Goulding {et~al}\mbox{.}(2012)Goulding, Alexander, Bauer, Forman,
  Hickox, Jones, Mullaney, \& Trichas}]{2012ApJ...755....5G}
Goulding A.~D., Alexander D.~M., Bauer F.~E., Forman W.~R., Hickox R.~C., Jones
  C., Mullaney J.~R., Trichas M., 2012, ApJ, 755, 5

\bibitem[{Goulding {et~al}\mbox{.}(2014)Goulding, Forman, Hickox, Jones,
  Murray, Paggi, Ashby, Coil, Cooper, Huang, Kraft, Newman, Weiner, \&
  Willner}]{2014ApJ...783...40G}
Goulding A.~D. {et~al.}, 2014, ApJ, 783, 40

\bibitem[{Guo {et~al}\mbox{.}(2010)Guo, White, Li, \&
  Boylan-Kolchin}]{2010MNRAS.404.1111G}
Guo Q., White S., Li C., Boylan-Kolchin M., 2010, MNRAS, 404, 1111

\bibitem[{Hainline {et~al}\mbox{.}(2014)Hainline, Hickox, Carroll, Myers,
  DiPompeo, \& Trouille}]{2014ApJ...795..124H}
Hainline K.~N., Hickox R.~C., Carroll C.~M., Myers A.~D., DiPompeo M.~A.,
  Trouille L., 2014, ApJ, 795, 124

\bibitem[{Hamilton \& Tegmark(2004)}]{2004MNRAS.349..115H}
Hamilton A. J.~S., Tegmark M., 2004, MNRAS, 349, 115

\bibitem[{Helfand, White \& Becker(2015)Helfand, White, \&
  Becker}]{2015ApJ...801...26H}
Helfand D.~J., White R.~L., Becker R.~H., 2015, ApJ, 801, 26

\bibitem[{Hewett, Foltz \& Chaffee(1995)Hewett, Foltz, \&
  Chaffee}]{1995AJ....109.1498H}
Hewett P.~C., Foltz C.~B., Chaffee F.~H., 1995, Astronomical Journal (ISSN
  0004-6256), 109, 1498

\bibitem[{Hickox {et~al}\mbox{.}(2007)Hickox, Jones, Forman, Murray, Brodwin,
  Brown, Eisenhardt, Stern, Kochanek, Eisenstein, Cool, Jannuzi, Dey, Brand,
  Gorjian, \& Caldwell}]{2007ApJ...671.1365H}
Hickox R.~C. {et~al.}, 2007, ApJ, 671, 1365

\bibitem[{Hickox {et~al}\mbox{.}(2011)Hickox, Myers, Brodwin, Alexander,
  Forman, Jones, Murray, Brown, Cool, Kochanek, Dey, Jannuzi, Eisenstein,
  Assef, Eisenhardt, Gorjian, Stern, Le~Floc'h, Caldwell, Goulding, \&
  Mullaney}]{2011ApJ...731..117H}
Hickox R.~C. {et~al.}, 2011, ApJ, 731, 117

\bibitem[{Ho {et~al}\mbox{.}(2015)Ho, Agarwal, Myers, Lyons, Disbrow, Seo,
  Ross, Hirata, Padmanabhan, O'Connell, Huff, Schlegel, Slosar, Weinberg,
  Strauss, Ross, Schneider, Bahcall, Brinkmann, Palanque-Delabrouille, \&
  Y{\`e}che}]{2015JCAP...05..040H}
Ho S. {et~al.}, 2015, JCAP, 05, 040

\bibitem[{Ho {et~al}\mbox{.}(2012)Ho, Cuesta, Seo, de~Putter, Ross, White,
  Padmanabhan, Saito, Schlegel, Schlafly, Seljak, Hernandez-Monteagudo,
  Sanchez, Percival, Blanton, Skibba, Schneider, Reid, Mena, Viel, Eisenstein,
  Prada, Weaver, Bahcall, Bizyaev, Brewinton, Brinkman, Nicolaci~da Costa,
  Gott, Malanushenko, Malanushenko, Nichol, Oravetz, Pan,
  Palanque-Delabrouille, Ross, Simmons, de~Simoni, Snedden, \&
  Yeche}]{2012ApJ...761...14H}
Ho S. {et~al.}, 2012, ApJ, 761, 14

\bibitem[{H{\o}g {et~al}\mbox{.}(2000)H{\o}g, Fabricius, Makarov, Urban,
  Corbin, Wycoff, Bastian, Schwekendiek, \& Wicenec}]{2000A&A...355L..27H}
H{\o}g E. {et~al.}, 2000, A{\&}A, 355, L27

\bibitem[{Hopkins {et~al}\mbox{.}(2008)Hopkins, Hernquist, Cox, \& Kere{\v
  s}}]{2008ApJS..175..356H}
Hopkins P.~F., Hernquist L., Cox T.~J., Kere{\v s} D., 2008, ApJS, 175, 356

\bibitem[{Hopkins, Richards \& Hernquist(2007)Hopkins, Richards, \&
  Hernquist}]{2007ApJ...654..731H}
Hopkins P.~F., Richards G.~T., Hernquist L., 2007, ApJ, 654, 731

\bibitem[{Hu(2001)}]{2001ApJ...557L..79H}
Hu W., 2001, ApJ, 557, L79

\bibitem[{Huterer {et~al}\mbox{.}(2006)Huterer, Takada, Bernstein, \&
  Jain}]{2006MNRAS.366..101H}
Huterer D., Takada M., Bernstein G., Jain B., 2006, MNRAS, 366, 101

\bibitem[{Kaiser(1984)}]{1984ApJ...284L...9K}
Kaiser N., 1984, ApJ, 284, L9

\bibitem[{Kochanek {et~al}\mbox{.}(2012)Kochanek, Eisenstein, Cool, Caldwell,
  Assef, Jannuzi, Jones, Murray, Forman, Dey, Brown, Eisenhardt, Gonzalez,
  Green, \& Stern}]{2012ApJS..200....8K}
Kochanek C.~S. {et~al.}, 2012, ApJS, 200, 8

\bibitem[{Komatsu {et~al}\mbox{.}(2011)Komatsu, Smith, Dunkley, Bennett, Gold,
  Hinshaw, Jarosik, Larson, Nolta, Page, Spergel, Halpern, Hill, Kogut, Limon,
  Meyer, Odegard, Tucker, Weiland, Wollack, \& Wright}]{2011ApJS..192...18K}
Komatsu E. {et~al.}, 2011, ApJS, 192, 18

\bibitem[{Kravtsov \& Klypin(1999)}]{1999ApJ...520..437K}
Kravtsov A.~V., Klypin A.~A., 1999, ApJ, 520, 437

\bibitem[{Krolewski \& Eisenstein(2015)}]{Krolewski:2015wb}
Krolewski A.~G., Eisenstein D.~J., 2015, arXiv

\bibitem[{Krumpe, Miyaji \& Coil(2010)Krumpe, Miyaji, \&
  Coil}]{2010ApJ...713..558K}
Krumpe M., Miyaji T., Coil A.~L., 2010, ApJ, 713, 558

\bibitem[{Krumpe, Miyaji \& Coil(2013)Krumpe, Miyaji, \& Coil}]{Krumpe:2013tv}
Krumpe M., Miyaji T., Coil A.~L., 2013, arXiv

\bibitem[{Lacy {et~al}\mbox{.}(2013)Lacy, Ridgway, Gates, Nielsen, Petric,
  Sajina, Urrutia, Cox~Drews, Harrison, Seymour, \&
  Storrie-Lombardi}]{2013ApJS..208...24L}
Lacy M. {et~al.}, 2013, ApJS, 208, 24

\bibitem[{Lacy {et~al}\mbox{.}(2015)Lacy, Ridgway, Sajina, Petric, Gates,
  Urrutia, \& Storrie-Lombardi}]{2015arXiv150104118L}
Lacy M., Ridgway S.~E., Sajina A., Petric A.~O., Gates E.~L., Urrutia T.,
  Storrie-Lombardi L.~J., 2015, arXiv, 4118

\bibitem[{Lacy {et~al}\mbox{.}(2004)Lacy, Storrie-Lombardi, Sajina, Appleton,
  Armus, Chapman, Choi, Fadda, Fang, Frayer, Heinrichsen, Helou, Im, Marleau,
  Masci, Shupe, Soifer, Surace, Teplitz, Wilson, \& Yan}]{2004ApJS..154..166L}
Lacy M. {et~al.}, 2004, ApJS, 154, 166

\bibitem[{Landy \& Szalay(1993)}]{1993ApJ...412...64L}
Landy S.~D., Szalay A.~S., 1993, ApJ, 412, 64

\bibitem[{Leistedt \& Peiris(2014)}]{2014MNRAS.444....2L}
Leistedt B., Peiris H.~V., 2014, MNRAS, 444, 2

\bibitem[{Lewis, Challinor \& Lasenby(2000)Lewis, Challinor, \&
  Lasenby}]{2000ApJ...538..473L}
Lewis A., Challinor A., Lasenby A., 2000, ApJ, 538, 473

\bibitem[{Lilly {et~al}\mbox{.}(2007)Lilly, Le~Fevre, Renzini, Zamorani,
  Scodeggio, Contini, Carollo, Hasinger, Kneib, Iovino, Le~Brun, Maier,
  Mainieri, Mignoli, Silverman, Tasca, Bolzonella, Bongiorno, Bottini, Capak,
  Caputi, Cimatti, Cucciati, Daddi, Feldmann, Franzetti, Garilli, Guzzo,
  Ilbert, Kampczyk, Kova{\v c}, Lamareille, Leauthaud, Borgne, McCracken,
  Marinoni, Pell{\`o}, Ricciardelli, Scarlata, Vergani, Sanders, Schinnerer,
  Scoville, Taniguchi, Arnouts, Aussel, Bardelli, Brusa, Cappi, Ciliegi,
  Finoguenov, Foucaud, Franceschini, Halliday, Impey, Knobel, Koekemoer, Kurk,
  Maccagni, Maddox, Marano, Marconi, Meneux, Mobasher, Moreau, Peacock,
  Porciani, Pozzetti, Scaramella, Schiminovich, Shopbell, Smail, Thompson,
  Tresse, Vettolani, Zanichelli, \& Zucca}]{2007ApJS..172...70L}
Lilly S.~J. {et~al.}, 2007, The Astrophysical Journal Supplement Series, 172,
  70

\bibitem[{Limber(1953)}]{1953ApJ...117..134L}
Limber D.~N., 1953, ApJ, 117, 134

\bibitem[{Mainzer {et~al}\mbox{.}(2011)Mainzer, Grav, Bauer, Masiero, McMillan,
  Cutri, Walker, Wright, Eisenhardt, Tholen, Spahr, Jedicke, Denneau, DeBaun,
  Elsbury, Gautier, Gomillion, Hand, Mo, Watkins, Wilkins, Bryngelson, Del
  Pino~Molina, Desai, G{\'o}mez~Camus, Hidalgo, Konstantopoulos, Larsen,
  Maleszewski, Malkan, Mauduit, Mullan, Olszewski, Pforr, Saro, Scotti, \&
  Wasserman}]{2011ApJ...743..156M}
Mainzer A. {et~al.}, 2011, ApJ, 743, 156

\bibitem[{Mateos {et~al}\mbox{.}(2013)Mateos, Alonso-Herrero, Carrera, Blain,
  Severgnini, Caccianiga, \& Ruiz}]{2013MNRAS.434..941M}
Mateos S., Alonso-Herrero A., Carrera F.~J., Blain A., Severgnini P.,
  Caccianiga A., Ruiz A., 2013, MNRAS, 434, 941

\bibitem[{Mateos {et~al}\mbox{.}(2012)Mateos, Alonso-Herrero, Carrera, Blain,
  Watson, Barcons, Braito, Severgnini, Donley, \& Stern}]{2012MNRAS.426.3271M}
Mateos S. {et~al.}, 2012, MNRAS, 426, 3271

\bibitem[{Mendez {et~al}\mbox{.}(2015)Mendez, Coil, Aird, Skibba,
  Diamond-Stanic, Moustakas, Blanton, Cool, Eisenstein, Wong, \&
  Zhu}]{2015arXiv150406284M}
Mendez A.~J. {et~al.}, 2015, arXiv, 6284

\bibitem[{Myers {et~al}\mbox{.}(2007)Myers, Brunner, Nichol, Richards,
  Schneider, \& Bahcall}]{2007ApJ...658...85M}
Myers A.~D., Brunner R.~J., Nichol R.~C., Richards G.~T., Schneider D.~P.,
  Bahcall N.~A., 2007, ApJ, 658, 85

\bibitem[{Myers {et~al}\mbox{.}(2006)Myers, Brunner, Richards, Nichol,
  Schneider, Vanden~Berk, Scranton, Gray, \& Brinkmann}]{2006ApJ...638..622M}
Myers A.~D. {et~al.}, 2006, ApJ, 638, 622

\bibitem[{Myers {et~al}\mbox{.}(2005)Myers, Outram, Shanks, Boyle, Croom,
  Loaring, Miller, \& Smith}]{2005MNRAS.359..741M}
Myers A.~D., Outram P.~J., Shanks T., Boyle B.~J., Croom S.~M., Loaring N.~S.,
  Miller L., Smith R.~J., 2005, MNRAS, 359, 741

\bibitem[{Myers {et~al}\mbox{.}(2015)Myers, Palanque-Delabrouille, Prakash,
  P{\^a}ris, Yeche, Dawson, Bovy, Lang, Schlegel, Newman, Petitjean, Kneib,
  Laurent, Percival, Ross, Seo, Tinker, Armengaud, Brownstein, Burtin, Cai,
  Comparat, Kasliwal, Kulkarni, Laher, Levitan, McBride, McGreer, Miller,
  Nugent, Ofek, Rossi, Ruan, Schneider, Sesar, Streblyanska, Surace, \&
  collaboration}]{2015arXiv150804472M}
Myers A.~D. {et~al.}, 2015, arXiv, 4472

\bibitem[{Navarro, Frenk \& White(1997)Navarro, Frenk, \&
  White}]{1997ApJ...490..493N}
Navarro J.~F., Frenk C.~S., White S. D.~M., 1997, ApJ, 490, 493

\bibitem[{Netzer(2015)}]{2015ARA&A..53..365N}
Netzer H., 2015, ARA{\&}A, 53, 365

\bibitem[{Padmanabhan {et~al}\mbox{.}(2009)Padmanabhan, White, Norberg, \&
  Porciani}]{2009MNRAS.397.1862P}
Padmanabhan N., White M., Norberg P., Porciani C., 2009, MNRAS, 397, 1862

\bibitem[{Peacock(1991)}]{1991MNRAS.253P...1P}
Peacock J.~A., 1991, MNRAS, 253, 1P

\bibitem[{Peebles(1980)}]{Peebles:1980vn}
Peebles P. J.~E., 1980, {The Large-Scale Structure of the Universe}. Princeton
  University Press, Princeton University

\bibitem[{Planck\hspace{0.1cm}Collaboration
  {et~al}\mbox{.}(2015{\natexlab{a}})Planck\hspace{0.1cm}Collaboration, Adam,
  Ade, Aghanim, Arnaud, Ashdown, Aumont, Baccigalupi, Banday, Barreiro,
  Bartlett, Bartolo, Basak, Battaner, Benabed, Beno{\^\i}t, Benoit-L{\'e}vy,
  Bernard, Bersanelli, Bielewicz, Bonaldi, Bonavera, Bond, Borrill, Bouchet,
  Boulanger, Bucher, Burigana, Butler, Calabrese, Cardoso, Casaponsa, Castex,
  Catalano, Challinor, Chamballu, Chary, Chiang, Christensen, Clements,
  Colombi, Colombo, Combet, Couchot, Coulais, Crill, Curto, Cuttaia, Danese,
  Davies, Davis, de~Bernardis, de~Rosa, De~Zotti, Delabrouille, D{\'e}sert,
  Dickinson, Diego, Dole, Donzelli, Dor{\'e}, Douspis, Ducout, Dupac,
  Efstathiou, Elsner, En{\ss}lin, Eriksen, Falgarone, Fantaye, Fergusson,
  Finelli, Forni, Frailis, Fraisse, Franceschi, Frejsel, Galeotta, Galli,
  Ganga, Ghosh, Giard, Giraud-H{\'e}raud, Gjerl{\o}w, Gonz{\'a}lez-Nuevo,
  G{\'o}rski, Gratton, Gregorio, Gruppuso, Gudmundsson, Hansen, Hanson,
  Harrison, Helou, Henrot-Versill{\'e}, Hern{\'a}ndez-Monteagudo, Herranz,
  Hildebrandt, Hivon, Hobson, Holmes, Hornstrup, Hovest, Huffenberger, Hurier,
  Jaffe, Jaffe, Jones, Juvela, Keih{\"a}nen, Keskitalo, Kisner, Kneissl,
  Knoche, Krachmalnicoff, Kunz, Kurki-Suonio, Lagache, Lamarre, Lasenby,
  Lattanzi, Lawrence, Le~Jeune, Leonardi, Lesgourgues, Levrier, Liguori, Lilje,
  Linden-V{\o}rnle, L{\'o}pez-Caniego, Lubin, Mac{\'\i}as-P{\'e}rez, Maggio,
  Maino, Mandolesi, Mangilli, Marshall, Martin, Mart{\'\i}nez-Gonz{\'a}lez,
  Masi, Matarrese, Mazzotta, McGehee, Meinhold, Melchiorri, Mendes, Mennella,
  Migliaccio, Mitra, Miville-Desch{\^e}nes, Molinari, Moneti, Montier,
  Morgante, Mortlock, Moss, Munshi, Murphy, Naselsky, Nati, Natoli,
  Netterfield, N{\o}rgaard-Nielsen, Noviello, Novikov, Novikov, Oxborrow, Paci,
  Pagano, Pajot, Paladini, Paoletti, Pasian, Patanchon, Pearson, Perdereau,
  Perotto, Perrotta, Pettorino, Piacentini, Piat, Pierpaoli, Pietrobon,
  Plaszczynski, Pointecouteau, Polenta, Pratt, Pr{\'e}zeau, Prunet, Puget,
  Rachen, Racine, Reach, Rebolo, Reinecke, Remazeilles, Renault, Renzi,
  Ristorcelli, Rocha, Rosset, Rossetti, Roudier, Rubi{\~n}o-Mart{\'\i}n,
  Rusholme, Sandri, Santos, Savelainen, Savini, Scott, Seiffert, Shellard,
  Spencer, Stolyarov, Stompor, Sudiwala, Sunyaev, Sutton, Suur-Uski, Sygnet,
  Tauber, Terenzi, Toffolatti, Tomasi, Tristram, Trombetti, Tucci, Tuovinen,
  Valenziano, Valiviita, Van~Tent, Vielva, Villa, Wade, Wandelt, Wehus, Yvon,
  Zacchei, \& Zonca}]{PlanckCollaboration:2015ue}
Planck\hspace{0.1cm}Collaboration {et~al.}, 2015{\natexlab{a}}, arXiv

\bibitem[{Planck\hspace{0.1cm}Collaboration
  {et~al}\mbox{.}(2014{\natexlab{a}})Planck\hspace{0.1cm}Collaboration, Ade,
  Aghanim, Alves, Armitage-Caplan, Arnaud, Ashdown, Atrio-Barandela, Aumont,
  Aussel, Baccigalupi, Banday, Barreiro, Barrena, Bartelmann, Bartlett,
  Bartolo, Basak, Battaner, Battye, Benabed, Beno{\^\i}t, Benoit-L{\'e}vy,
  Bernard, Bersanelli, Bertincourt, B{\'e}thermin, Bielewicz, Bikmaev,
  Blanchard, Bobin, Bock, B{\"o}hringer, Bonaldi, Bonavera, Bond, Borrill,
  Bouchet, Boulanger, Bourdin, Bowyer, Bridges, Brown, Bucher, Burenin,
  Burigana, Butler, Calabrese, Cappellini, Cardoso, Carr, Carvalho, Casale,
  Castex, Catalano, Challinor, Chamballu, Chary, Chen, Chiang, Chiang, Chon,
  Christensen, Churazov, Church, Clemens, Clements, Colombi, Colombo, Combet,
  Comis, Couchot, Coulais, Crill, Cruz, Curto, Cuttaia, Da~Silva, Dahle,
  Danese, Davies, Davis, de~Bernardis, de~Rosa, De~Zotti, D{\'e}chelette,
  Delabrouille, Delouis, D{\'e}mocl{\`e}s, D{\'e}sert, Dick, Dickinson, Diego,
  Dolag, Dole, Donzelli, Dor{\'e}, Douspis, Ducout, Dunkley, Dupac, Efstathiou,
  Elsner, En{\ss}lin, Eriksen, Fabre, Falgarone, Falvella, Fantaye, Fergusson,
  Filliard, Finelli, Flores-Cacho, Foley, Forni, Fosalba, Frailis, Fraisse,
  Franceschi, Freschi, Fromenteau, Frommert, Gaier, Galeotta, Gallegos, Galli,
  Gandolfo, Ganga, Gauthier, G{\'e}nova-Santos, Ghosh, Giard, Giardino,
  Gilfanov, Girard, Giraud-H{\'e}raud, Gjerl{\o}w, Gonz{\'a}lez-Nuevo,
  G{\'o}rski, Gratton, Gregorio, Gruppuso, Gudmundsson, Haissinski, Hamann,
  Hansen, Hansen, Hanson, Harrison, Heavens, Helou, Hempel,
  Henrot-Versill{\'e}, Hern{\'a}ndez-Monteagudo, Herranz, Hildebrandt, Hivon,
  Ho, Hobson, Holmes, Hornstrup, Hou, Hovest, Huey, Huffenberger, Hurier,
  Ili{\'c}, Jaffe, Jaffe, Jasche, Jewell, Jones, Juvela, Kalberla, Kangaslahti,
  Keih{\"a}nen, Kerp, Keskitalo, Khamitov, Kiiveri, Kim, Kisner, Kneissl,
  Knoche, Knox, Kunz, Kurki-Suonio, Lacasa, Lagache, L{\"a}hteenm{\"a}ki,
  Lamarre, Langer, Lasenby, Lattanzi, Laureijs, Lavabre, Lawrence, Le~Jeune,
  Leach, Leahy, Leonardi, Le{\'o}n-Tavares, Leroy, Lesgourgues, Lewis, Li,
  Liddle, Liguori, Lilje, Linden-V{\o}rnle, Lindholm, L{\'o}pez-Caniego, Lowe,
  Lubin, Mac{\'\i}as-P{\'e}rez, MacTavish, Maffei, Maggio, Maino, Mandolesi,
  Mangilli, Marcos-Caballero, Marinucci, Maris, Marleau, Marshall, Martin,
  Mart{\'\i}nez-Gonz{\'a}lez, Masi, Massardi, Matarrese, Matsumura, Matthai,
  Maurin, Mazzotta, McDonald, McEwen, McGehee, Mei, Meinhold, Melchiorri,
  Melin, Mendes, Menegoni, Mennella, Migliaccio, Mikkelsen, Millea, Miniscalco,
  Mitra, Miville-Desch{\^e}nes, Molinari, Moneti, Montier, Morgante, Morisset,
  Mortlock, Moss, Munshi, Murphy, Naselsky, Nati, Natoli, Negrello, Nesvadba,
  Netterfield, N{\o}rgaard-Nielsen, North, Noviello, Novikov, Novikov, O'Dwyer,
  Orieux, Osborne, O'Sullivan, Oxborrow, Paci, Pagano, Pajot, Paladini,
  Pandolfi, Paoletti, Partridge, Pasian, Patanchon, Paykari, Pearson, Pearson,
  Peel, Peiris, Perdereau, Perotto, \& Perrot...}]{2014A&A...571A...1P}
Planck\hspace{0.1cm}Collaboration {et~al.}, 2014{\natexlab{a}}, A{\&}A, 571, A1

\bibitem[{Planck\hspace{0.1cm}Collaboration
  {et~al}\mbox{.}(2014{\natexlab{b}})Planck\hspace{0.1cm}Collaboration, Ade,
  Aghanim, Armitage-Caplan, Arnaud, Ashdown, Atrio-Barandela, Aumont,
  Baccigalupi, Banday, Barreiro, Bartlett, Basak, Battaner, Benabed,
  Beno{\^\i}t, Benoit-L{\'e}vy, Bernard, Bersanelli, Bielewicz, Bobin, Bock,
  Bonaldi, Bonavera, Bond, Borrill, Bouchet, Bridges, Bucher, Burigana, Butler,
  Cardoso, Catalano, Challinor, Chamballu, Chiang, Chiang, Christensen, Church,
  Clements, Colombi, Colombo, Couchot, Coulais, Crill, Curto, Cuttaia, Danese,
  Davies, Davis, de~Bernardis, de~Rosa, De~Zotti, D{\'e}chelette, Delabrouille,
  Delouis, D{\'e}sert, Dickinson, Diego, Dole, Donzelli, Dor{\'e}, Douspis,
  Dunkley, Dupac, Efstathiou, En{\ss}lin, Eriksen, Finelli, Forni, Frailis,
  Franceschi, Galeotta, Ganga, Giard, Giardino, Giraud-H{\'e}raud,
  Gonz{\'a}lez-Nuevo, G{\'o}rski, Gratton, Gregorio, Gruppuso, Gudmundsson,
  Hansen, Hanson, Harrison, Henrot-Versill{\'e}, Hern{\'a}ndez-Monteagudo,
  Herranz, Hildebrandt, Hivon, Ho, Hobson, Holmes, Hornstrup, Hovest,
  Huffenberger, Jaffe, Jaffe, Jones, Juvela, Keih{\"a}nen, Keskitalo, Kisner,
  Kneissl, Knoche, Knox, Kunz, Kurki-Suonio, Lagache, L{\"a}hteenm{\"a}ki,
  Lamarre, Lasenby, Laureijs, Lavabre, Lawrence, Leahy, Leonardi,
  Le{\'o}n-Tavares, Lesgourgues, Lewis, Liguori, Lilje, Linden-V{\o}rnle,
  L{\'o}pez-Caniego, Lubin, Mac{\'\i}as-P{\'e}rez, Maffei, Maino, Mandolesi,
  Mangilli, Maris, Marshall, Martin, Mart{\'\i}nez-Gonz{\'a}lez, Masi,
  Massardi, Matarrese, Matthai, Mazzotta, Melchiorri, Mendes, Mennella,
  Migliaccio, Mitra, Miville-Desch{\^e}nes, Moneti, Montier, Morgante,
  Mortlock, Moss, Munshi, Murphy, Naselsky, Nati, Natoli, Netterfield,
  N{\o}rgaard-Nielsen, Noviello, Novikov, Novikov, Osborne, Oxborrow, Paci,
  Pagano, Pajot, Paoletti, Partridge, Pasian, Patanchon, Perdereau, Perotto,
  Perrotta, Piacentini, Piat, Pierpaoli, Pietrobon, Plaszczynski,
  Pointecouteau, Polenta, Ponthieu, Popa, Poutanen, Pratt, Pr{\'e}zeau, Prunet,
  Puget, Pullen, Rachen, Rebolo, Reinecke, Remazeilles, Renault, Ricciardi,
  Riller, Ristorcelli, Rocha, Rosset, Roudier, Rowan-Robinson,
  Rubi{\~n}o-Mart{\'\i}n, Rusholme, Sandri, Santos, Savini, Scott, Seiffert,
  Shellard, Smith, Spencer, Starck, Stolyarov, Stompor, Sudiwala, Sunyaev,
  Sureau, Sutton, Suur-Uski, Sygnet, Tauber, Tavagnacco, Terenzi, Toffolatti,
  Tomasi, Tristram, Tucci, Tuovinen, Umana, Valenziano, Valiviita, Van~Tent,
  Vielva, Villa, Vittorio, Wade, Wandelt, White, White, Yvon, Zacchei, \&
  Zonca}]{2014A&A...571A..17P}
Planck\hspace{0.1cm}Collaboration {et~al.}, 2014{\natexlab{b}}, A{\&}A, 571,
  A17

\bibitem[{Planck\hspace{0.1cm}Collaboration
  {et~al}\mbox{.}(2011)Planck\hspace{0.1cm}Collaboration, Ade, Aghanim, Arnaud,
  Ashdown, Aumont, Baccigalupi, Baker, Balbi, Banday, Barreiro, Bartlett,
  Battaner, Benabed, Bennett, Beno{\^\i}t, Bernard, Bersanelli, Bhatia, Bock,
  Bonaldi, Bond, Borrill, Bouchet, Bradshaw, Bremer, Bucher, Burigana, Butler,
  Cabella, Cantalupo, Cappellini, Cardoso, Carr, Casale, Catalano, Cay{\'o}n,
  Challinor, Chamballu, Charra, Chary, Chiang, Chiang, Christensen, Clements,
  Colombi, Couchot, Coulais, Crill, Crone, Crook, Cuttaia, Danese, D'Arcangelo,
  Davies, Davis, de~Bernardis, de~Bruin, de~Gasperis, de~Rosa, De~Zotti,
  Delabrouille, Delouis, D{\'e}sert, Dick, Dickinson, Dolag, Dole, Donzelli,
  Dor{\'e}, D{\"o}rl, Douspis, Dupac, Efstathiou, En{\ss}lin, Eriksen, Finelli,
  Foley, Forni, Fosalba, Frailis, Franceschi, Freschi, Gaier, Galeotta,
  Gallegos, Gandolfo, Ganga, Giard, Giardino, Gienger, Giraud-H{\'e}raud,
  Gonz{\'a}lez, Gonz{\'a}lez-Nuevo, G{\'o}rski, Gratton, Gregorio, Gruppuso,
  Guyot, Haissinski, Hansen, Harrison, Helou, Henrot-Versill{\'e},
  Hern{\'a}ndez-Monteagudo, Herranz, Hildebrandt, Hivon, Hobson, Holmes,
  Hornstrup, Hovest, Hoyland, Huffenberger, Jaffe, Jagemann, Jones, Juillet,
  Juvela, Kangaslahti, Keih{\"a}nen, Keskitalo, Kisner, Kneissl, Knox,
  Krassenburg, Kurki-Suonio, Lagache, L{\"a}hteenm{\"a}ki, Lamarre, Lange,
  Lasenby, Laureijs, Lawrence, Leach, Leahy, Leonardi, Leroy, Lilje,
  Linden-V{\o}rnle, L{\'o}pez-Caniego, Lowe, Lubin, Mac{\'\i}as-P{\'e}rez,
  Maciaszek, MacTavish, Maffei, Maino, Mandolesi, Mann, Maris,
  Mart{\'\i}nez-Gonz{\'a}lez, Masi, Massardi, Matarrese, Matthai, Mazzotta,
  McDonald, McGehee, Meinhold, Melchiorri, Melin, Mendes, Mennella, Mevi,
  Miniscalco, Mitra, Miville-Desch{\^e}nes, Moneti, Montier, Morgante,
  Morisset, Mortlock, Munshi, Murphy, Naselsky, Natoli, Netterfield,
  N{\o}rgaard-Nielsen, Noviello, Novikov, Novikov, O'Dwyer, Ortiz, Osborne,
  Osuna, Oxborrow, Pajot, Paladini, Partridge, Pasian, Passvogel, Patanchon,
  Pearson, Pearson, Perdereau, Perotto, Perrotta, Piacentini, Piat, Pierpaoli,
  Plaszczynski, Platania, Pointecouteau, Polenta, Ponthieu, Popa, Poutanen,
  Pr{\'e}zeau, Prunet, Puget, Rachen, Reach, Rebolo, Reinecke, Reix, Renault,
  Ricciardi, Riller, Ristorcelli, Rocha, Rosset, Rowan-Robinson,
  Rubi{\~n}o-Mart{\'\i}n, Rusholme, Salerno, Sandri, Santos, Savini, Schaefer,
  Scott, Seiffert, Shellard, Simonetto, Smoot, Sozzi, Starck, Sternberg,
  Stivoli, Stolyarov, Stompor, Stringhetti, Sudiwala, Sunyaev, Sygnet,
  Tapiador, Tauber, Tavagnacco, Taylor, Terenzi, Texier, Toffolatti, Tomasi,
  Torre, Tristram, Tuovinen, T{\"u}rler, Tuttlebee, Umana, Valenziano,
  Valiviita, Varis, Vibert, Vielva, Villa, Vittorio, Wade, Wandelt, Watson,
  White, White, Wilkinson, Yvon, Zacchei, \& Zonca}]{2011A&A...536A...1P}
Planck\hspace{0.1cm}Collaboration {et~al.}, 2011, A{\&}A, 536, 1

\bibitem[{Planck\hspace{0.1cm}Collaboration
  {et~al}\mbox{.}(2015{\natexlab{b}})Planck\hspace{0.1cm}Collaboration, Ade,
  Aghanim, Arnaud, Ashdown, Aumont, Baccigalupi, Banday, Barreiro, Bartlett,
  Bartolo, Battaner, Benabed, Beno{\^\i}t, Benoit-L{\'e}vy, Bernard,
  Bersanelli, Bielewicz, Bonaldi, Bonavera, Bond, Borrill, Bouchet, Boulanger,
  Bucher, Burigana, Butler, Calabrese, Cardoso, Catalano, Challinor, Chamballu,
  Chiang, Christensen, Church, Clements, Colombi, Colombo, Combet, Couchot,
  Coulais, Crill, Curto, Cuttaia, Danese, Davies, Davis, de~Bernardis, de~Rosa,
  De~Zotti, Delabrouille, D{\'e}sert, Diego, Dole, Donzelli, Dor{\'e}, Douspis,
  Ducout, Dunkley, Dupac, Efstathiou, Elsner, En{\ss}lin, Eriksen, Fergusson,
  Finelli, Forni, Frailis, Fraisse, Franceschi, Frejsel, Galeotta, Galli,
  Ganga, Giard, Giraud-H{\'e}raud, Gjerl{\o}w, Gonz{\'a}lez-Nuevo, G{\'o}rski,
  Gratton, Gregorio, Gruppuso, Gudmundsson, Hansen, Hanson, Harrison,
  Henrot-Versill{\'e}, Hern{\'a}ndez-Monteagudo, Herranz, Hildebrandt, Hivon,
  Hobson, Holmes, Hornstrup, Hovest, Huffenberger, Hurier, Jaffe, Jaffe, Jones,
  Juvela, Keih{\"a}nen, Keskitalo, Kisner, Kneissl, Knoche, Kunz, Kurki-Suonio,
  Lagache, L{\"a}hteenm{\"a}ki, Lamarre, Lasenby, Lattanzi, Lawrence, Leonardi,
  Lesgourgues, Levrier, Lewis, Liguori, Lilje, Linden-V{\o}rnle,
  L{\'o}pez-Caniego, Lubin, Mac{\'\i}as-P{\'e}rez, Maggio, Maino, Mandolesi,
  Mangilli, Martin, Mart{\'\i}nez-Gonz{\'a}lez, Masi, Matarrese, Mazzotta,
  McGehee, Meinhold, Melchiorri, Mendes, Mennella, Migliaccio, Mitra,
  Miville-Desch{\^e}nes, Moneti, Montier, Morgante, Mortlock, Moss, Munshi,
  Murphy, Naselsky, Nati, Natoli, Netterfield, N{\o}rgaard-Nielsen, Noviello,
  Novikov, Novikov, Oxborrow, Paci, Pagano, Pajot, Paoletti, Pasian, Patanchon,
  Perdereau, Perotto, Perrotta, Pettorino, Piacentini, Piat, Pierpaoli,
  Pietrobon, Plaszczynski, Pointecouteau, Polenta, Popa, Pratt, Pr{\'e}zeau,
  Prunet, Puget, Rachen, Reach, Rebolo, Reinecke, Remazeilles, Renault, Renzi,
  Ristorcelli, Rocha, Rosset, Rossetti, Roudier, Rowan-Robinson,
  Rubi{\~n}o-Mart{\'\i}n, Rusholme, Sandri, Santos, Savelainen, Savini, Scott,
  Seiffert, Shellard, Spencer, Stolyarov, Stompor, Sudiwala, Sunyaev, Sutton,
  Suur-Uski, Sygnet, Tauber, Terenzi, Toffolatti, Tomasi, Tristram, Tucci,
  Tuovinen, Valenziano, Valiviita, Van~Tent, Vielva, Villa, Wade, Wandelt,
  Wehus, White, Yvon, Zacchei, \& Zonca}]{PlanckCollaboration:2015tp}
Planck\hspace{0.1cm}Collaboration {et~al.}, 2015{\natexlab{b}}, arXiv

\bibitem[{Porciani, Magliocchetti \& Norberg(2004)Porciani, Magliocchetti, \&
  Norberg}]{2004MNRAS.355.1010P}
Porciani C., Magliocchetti M., Norberg P., 2004, MNRAS, 355, 1010

\bibitem[{Richards {et~al}\mbox{.}(2005)Richards, Croom, Anderson,
  Bland-Hawthorn, Boyle, De~Propris, Drinkwater, Fan, Gunn, Ivezi{\'c}, Jester,
  Loveday, Meiksin, Miller, Myers, Nichol, Outram, Pimbblet, Roseboom, Ross,
  Schneider, Shanks, Sharp, Stoughton, Strauss, Szalay, Vanden~Berk, \&
  York}]{2005MNRAS.360..839R}
Richards G.~T. {et~al.}, 2005, MNRAS, 360, 839

\bibitem[{Richards {et~al}\mbox{.}(2006)Richards, Lacy, Storrie-Lombardi, Hall,
  Gallagher, Hines, Fan, Papovich, Vanden~Berk, Trammell, Schneider,
  Vestergaard, York, Jester, Anderson, Budav{\'a}ri, \&
  Szalay}]{Richards:2006p3932}
Richards G.~T. {et~al.}, 2006, ApJS, 166, 470

\bibitem[{Richardson {et~al}\mbox{.}(2012)Richardson, Zheng, Chatterjee, Nagai,
  \& Shen}]{2012ApJ...755...30R}
Richardson J., Zheng Z., Chatterjee S., Nagai D., Shen Y., 2012, ApJ, 755, 30

\bibitem[{Ross {et~al}\mbox{.}(2011)Ross, Ho, Cuesta, Tojeiro, Percival, Wake,
  Masters, Nichol, Myers, de~Simoni, Seo, Hernandez-Monteagudo, Crittenden,
  Blanton, Brinkmann, da~Costa, Guo, Kazin, Maia, Maraston, Padmanabhan, Prada,
  Ramos, Sanchez, Schlafly, Schlegel, Schneider, Skibba, Thomas, Weaver, White,
  \& Zehavi}]{2011MNRAS.417.1350R}
Ross A.~J. {et~al.}, 2011, MNRAS, 417, 1350

\bibitem[{Ross {et~al}\mbox{.}(2009)Ross, Shen, Strauss, Vanden~Berk, Connolly,
  Richards, Schneider, Weinberg, Hall, Bahcall, \&
  Brunner}]{2009ApJ...697.1634R}
Ross N.~P. {et~al.}, 2009, ApJ, 697, 1634

\bibitem[{Sabra {et~al}\mbox{.}(2015)Sabra, Saliba, Akl, \&
  Chahine}]{2015arXiv150200775S}
Sabra B.~M., Saliba C., Akl M.~A., Chahine G., 2015, arXiv, 775

\bibitem[{Sanders {et~al}\mbox{.}(1988)Sanders, Soifer, Elias, Madore,
  Matthews, Neugebauer, \& Scoville}]{1988ApJ...325...74S}
Sanders D.~B., Soifer B.~T., Elias J.~H., Madore B.~F., Matthews K., Neugebauer
  G., Scoville N.~Z., 1988, ApJ, 325, 74

\bibitem[{Satyapal {et~al}\mbox{.}(2014)Satyapal, Ellison, McAlpine, Hickox,
  Patton, \& Mendel}]{2014MNRAS.441.1297S}
Satyapal S., Ellison S.~L., McAlpine W., Hickox R.~C., Patton D.~R., Mendel
  J.~T., 2014, MNRAS, 441, 1297

\bibitem[{Schlegel, Finkbeiner \& Davis(1998)Schlegel, Finkbeiner, \&
  Davis}]{1998ApJ...500..525S}
Schlegel D.~J., Finkbeiner D.~P., Davis M., 1998, ApJ, 500, 525

\bibitem[{Scranton {et~al}\mbox{.}(2002)Scranton, Johnston, Dodelson, Frieman,
  Connolly, Eisenstein, Gunn, Hui, Jain, Kent, Loveday, Narayanan, Nichol,
  O'Connell, Scoccimarro, Sheth, Stebbins, Strauss, Szalay, Szapudi, Tegmark,
  Vogeley, Zehavi, Annis, Bahcall, Brinkman, Csabai, Hindsley, Ivezi{\'c}, Kim,
  Knapp, Lamb, Lee, Lupton, McKay, Munn, Peoples, Pier, Richards, Rockosi,
  Schlegel, Schneider, Stoughton, Tucker, Yanny, \& York}]{2002ApJ...579...48S}
Scranton R. {et~al.}, 2002, ApJ, 579, 48

\bibitem[{Secrest {et~al}\mbox{.}(2015)Secrest, Dudik, Dorland, Zacharias,
  Makarov, Fey, Frouard, \& Finch}]{Secrest:2015tt}
Secrest N., Dudik R., Dorland B., Zacharias N., Makarov V., Fey A., Frouard J.,
  Finch C., 2015, arXiv

\bibitem[{Seljak \& Zaldarriaga(1999)}]{1999PhRvL..82.2636S}
Seljak U., Zaldarriaga M., 1999, Physical Review Letters, 82, 2636

\bibitem[{Setti \& Woltjer(1989)}]{1989A&A...224L..21S}
Setti G., Woltjer L., 1989, A{\&}A, 224, L21

\bibitem[{Shankar {et~al}\mbox{.}(2006)Shankar, Shankar, Lapi, Lapi, Salucci,
  Salucci, De~Zotti, De~Zotti, Danese, \& Danese}]{2006ApJ...643...14S}
Shankar F. {et~al.}, 2006, ApJ, 643, 14

\bibitem[{Shen {et~al}\mbox{.}(2013)Shen, McBride, White, Zheng, Myers, Guo,
  Kirkpatrick, Padmanabhan, Parejko, Ross, Schlegel, Schneider, Streblyanska,
  Swanson, Zehavi, Pan, Bizyaev, Brewington, Ebelke, Malanushenko,
  Malanushenko, Oravetz, Simmons, \& Snedden}]{2013ApJ...778...98S}
Shen Y. {et~al.}, 2013, ApJ, 778, 98

\bibitem[{Shen {et~al}\mbox{.}(2007)Shen, Strauss, Oguri, Hennawi, Fan,
  Richards, Hall, Gunn, Schneider, Szalay, Thakar, Vanden~Berk, Anderson,
  Bahcall, Connolly, \& Knapp}]{2007AJ....133.2222S}
Shen Y. {et~al.}, 2007, AJ, 133, 2222

\bibitem[{Shen {et~al}\mbox{.}(2009)Shen, Strauss, Ross, Hall, Lin, Richards,
  Schneider, Weinberg, Connolly, Fan, Hennawi, Shankar, Vanden~Berk, Bahcall,
  \& Brunner}]{2009ApJ...697.1656S}
Shen Y. {et~al.}, 2009, ApJ, 697, 1656

\bibitem[{Sherwin {et~al}\mbox{.}(2012)Sherwin, Das, Hajian, Addison, Bond,
  Crichton, Devlin, Dunkley, Gralla, Halpern, Hill, Hincks, Hughes,
  Huffenberger, Hlozek, Kosowsky, Louis, Marriage, Marsden, Menanteau, Moodley,
  Niemack, Page, Reese, Sehgal, Sievers, Sif{\'o}n, Spergel, Staggs, Switzer,
  \& Wollack}]{2012PhRvD..86h3006S}
Sherwin B.~D. {et~al.}, 2012, PhRvD, 86, 83006

\bibitem[{Sheth, Mo \& Tormen(2001)Sheth, Mo, \& Tormen}]{2001MNRAS.323....1S}
Sheth R.~K., Mo H.~J., Tormen G., 2001, MNRAS, 323, 1

\bibitem[{Smith, Koss \& Mushotzky(2014)Smith, Koss, \&
  Mushotzky}]{Smith:2014vd}
Smith K.~L., Koss M., Mushotzky R.~F., 2014, arXiv

\bibitem[{Song {et~al}\mbox{.}(2003)Song, Song, Cooray, Cooray, Knox, Knox,
  Zaldarriaga, \& Zaldarriaga}]{2003ApJ...590..664S}
Song Y.-S., Song Y.-S., Cooray A., Cooray A., Knox L., Knox L., Zaldarriaga M.,
  Zaldarriaga M., 2003, ApJ, 590, 664

\bibitem[{Stern {et~al}\mbox{.}(2012)Stern, Assef, Benford, Blain, Cutri, Dey,
  Eisenhardt, Griffith, Jarrett, Lake, Masci, Petty, Stanford, Tsai, Wright,
  Yan, Harrison, \& Madsen}]{2012ApJ...753...30S}
Stern D. {et~al.}, 2012, ApJ, 753, 30

\bibitem[{Stern {et~al}\mbox{.}(2005)Stern, Eisenhardt, Gorjian, Kochanek,
  Caldwell, Eisenstein, Brodwin, Brown, Cool, Dey, Green, Jannuzi, Murray,
  Pahre, \& Willner}]{Stern:2005p2563}
Stern D. {et~al.}, 2005, ApJ, 631, 163

\bibitem[{Stoughton {et~al}\mbox{.}(2002)Stoughton, Lupton, Bernardi, Blanton,
  Burles, Castander, Connolly, Eisenstein, Frieman, Hennessy, Hindsley,
  Ivezi{\'c}, Kent, Kunszt, Lee, Meiksin, Munn, Newberg, Nichol, Nicinski,
  Pier, Richards, Richmond, Schlegel, Smith, Strauss, SubbaRao, Szalay, Thakar,
  Tucker, Vanden~Berk, Yanny, Adelman, Anderson, Anderson, Annis, Bahcall,
  Bakken, Bartelmann, Bastian, Bauer, Berman, B{\"o}hringer, Boroski, Bracker,
  Briegel, Briggs, Brinkmann, Brunner, Carey, Carr, Chen, Christian, Colestock,
  Crocker, Csabai, Czarapata, Dalcanton, Davidsen, Davis, Dehnen, Dodelson,
  Doi, Dombeck, Donahue, Ellman, Elms, Evans, Eyer, Fan, Federwitz, Friedman,
  Fukugita, Gal, Gillespie, Glazebrook, Gray, Grebel, Greenawalt, Greene, Gunn,
  de~Haas, Haiman, Haldeman, Hall, Hamabe, Hansen, Harris, Harris, Harvanek,
  Hawley, Hayes, Heckman, Helmi, Henden, Hogan, Hogg, Holmgren, Holtzman,
  Huang, Hull, Ichikawa, Ichikawa, Johnston, Kauffmann, Kim, Kimball, Kinney,
  Klaene, Kleinman, Klypin, Knapp, Korienek, Krolik, Kron, Krzesinski, Lamb,
  Leger, Limmongkol, Lindenmeyer, Long, Loomis, Loveday, MacKinnon, Mannery,
  Mantsch, Margon, McGehee, McKay, McLean, Menou, Merelli, Mo, Monet, Nakamura,
  Narayanan, Nash, Neilsen, Newman, Nitta, Odenkirchen, Okada, Okamura,
  Ostriker, Owen, Pauls, Peoples, Peterson, Petravick, Pope, Pordes, Postman,
  Prosapio, Quinn, Rechenmacher, Rivetta, Rix, Rockosi, Rosner, Ruthmansdorfer,
  Sandford, Schneider, Scranton, Sekiguchi, Sergey, Sheth, Shimasaku, Smee,
  Snedden, Stebbins, Stubbs, Szapudi, Szkody, Szokoly, Tabachnik, Tsvetanov,
  Uomoto, Vogeley, Voges, Waddell, Walterbos, Wang, Watanabe, Weinberg, White,
  White, Wilhite, Wolfe, Yasuda, York, Zehavi, \& Zheng}]{2002AJ....123..485S}
Stoughton C. {et~al.}, 2002, AJ, 123, 485

\bibitem[{Swanson {et~al}\mbox{.}(2008)Swanson, Tegmark, Hamilton, \&
  Hill}]{2008MNRAS.387.1391S}
Swanson M. E.~C., Tegmark M., Hamilton A. J.~S., Hill J.~C., 2008, MNRAS, 387,
  1391

\bibitem[{Tinker {et~al}\mbox{.}(2010)Tinker, Robertson, Kravtsov, Klypin,
  Warren, Yepes, \& Gottlober}]{2010ApJ...724..878T}
Tinker J.~L., Robertson B.~E., Kravtsov A.~V., Klypin A., Warren M.~S., Yepes
  G., Gottlober S., 2010, ApJ, 724, 878

\bibitem[{Totsuji \& Kihara(1969)}]{1969PASJ...21..221T}
Totsuji H., Kihara T., 1969, PASJ, 21, 221

\bibitem[{Vale \& Ostriker(2004)}]{2004MNRAS.353..189V}
Vale A., Ostriker J.~P., 2004, MNRAS, 353, 189

\bibitem[{van Engelen {et~al}\mbox{.}(2012)van Engelen, Keisler, Zahn, Aird,
  Benson, Bleem, Carlstrom, Chang, Cho, Crawford, Crites, de~Haan, Dobbs,
  Dudley, George, Halverson, Holder, Holzapfel, Hoover, Hou, Hrubes, Joy, Knox,
  Lee, Leitch, Lueker, Luong-Van, McMahon, Mehl, Meyer, Millea, Mohr, Montroy,
  Natoli, Padin, Plagge, Pryke, Reichardt, Ruhl, Sayre, Schaffer, Shaw,
  Shirokoff, Spieler, Staniszewski, Stark, Story, Vanderlinde, Vieira, \&
  Williamson}]{2012ApJ...756..142V}
van Engelen A. {et~al.}, 2012, ApJ, 756, 142

\bibitem[{Volonteri, Natarajan \& G{\"u}ltekin(2011)Volonteri, Natarajan, \&
  G{\"u}ltekin}]{2011ApJ...737...50V}
Volonteri M., Natarajan P., G{\"u}ltekin K., 2011, ApJ, 737, 50

\bibitem[{Warren \& Hoffleit(1987)}]{1987BAAS...19..733W}
Warren W. H.~J., Hoffleit D., 1987, Bulletin of the American Astronomical
  Society, 19, 733

\bibitem[{Werner {et~al}\mbox{.}(2004)Werner, Roellig, Low, Rieke, Rieke,
  Hoffmann, Young, Houck, Brandl, Fazio, Hora, Gehrz, Helou, Soifer, Stauffer,
  Keene, Eisenhardt, Gallagher, Gautier, Irace, Lawrence, Simmons, Van~Cleve,
  Jura, Wright, \& Cruikshank}]{2004ApJS..154....1W}
Werner M.~W. {et~al.}, 2004, The Astrophysical Journal Supplement Series, 154,
  1

\bibitem[{White {et~al}\mbox{.}(2011)White, Blanton, Bolton, Schlegel, Tinker,
  Berlind, da~Costa, Kazin, Lin, Maia, McBride, Padmanabhan, Parejko, Percival,
  Prada, Ramos, Sheldon, de~Simoni, Skibba, Thomas, Wake, Zehavi, Zheng,
  Nichol, Schneider, Strauss, Weaver, \& Weinberg}]{2011ApJ...728..126W}
White M. {et~al.}, 2011, ApJ, 728, 126

\bibitem[{White {et~al}\mbox{.}(2012)White, Myers, Ross, Schlegel, Hennawi,
  Shen, McGreer, Strauss, Bolton, Bovy, Fan, Miralda-Escude,
  Palanque-Delabrouille, Paris, Petitjean, Schneider, Viel, Weinberg, Yeche,
  Zehavi, Pan, Snedden, Bizyaev, Brewington, Brinkmann, Malanushenko,
  Malanushenko, Oravetz, Simmons, Sheldon, \& Weaver}]{2012MNRAS.424..933W}
White M. {et~al.}, 2012, MNRAS, 424, 933

\bibitem[{Wright {et~al}\mbox{.}(2010)Wright, Eisenhardt, Mainzer, Ressler,
  Cutri, Jarrett, Kirkpatrick, Padgett, McMillan, Skrutskie, Stanford, Cohen,
  Walker, Mather, Leisawitz, Gautier, McLean, Benford, Lonsdale, Blain, Mendez,
  Irace, Duval, Liu, Royer, Heinrichsen, Howard, Shannon, Kendall, Walsh,
  Larsen, Cardon, Schick, Schwalm, Abid, Fabinsky, Naes, \&
  Tsai}]{2010AJ....140.1868W}
Wright E.~L. {et~al.}, 2010, AJ, 140, 1868

\bibitem[{York {et~al}\mbox{.}(2000)York, Adelman, Anderson, Anderson, Annis,
  Bahcall, Bakken, Barkhouser, Bastian, Berman, Boroski, Bracker, Briegel,
  Briggs, Brinkmann, Brunner, Burles, Carey, Carr, Castander, Chen, Colestock,
  Connolly, Crocker, Csabai, Czarapata, Davis, Doi, Dombeck, Eisenstein,
  Ellman, Elms, Evans, Fan, Federwitz, Fiscelli, Friedman, Frieman, Fukugita,
  Gillespie, Gunn, Gurbani, de~Haas, Haldeman, Harris, Hayes, Heckman,
  Hennessy, Hindsley, Holm, Holmgren, Huang, Hull, Husby, Ichikawa, Ichikawa,
  Ivezi{\'c}, Kent, Kim, Kinney, Klaene, Kleinman, Kleinman, Knapp, Korienek,
  Kron, Kunszt, Lamb, Lee, Leger, Limmongkol, Lindenmeyer, Long, Loomis,
  Loveday, Lucinio, Lupton, MacKinnon, Mannery, Mantsch, Margon, McGehee,
  McKay, Meiksin, Merelli, Monet, Munn, Narayanan, Nash, Neilsen, Neswold,
  Newberg, Nichol, Nicinski, Nonino, Okada, Okamura, Ostriker, Owen, Pauls,
  Peoples, Peterson, Petravick, Pier, Pope, Pordes, Prosapio, Rechenmacher,
  Quinn, Richards, Richmond, Rivetta, Rockosi, Ruthmansdorfer, Sandford,
  Schlegel, Schneider, Sekiguchi, Sergey, Shimasaku, Siegmund, Smee, Smith,
  Snedden, Stone, Stoughton, Strauss, Stubbs, SubbaRao, Szalay, Szapudi,
  Szokoly, Thakar, Tremonti, Tucker, Uomoto, Vanden~Berk, Vogeley, Waddell,
  Wang, Watanabe, Weinberg, Yanny, Yasuda, \&
  collaboration}]{2000AJ....120.1579Y}
York D.~G. {et~al.}, 2000, AJ, 120, 1579

\bibitem[{Zheng {et~al}\mbox{.}(2005)Zheng, Berlind, Weinberg, Benson, Baugh,
  Cole, Dav{\'e}, Frenk, Katz, \& Lacey}]{2005ApJ...633..791Z}
Zheng Z. {et~al.}, 2005, ApJ, 633, 791

\bibitem[{Zheng \& Weinberg(2007)}]{2007ApJ...659....1Z}
Zheng Z., Weinberg D.~H., 2007, ApJ, 659, 1

\end{thebibliography}
